\documentclass[useAMS,usenatbib]{mn2e}

\usepackage{natbib}
\usepackage{graphicx}
\bibpunct{(}{)}{;}{a}{}{,}
\usepackage{txfonts}
\usepackage[T1]{fontenc} 
\usepackage{aecompl}

\title[Deuteration in IRAS16293-2422]{Chemical modeling of water deuteration in IRAS16293-2422}
\author[V. Wakelam et al.]{
V. Wakelam$^{1,2}$\thanks{E-mail:
wakelam@obs.u-bordeaux1.fr}, C. Vastel$^{3,4}$, Y. Aikawa$^{5}$, A. Coutens$^{6,7}$, S. Bottinelli$^{3,4}$, E. Caux$^{3,4}$\\
$^{1}$Univ. Bordeaux, LAB, UMR 5804, F-33270, Floirac, France\\
$^{2}$CNRS, LAB, UMR 5804, F-33270, Floirac, France\\
$^3$ Universit\'e de Toulouse; UPS-OMP; Institut de Recherche en Astrophysique et Plan\'etologie (IRAP) ; UMR 5277 ; Toulouse, France\\
$^4$ CNRS; IRAP; UMR 5277 ; 9 Av. colonel Roche, BP 44346, F-31028 Toulouse cedex 4, France\\
$^5$ Department of Earth and Planetary Sciences, Kobe University, Kobe 657-8501, Japan\\
$^6$ Niels Bohr Institute, University of Copenhagen, Juliane Maries Vej 30, DK-2100 Copenhagen {\O}, Denmark \\
$^7$ Centre for Star and Planet Formation and Natural History Museum of Denmark, University of Copenhagen,  {\O}ster Voldgade 5-7, DK-1350 Copenhagen K., Denmark 
}
\begin{document}

\date{xx}

\pagerange{\pageref{firstpage}--\pageref{lastpage}} \pubyear{2002}

\maketitle

\label{firstpage}

\begin{abstract}

IRAS 16293-2422 is a well studied low-mass protostar characterized by a strong level of deuterium fractionation. In the line of sight of the protostellar envelope, an additional absorption layer, rich in singly and doubly deuterated water has been discovered by a detailed multiline analysis of HDO. To model the chemistry in this source, the gas-grain chemical code Nautilus has been used with an extended deuterium network. For the protostellar envelope, we solve the chemical reaction network in infalling fluid parcels in a protostellar core model. For the foreground cloud, we explored several physical conditions (density, cosmic ionization rate, C/O ratio). \\
The main results of the paper are that gas-phase abundances of H$_2$O, HDO and D$_2$O observed in the inner regions of IRAS16293-2422 are lower than those predicted by a 1D dynamical/chemical (hot corino) model in which the ices are fully evaporated. The abundance in the outer part of the envelope present chaotic profiles due to adsorption/evaporation competition, very different from the constant abundance assumed for the analysis of the observations.  We also found that the large abundances of gas-phase H$_2$O, HDO and D$_2$O observed in the absorption layer are more likely explained by exothermic surface reactions rather than photodesorption processes.

\end{abstract}


\section{Introduction}

Water in the interstellar medium has been widely observed and analyzed in the past few years thanks to the success of the Herschel Space Observatory \citep{2010A&A...518L...1P} and the high spectral resolution of its Heterodyne Instrument for the Far-Infrared \citep{2010A&A...518L...6D}. It has been detected in various environments like comets \citep{2011Natur.478..218H,2012A&A...544L..15B}, low-mass star-forming regions \citep{2012A&A...542A...8K}, high-mass star-forming regions \citep{2013A&A...554A..83V}, protoplanetary disks \citep{2013ApJ...766L...5P} and one prestellar core \citep{2012ApJ...759L..37C}.  See \citet{2013ChRv..113.9043V} for a full review on interstellar water. Water can be formed in the gas-phase, mostly by ion-molecular reactions starting from the atomic oxygen and ending with the dissociative recombination of H$_3$O$^+$ in dense regions but it is most efficiently formed at the surface of interstellar grains by the successive hydrogenation of atomic and molecular oxygen \citep{2008CPL...456...27M,2010A&A...512A..30D,2010PCCP...1212065I,2010PCCP...1212077C}. Water ice formation starts at a threshold extinction of A$_V$ $\sim$ 3-5 magnitudes, well before the cloud collapses \citep{1993dca..book....9W,2003A&G....44b..35H}. \\
In very cold regions, a small fraction of the H$_2$O ice formed by surface reactions can be desorbed back into the gas phase through non-thermal processes.
For example the cosmic rays have the ability to penetrate even the densest clouds and therefore maintain a low level of UV radiation by interacting with molecular hydrogen. The indirect cosmic ray induced desorption is then an important mechanism \citep{1985A&A...144..147L,1993MNRAS.261...83H}. In the same way, the far UV starlight pervading interstellar space photodesorbs water molecules from grains in dense regions well below the evaporation temperature  \citep{2009ApJ...690.1497H}. 
The efficiency of the photodesorption has been quantified by laboratory experiments to be $\sim 10^{-3}$ per incident UV photon with a dependence on the ice thickness and temperature \citep{2009ApJ...693.1209O}. Similar results were found theoretically by \citet{2006JChPh.124f4715A} and \citet{2008A&A...491..907A}. In particular, the detection of water vapor inside pre-stellar cores seems to indicate that the photo evaporation processes of the ices by the cosmic-ray induced FUV photons can be efficient \citep{2012ApJ...759L..37C}. \\
The start of the collapse of a cloud triggers the increase of luminosity at the centre of the protostellar object, which eventually increases the temperatures well above 100~K, then H$_2$O ices are thermally desorbed on very short timescales \citep{2001MNRAS.327.1165F}. This evaporation results in {\bf theoretical} gas-phase abundances of H$_2$O as high as $\sim$ 10$^{-4}$, relative to hydrogen. Together with water, other species desorb at the same time characterizing these hot cores (in the case of high-mass young stellar objects) or hot corinos (in the case of low-mass young stellar objects). Other regions in star forming regions associated with shocks (created by the interaction of molecular outflows with the quiescent surrounding cloud) show large enhancement of H$_2$O abundances due to temperatures of several thousand degrees Kelvin. 
Detection of cold water vapor in a disk around the young star TW Hydrae suggests a water ice reservoir equivalent up to several thousand Earth oceans in mass, with implications on the origin of our Earth's oceans \citep{2011Sci...334..338H}.  \\
Water deuteration is unique compared to other species. 
In the first stages of star formation, low temperatures and the disappearance of most molecules, and particularly of CO, from the gas phase trigger a peculiar chemistry \citep{1989ApJ...340..906M}. An extreme molecular deuteration has been extensively discovered over the past 10 years in dense prestellar cores and low-mass protostars, with deuteration fractionation as high as 100\% (see Ceccarelli et al. 2014, PP6, for a review). The D/H ratio of interstellar water ice is, however, not well constrained. HDO ice has not been detected, and upper limits in the range from 0.002 to 0.02 have been estimated in protostar samples \citep{2003A&A...399.1009D,2003A&A...410..897P,2012A&A...538A..57A}. Since water ice is sublimated to the gas phase in hot corinos, gas-phase HDO has been extensively investigated in recent years \citep{2012A&A...539A.132C,2013A&A...550A.127T,2010ApJ...725L.172J}. The deuteration fraction of water is much lower (less than 1$\%$) for both high-mass and low-mass protostars than for other species that show sometimes a D/H ratio several orders of magnitude higher compared to the cosmic value. Furthermore, no HDO has been detected yet in the prestellar core phase, although many deuterated species appear with a high deuteration fractionation \citep{2003cdsf.conf....3P,2004ApJ...606L.127V}. {\bf If deuterated water is mostly formed by surface reactions, the level of deuteration of water may be much lower compared to other species formed in the gas-phase in cold dense clouds \citep{2003ApJ...591L..41R}. Comparing the HDO/H$_2$O ratios observed in protostellar envelopes to the ones in comets and the Earth's oceans (with a Vienna Standard Mean Ocean Water value of $1.558\times10^{-4}$) could bring new insight on the origin of water and life on planets.  }
The recent observations of a Jupiter-family comet is consistent with the VSMOW value, suggesting that the ocean water could have been delivered by these external sources. \\

\begin{table*}
\caption{Observed abundance of H$_2$O, HDO and D$_2$O towards the line of sight of IRAS16293 from \citet{2013A&A...553A..75C}.}
\begin{center}
\begin{tabular}{l|c|c|c}
\hline
\hline
& Hot corino & Outer envelope & foreground cloud$^*$ \\
& (r $\le 75$~AU) &  (r $> 75$~AU) & (A$_V$ = 4 - 3)\\
\hline
H$_2$O & $4.7-40 \times 10^{-6}$ & $7.0-22.5 \times 10^{-9}$ & $1.3-1.8\times 10^{-7}$\\
HDO &  $1.4-2.4 \times 10^{-7}$ & $5.5-10.6 \times 10^{-11}$ & $6-8\times 10^{-9}$\\
D$_2$O &  $\le 1.3\times 10^{-9}$ & $\le 1.3 \times 10^{-11}$ & $6.6-8.9\times 10^{-10}$\\
HDO/H$_2$O & $4\times 10^{-3} - 5.1\times 10^{-2}$ & $3\times 10^{-3} - 1.5\times 10^{-2}$ & $\sim 4\times 10^{-2}$ \\
D$_2$O/HDO & $\le 9\times 10^{-3}$ & $\le 0.23$ & $\sim 0.11$ \\
D$_2$O/H$_2$O & $\le 3\times 10^{-4}$ & $\le 2\times 10^{-3}$ & $\sim 5\times 10^{-3}$ \\
\hline
\end{tabular}
\\
$^*$ The abundances are given for A$_V$ = 4 and 3 respectively.
\end{center}
\label{obs_tab}
\end{table*}%

We concentrate, in this paper, on water vapor in the IRAS 16293-2422 low-mass protostar (hereafter IRAS16293), which remains the first source in which doubly-deuterated water has been detected \citep{1997AAS...191.2007B,2010A&A...521L..31V} and towards which the most complete study of water and deuterated water has been performed to date \citep{2012A&A...539A.132C,2013A&A...553A..75C}. Observed data for 16 HDO, 7 para-H$_2^{18}$O, 8 ortho-H$_2^{18}$O, 1 ortho- and 1 para- H$_2^{17}$O, 1 HD$^{18}$O, 1 ortho-D$_2$O and 2 para-D$_2$O transitions have been used simultaneously with the spherical Monte Carlo radiative transfer code RATRAN \citep{2000A&A...362..697H} assuming an abundance jump at 100 K \citep{2001MNRAS.327.1165F}, which corresponds to an approximate radius of 70~AU. A water-rich absorbing layer is required in front of the envelope to reproduce the HDO absorption components observed at 465 and 894 GHz. \citet{2012A&A...539A.132C} proposed that this layer  results from the photodesorption of water ice induced by the external UV radiation field, as predicted by \citet{2009ApJ...690.1497H} in molecular clouds at a visual extinction, A$_V$, between 1 and 4. The only physical constraints on this foreground cloud is that the temperature is lower than 30~K and the H$_2$ density lower than $10^5$~cm$^{-3}$. The observed abundances in the different components in the line of sight of the protostar are summarized in Table~\ref{obs_tab}. Note that for the foreground cloud, we have only reported abundances computed assuming A$_V$ of 3 and 4 since we show in this paper that at smaller A$_V$ the model predicts very low abundances. \\
The present paper aims at studying the deuteration of water in the envelope of IRAS16293 and the foreground cloud in order to reproduce the observed large abundances of deuterated water. The model (with a different network) and approach used in this paper are the same as in \citet{2014MNRAS.441.1964B} for the analysis of CH lines in the line of sight of IRAS16293. The modeling for the two different components is presented in separate sections: Section~\ref{proto} for the protostellar envelope and Section~\ref{cloud} for the foreground cloud. In each of the two sections, the model is first presented. Then the model results are shown with some studies of the results sensitivity to the model parameters.  In Section~\ref{discussion}, we discuss the effect of a number of parameters and the efficiency of the photo evaporation processes in the foreground cloud. We conclude in the last section.

\section{The protostellar envelope}\label{proto}

\subsection{Chemical and physical model description}

\subsubsection{The Nautilus chemical model}

To theoretically study the deuteration of water in the envelope of IRAS16293, the chemical model Nautilus has been used with the spherical protostellar core model similar to \citet{2008ApJ...674..984A,2012ApJ...760...40A} (see section~\ref{physics}). Nautilus is a chemical model that computes the evolution of the species abundances as a function of time in the gas-phase and at grain surfaces. The code has been used for a variety of environments such as dense clouds \citep{2014MNRAS.437..930L}, low mass protostellar envelopes \citep{2014MNRAS.441.1964B} and the outer regions of protoplanetary disks \citep{2011A&A...535A.104D}. A large number of gas-phase processes are included in the network: bimolecular reactions (between neutral species, between charged species and between neutral and charged species) and unimolecular reactions, i.e. photo reactions with direct UV photons and UV photons produced by the deexcitation of H$_2$ excited by cosmic-ray particles (Pratap \& Tarafdar mechanism), and direct ionization and dissociation by cosmic-ray particles. The interactions of the gas-phase species with the interstellar grains are: sticking of neutral gas-phase species to the grain surfaces, evaporation of the species from the surfaces due to the temperature, the cosmic-ray heating and the exothermicity of the reactions at the surface of the grains. The species can diffuse and undergo reactions using the rate equation approximation at the surface of the grains \citep{1992ApJS...82..167H}. Details on the processes included in the model can be found in \citet{2010A&A...522A..42S} and \citet{2014MNRAS.437..930L}. 

\subsubsection{The deuterated network}\label{deut_network}

 We adopt the reaction network model of \citet{2012ApJ...760...40A}, which we briefly describe in the following.
\citet{2012ApJ...760...40A} extended the gas-grain reaction network of \citet{2006A&A...457..927G} to include
mono-, doubly-, and triply-deuterated species. The reaction rate coefficients of deuterated species are set
to be the same as that of the original reactions \citep{2006A&A...457..927G}.
If there are more than one set of possible products (e.g. XH+YD and XD+YH), statistical branching ratio is assumed.
Reactions listed in \citet{1989ApJ...340..906M} and \citet{2004A&A...424..905R} are included in the network;
these are mostly exchange reactions that trigger the isotopic fractionation.

At temperatures up to several tens of Kelvin, deuterium fractionation is mainly triggered by an exothermic
exchange reaction H$_3^+$ + HD $\rightarrow$ H$_2$D$^+$ + H$_2$ and its multi-deuterated analogues. Since the reactions
are exothermic, the backward reaction is much less efficient, and H$_3^+$ is highly deuterated.
Since CO is the main  reactant of H$_2$D$^+$, CO depletion in prestellar cores further enhances the D/H ratio of H$_3^+$ \citep{2003ApJ...593L..97V}. However, such enhancement can be limited by ortho-H$_2$, which can react with the deuterated trihydrogen cation and remove the deuterium. Indeed, the 170K internal energy of the lowest ortho-H$_2$ level (J=1) is large compared to the temperatures considered in the cold medium and reactions which are endothermic with para-H$_2$ can have much smaller endothermicity with ortho-H$_2$. Considering the complexity of the network, we did not consider the ortho, para or meta forms of any species. We discuss the importance of this approximation in section~\ref{dis_opH2}.

In molecular clouds, water is mainly formed via three chemical paths. The first path is a series of ion-molecule reactions
and recombination:
\begin{equation}\label{ion_mol}
{\rm O} + {\rm H}_3^+ \rightarrow {\rm OH}^+ + {\rm H}_2 
\end{equation}
\begin{equation}
{\rm OH}^+ + {\rm H}_2 \rightarrow {\rm H}_2{\rm O}^+ + {\rm H} 
\end{equation}
\begin{equation}
{\rm H}_2{\rm O}^+ + {\rm H}_2 \rightarrow {\rm H}_3{\rm O}^+ + {\rm H}  
\end{equation}
\begin{equation}
{\rm H}_3{\rm O}^+ + {\rm e} \rightarrow {\rm H}_2{\rm O} + {\rm H}
\end{equation}
If deuterated H$_3^+$, e.g. H$_2$D$^+$, reacts with O atom in the first step (reaction \ref{ion_mol}), HDO can be
formed with a statistical branching ratio. The D/H ratio of water formed by this ion-molecule path is similar to
(although slightly smaller than) that of H$_3^+$.\\
The second path is a hydrogenation of O atom on grain surface:
\begin{equation}
{\rm O} + {\rm H} \rightarrow {\rm OH} 
\end{equation}
\begin{equation}
{\rm OH} + {\rm H} \rightarrow {\rm H}_2{\rm O}
\end{equation}
The recombination of deuterated H$_3^+$ enhances the atomic D/H ratio in the gas phase. These atoms then collide with
grains and participate in the hydrogenation on grain surfaces. Water formed on grains are thus deuterated as well.\\
The third path is neutral-neutral reactions in the gas-phase:
\begin{equation}
{\rm O} + {\rm H}_2 \rightarrow {\rm OH} + {\rm H}  
\end{equation}
\begin{equation}
{\rm OH} + {\rm H}_2 \rightarrow {\rm H}_2{\rm O} + {\rm H}
\end{equation} 
These reactions are endothermic by $\sim 3000$ K for the first one and exothermic but with an activation barrier of $\sim 1000$ K for the second one, and thus are efficient only at temperatures higher than a few hundred K. 

\subsubsection{The protostellar physical model}\label{physics}

\begin{figure}
\includegraphics[width=0.8\linewidth]{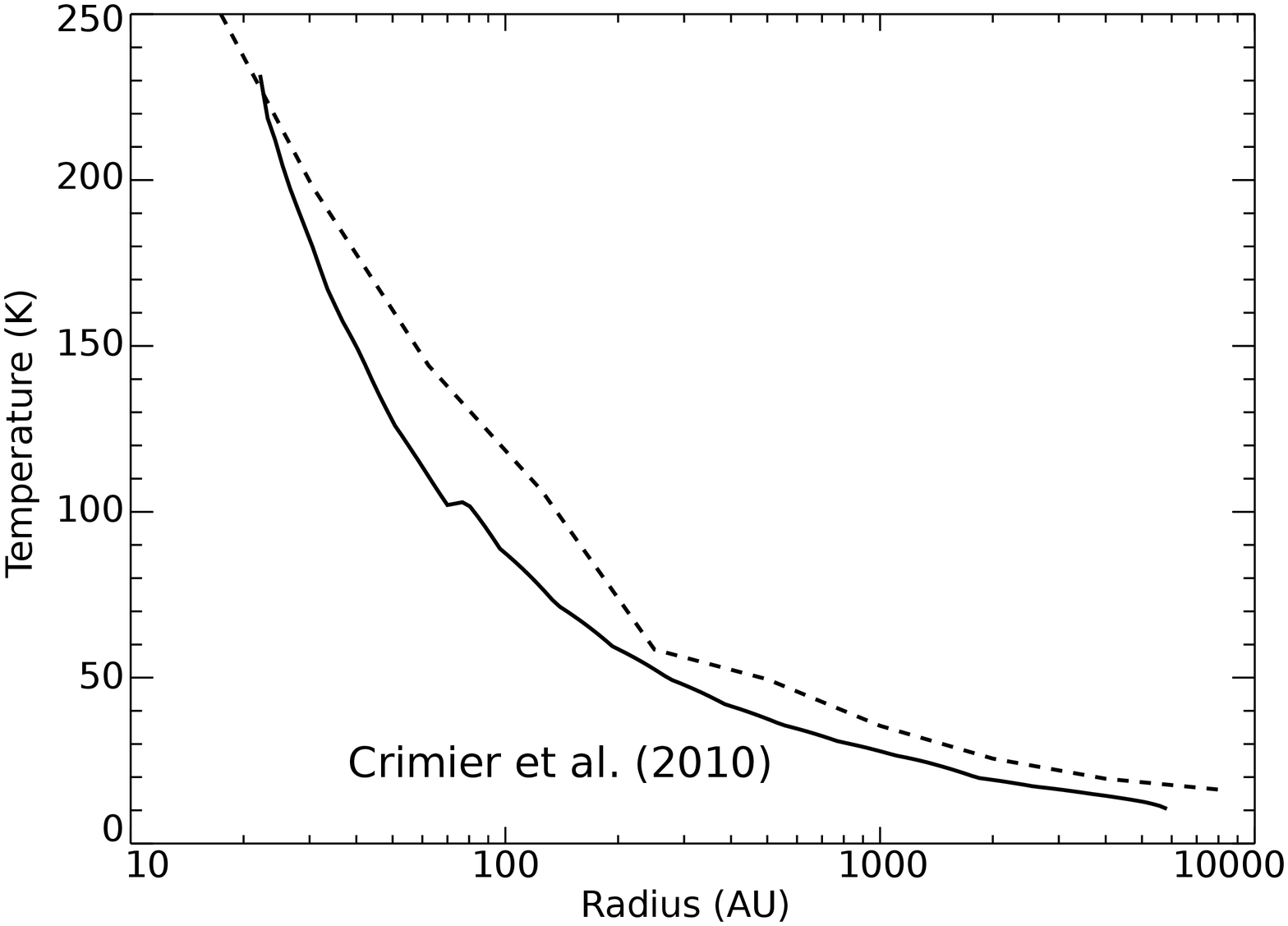}
\includegraphics[width=0.8\linewidth]{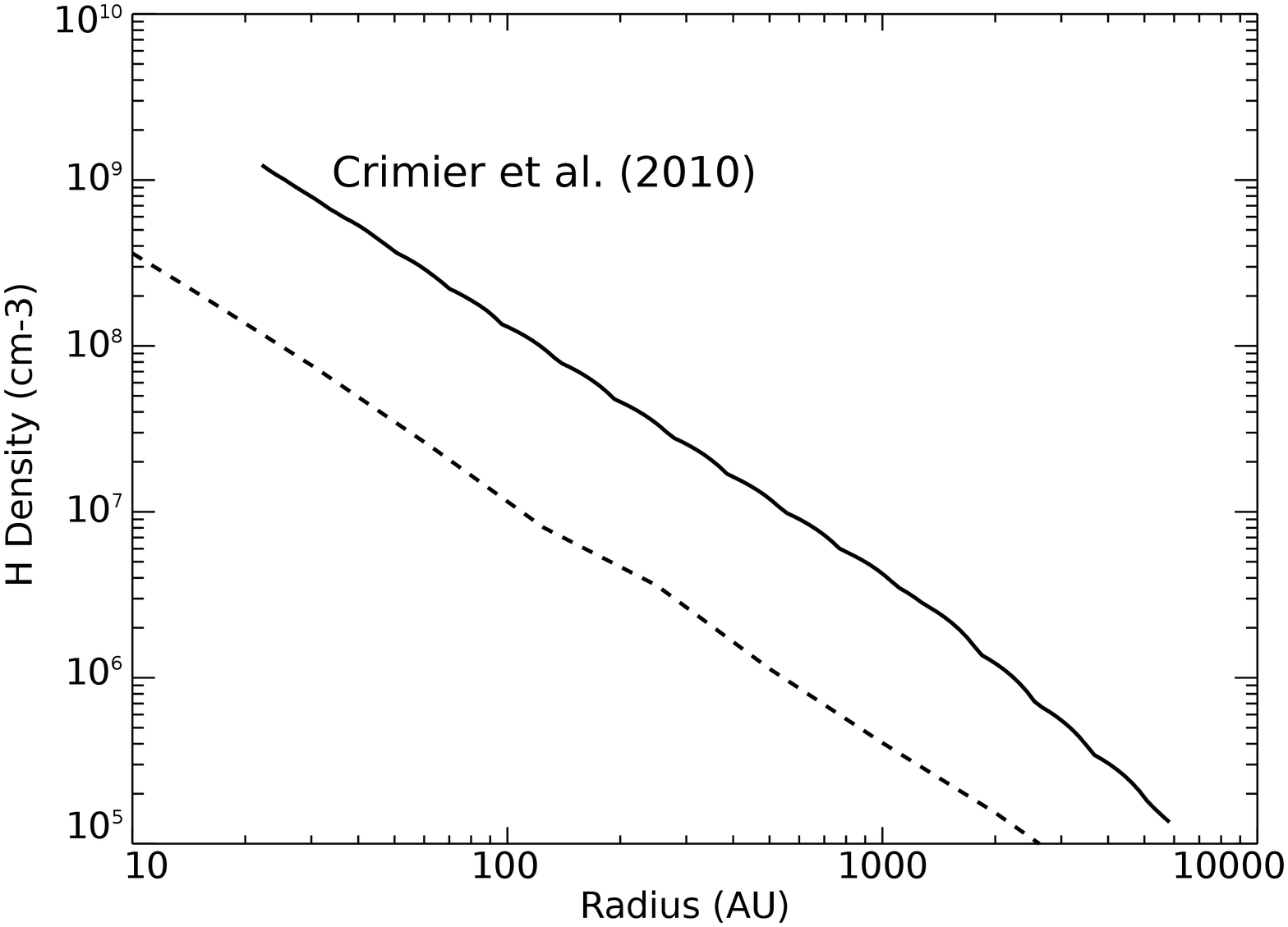}
\caption{Temperature and density profiles in the protostellar envelope. The dashed curve represents the protostellar core model from \citet{2000ApJ...531..350M} and the solid one the profiles derived by \citet{2010A&A...519A..65C} in the envelope of IRAS16293-2422. \label{comp_struct}}
\end{figure}

As in \citet{2008ApJ...674..984A,2012ApJ...760...40A}, the physical structure is based on the radiation hydrodynamical model from \citet{2000ApJ...531..350M}. The model starts from the dense molecular cloud core with a central density of 
$n$(H$_2$) $\sim 3 \times 10^4$ cm$^{-3}$. The core extends to $r=4 \times 10^4$ AU, and the total mass is 3.852 $M_{\odot}$.
The contraction is almost isothermal as long as the cooling is efficient.
Eventually the compressional heating overwhelms the cooling, and the temperature rises
in the central region. Then the core goes through the first core stage and second collapse to form a protostar at the center.
In the model we adopt, the prestellar core evolves to the protostellar core in $2.5 \times 10^5$ yr.
After the birth of the protostar, the model further follows the
evolution for $9.3 \times 10^4$ yr, during which the protostar grows by mass accretion from
the envelope.

The physical structure of the envelope at the final time of the simulation ($9.3\times 10^4$~yr) are similar to the one constrained in the envelope of IRAS16293 by \citet{2010A&A...519A..65C} from multi-wavelenght dust and molecular observations. Fig.~\ref{comp_struct} shows a comparison between the model of \citet{2000ApJ...531..350M} and observed temperature and density profiles as a function of radius. The temperature profiles are quite similar whereas the observed density profile is about ten times larger in the calculated one. To better match the observations and be consistent with the profile used to interpret the observations, we have multiplied by 10 all the densities in our simulations. The consequences of this modification of the physical model are discussed in section~\ref{disc_phys}.



\subsubsection{Model parameters}\label{proto_parameters}

The chemical composition is computed using the physical structure previously described. Each cell of material is assumed to be independent so that there is no mixing between cells. In total, there are 15 different cells that are allowed to dynamically evolve and which defines the physical and chemical structure of the envelope from 0.66 AU up to 8000 AU at the final time. As in \citet{2008ApJ...674..984A,2012ApJ...760...40A}, we assume that the object is embedded in a molecular cloud so that the minimum A$_V$ at the border of the envelope is approximately 4. To obtain the chemical composition prior to the collapse, we compute the chemistry, using the local physical conditions in a pre stellar dense core, for $10^6$~yr as in \citet{2008ApJ...674..984A}. Note that the density at this stage is also multiplied by a factor of 10. For this first step, the species are assumed to be initially in the atomic form, except for the hydrogen and the deuterium. The hydrogen is initially molecular H$_2$ and deuterium in the form of HD. The elemental abundances  used for the simulations are the same as in \citet{2011A&A...530A..61H}, with an oxygen elemental abundance of $3.3\times 10^{-4}$ (compared to the total H density, which gives a C/O elemental ratio of 0.5). 
  The deuterium elemental abundance is set to $1.5\times 10^{-5}$ compared to H, which is the average value found by absorption spectroscopy measurements towards stars in the vicinity of the Sun \citep[e.g.][]{2006ApJ...647.1106L}.
The cosmic-ray ionization rate is $1.3\times 10^{-17}$~s$^{-1}$, typical value used in chemical modeling and found in dense clouds \citep{ 1998ApJ...499..234C}. \\
In addition to this standard model, we have explored the importance of the three following parameters: the age of the parent cloud (Parameter 1: P1), the oxygen elemental abundance (P2) and the cosmic-ray ionization rate (P3). Instead of running the code for $10^6$~yr (P1) before the collapse, we run it for $10^4$ and $10^5$~yr. For the oxygen elemental abundance (P2), we also considered an abundance of $1.4\times 10^{-4}$ (corresponding to a C/O of 1.2). Finally, for the cosmic-ray ionization rate (P3), we used a ten times larger value of $10^{-16}$~s$^{-1}$, more typical of diffuse clouds observations \citep{2012ApJ...745...91I} but proposed by \citet{2004A&A...418.1021D} for the envelope of IRAS16293. All the parameters, which impact on the model predictions have been studied, are summarized in Table~\ref{explore_param}.


\begin{table*}
\caption{Parameters of the chemical model explored for the 1D model of the protostellar envelope and the 0D model of the foreground cloud}
\begin{center}
\begin{tabular}{lcc}
\hline\hline
\multicolumn{2}{c}{1D Modeling of the protostellar envelope} \\
\hline
\multicolumn{2}{c}{Parameters of the standard model} \\
Density structure & 10 $\times$ density computed by the MHD model \\
Free-fall timescale & Timescale computed by the MHD model \\
Time before collapse & $10^6$~yr \\
\hline
\multicolumn{2}{c}{Parameters tested} \\
Density structure & Density computed by the MHD model (section 4.5) \\
Free-fall timescale & Timescale computed by the MHD model / 3 (section 4.5) \\
Time before collapse & $10^4$ and $10^5$~yr (section 2.4) \\
\hline
\hline
\multicolumn{2}{c}{0D Modeling for the foreground cloud} \\
Gas and dust temperature & 15 and 30~K \\
H density & $2\times 10^4$ and $2\times 10^5$ cm$^{-3}$ \\
Visual extinction & 2, 3 and 4\\
\hline
\hline
\multicolumn{2}{c}{Parameters tested for both 0D and 1D modelings} \\
Cosmic-ray ionization rate & $10^{-17}$ and $10^{-16}$~s$^{-1}$ \\
C/O elemental ratio & 0.5 and 1.2 \\
Deuterium elemental abundance & $7.5\times 10^{-6}$, $1.5\times 10^{-5}$, $3\times 10^{-5}$ \\
Rate coefficient of the reaction H$_2$D$^+$ + H$_2$ $\rightarrow$ H$_3^+$ + HD & see section 4.4 \\
\hline
\end{tabular}
\end{center}
\label{explore_param}
\end{table*}%

\subsection{Modeling results}\label{proto_results}

\begin{figure}
\includegraphics[width=1\linewidth]{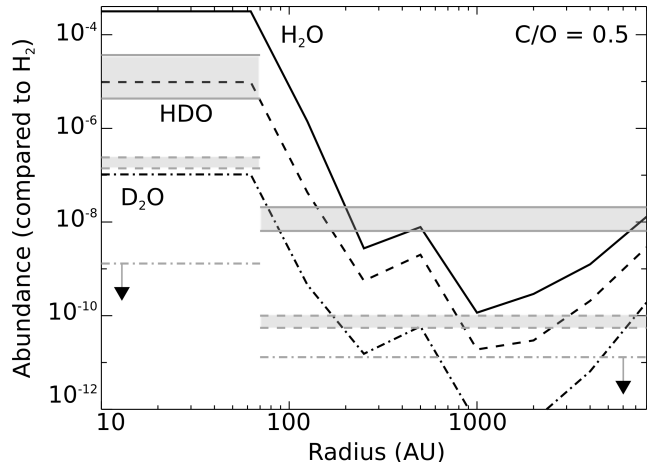}
\caption{Abundance profiles, in the gas-phase, of H$_2$O (solid line), HDO (dashed line) and D$_2$O (dash-dotted line) predicted by the standard model towards the envelope of IRAS16293. Grey lines and areas represent the observational constraints. Constraints on D$_2$O abundance are only upper limits.   \label{ab_proto_standard}}
\end{figure}

\begin{figure}
\includegraphics[width=1\linewidth]{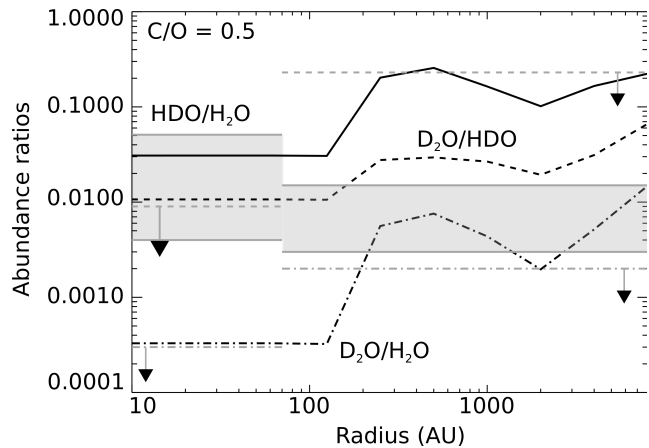}
\caption{HDO/H$_2$O (solid line), D$_2$O/H$_2$O (dash-dotted line) and D$_2$O/HDO (dashed line) abundance ratios as a function of radius towards the center of the protostellar envelope predicted by the standard model. Grey lines and areas represent the observational constraints (only upper limits for D$_2$O/HDO and D$_2$O/H$_2$O). \label{ratio_proto_standard}}
\end{figure}

 In this section, we present the predicted abundances for water and its deuterated forms in the envelope of IRAS16293 for the standard model, and compare them with the observed values. Note that the radius, at which the temperature is 100~K, defines the limit between the inner and outer regions of the protostar for the analysis of the observations. In \citet{2012A&A...539A.132C}, they have used the radius of 75~AU from \citet{2010A&A...519A..65C}. In our simulations, it would be 70~AU and we have chosen to use this value to plot the observed abundances on the figures.
Fig.~\ref{ab_proto_standard} shows the calculated abundances of H$_2$O, HDO and D$_2$O in the protostellar envelope as a function of radii, at the end of the simulations, i.e. for a protostellar age of $9.3\times 10^4$~yr. The gas-phase abundances of these species predicted by the model can be divided in three regions. The first one is defined by radii larger than 1000~AU and temperatures below 30~K. In this region, the gas-phase abundances decrease towards the center of the envelope due to the depletion of species because of the density increase. Between 1000 and about 70~AU, the gas-phase abundances increase. The cause of this increase is first the thermal desorption of precursors such as HDCO and H$_2$CO from the ices that react with OH and OD between 1000 and 250~AU (where the temperature is between 30 and 60~K). The second cause is the direct thermal evaporation of H$_2$O, HDO and D$_2$O from the mantles until complete evaporation at between 250 and 70~AU (where the temperature is between 60 and 100~K). 
The last region is defined by radii smaller than 70~AU where the gas-phase abundances are constant. In fact the inner abundances are equal to the abundances in the ices that are constant across the envelope before the collapse since H$_2$O formation and deuteration occurs at the surface of the grains during this phase. At the protostellar stage, the ice abundance of these species stays constant for radii larger than 70~AU. Contrary to \citet{2013A&A...558A.126M}, we have a smoother increase of the water gas-phase abundance towards the center of the protostar. The main difference with their work is that we are using a dynamical structure rather than a static one. In a static model, the gas and dust stay at the same radius and physical conditions for the entire time of the collapse. The depletion of molecules is then more important than in a dynamical model in which each cell of material stays less longer at high density. In addition, \citet{2013A&A...558A.126M} have used a simplified chemistry to compte the abundance of water, which very likely does not include the gas-phase formation of H$_2$O from H$_2$CO evaporated from the ices.  \\


\subsection{Comparison with the observations}

Compared to the observations \citep{2013A&A...553A..75C}, the most inner abundances are over-predicted by the chemical model, in particular for D$_2$O (see Fig.~\ref{ab_proto_standard}). This disagreement could be explained by the uncertainties on the structure of \citet{2010A&A...519A..65C}, which is not constrained at small scale ($\lesssim$ 500 AU). The abundances are however nicely reproduced for the three molecules for a radius of 90 to 100 AU, which could suggest that water is not totally desorbed in the inner regions. This could be expected in case of the presence of a disk, as most of the molecules are assumed to be trapped on the grain mantles in the cold midplane layer of the disk. In particular, \citet{2012A&A...544L...7P} suggested that a disk would surround the source A. The determination of the disk properties (H$_2$ density, temperature) would then be required in order that this 2D or 3D structure can be taken into account both to derive the observed abundances and to predict the chemical abundances. This would enable a more reliable comparison between the observations and the chemical predictions.
Note that the water abundances predicted by our chemical model in the gas-phase (which results for the ice evaporation) are in agreement with the abundances of H$_2$O observed in the ices of about $10^{-4}$ \citep{1996A&A...315L.333S}.
Fig.~\ref{ratio_proto_standard} shows the abundance ratios HDO/H$_2$O, D$_2$O/H$_2$O and D$_2$O/HDO predicted by the model as well as the observational constraints derived by \citet{2013A&A...553A..75C}. Abundance ratios in the inner region of the envelope are satisfactory reproduced by the model.
A low HDO/H$_2$O ratio (9.2 $\pm$ 2.6 $\times$ 10$^{-4}$) was however recently derived in the inner regions of IRAS16293, based on interferometric data of the HDO line at 226 GHz and the H$_2$$^{18}$O lines at 203 and 692 GHz \citep{2013A&A...549L...3P}, closer to the lower value of the ratio determined by Coutens et al. (see Table 1). The difference between the HDO/H$_2$O ratios derived by \citet{2013A&A...553A..75C} and \citet{2013A&A...549L...3P} could be due to the optically thickness of some H$_2^{18}$O lines observed with \textit{Herschel}/HIFI \citep{2013ApJ...769...19V}.  Other low-mass protostars (NGC1333 IRAS4A, IRAS4B and IRAS2A) seem to show a similar HDO/H$_2$O ratios of $\sim$ 10$^{-3}$ in the warm inner regions \citep{2013ApJ...769...19V,2013A&A...560A..39C,2014A&A...563A..74P}, although \citet{2013ApJ...768L..29T} do not exclude high HDO/H$_2$O ratios for these same sources. Such a low ratio cannot be reproduced by the chemical predictions except maybe if the ices are built during a translucent phase where the A$_V$ is smaller than 3 {or if water is mostly formed when the o/p ratio of H$_2$ is high} \citet{1999ApJ...510L.145B,2013A&A...550A.127T}. Very high spatial resolution observations of additional HDO and H$_2^{18}$O lines would then be helpful to firmly confirm that the HDO/H$_2$O ratio is about 10$^{-3}$ in the most inner regions.

The chaotic profiles of the predicted abundances in the outer regions (see Fig.~\ref{ab_proto_standard}) are difficult to compare with the observations, since the observed abundances have been determined using a simple jump model, in which the envelope is divided in two regions: an inner and an outer regions with both constant abundances with a smaller one in the outer region \citep{2012A&A...539A.132C,2013A&A...553A..75C}. Note that the outer region in these jump models includes the two ones that we have defined for radii larger than 70 AU in our model. Regarding the abundance ratios (see Fig.\ref{ratio_proto_standard}), the model strongly overproduces the HDO/H$_2$O ratio. No conclusion can be drawn from the two other abundance ratios since only upper limits on the observed abundance of D$_2$O have been obtained but the limits are in agreement with our model. If a part of the absorbing layer mentioned in \citet{2012A&A...539A.132C} is produced in the outermost regions of the envelope, the observed HDO/H$_2$O ratio could be slightly higher ($\leq$ 5\%) at the highest radii but it would not reach the 10\% predicted by the chemical model and it would not explain the high ratios between $\sim$ 100 and $\sim$ 1000 AU anyway. We conclude this subsection by emphasizing that a proper quantitative comparison of the model predictions with the observations would require the comparison between predicted and observed molecular lines. This way, we would avoid many steps in between the comparison method and inconsistencies between assumptions. Such a methodology would however require the development of a number of comparison tools, which will be done in the future.

\subsection{Sensitivity to the parameters}

\begin{figure}
\includegraphics[width=1\linewidth]{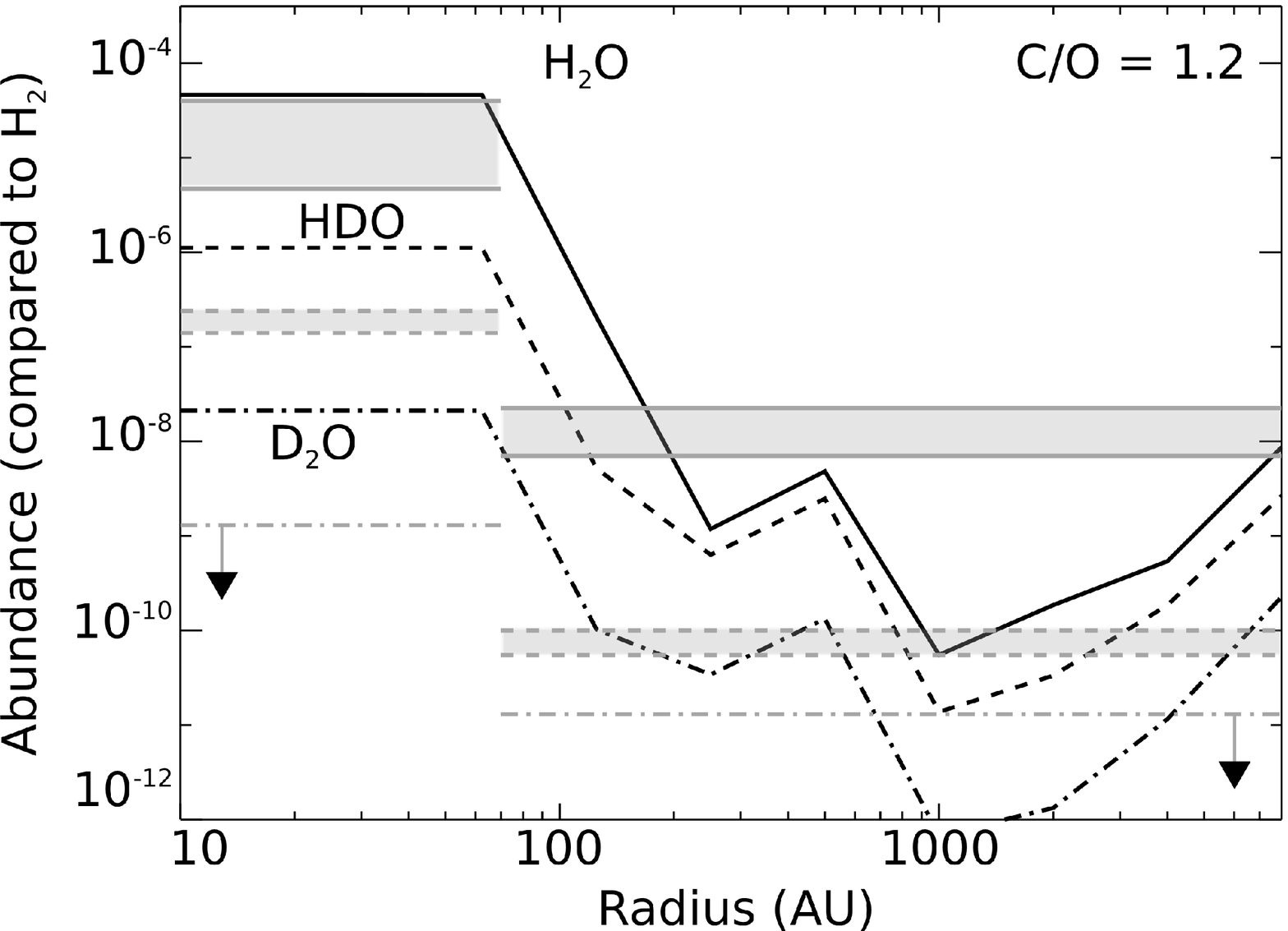}
\includegraphics[width=1\linewidth]{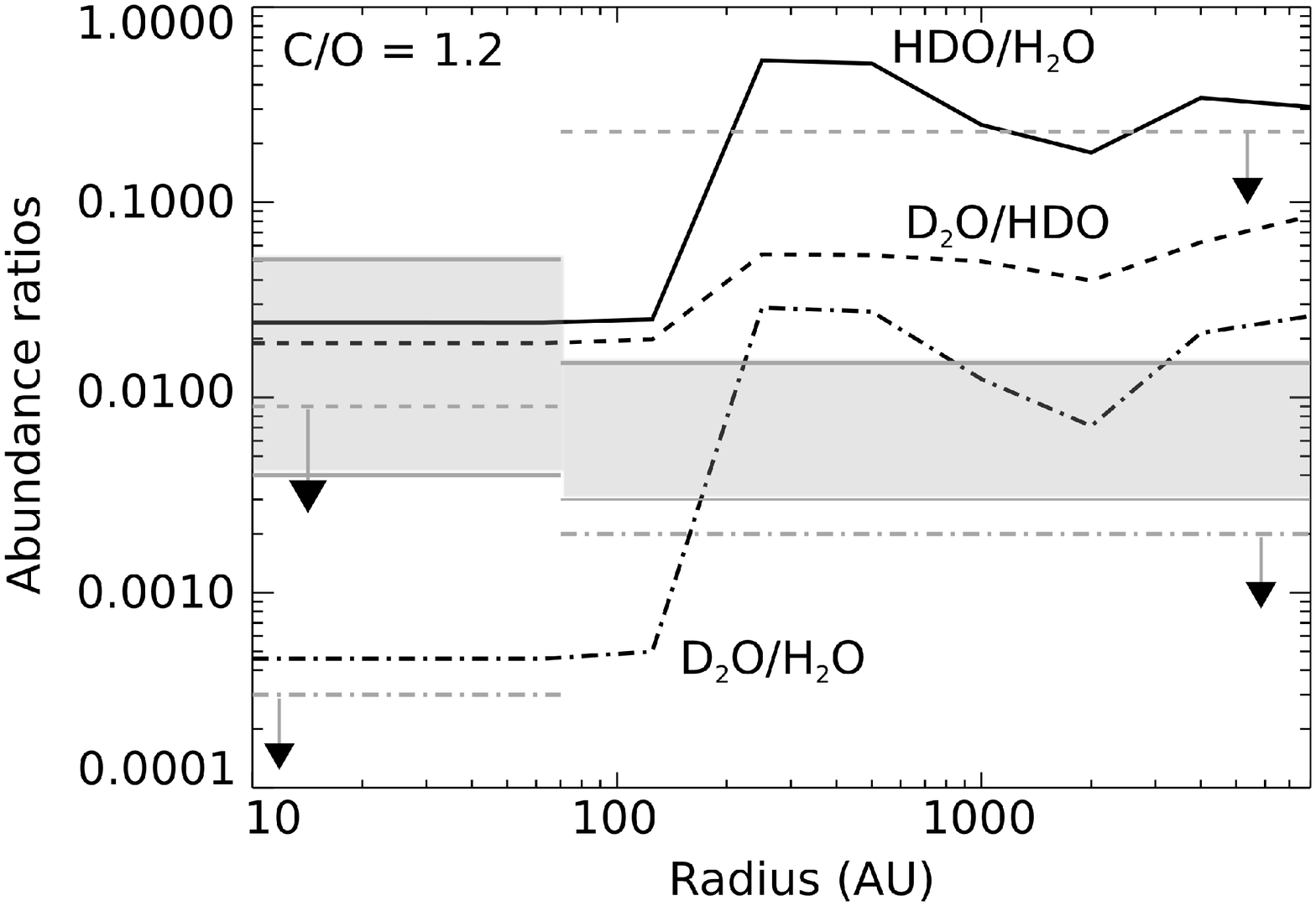}
\caption{Same as Figs.~\ref{ab_proto_standard} and \ref{ratio_proto_standard} but with a C/O elemental ratio of 1.2. \label{proto_1.2}}
\end{figure}

The model results presented in the previous section are not sensitive to the age of the parent cloud (P1). \\
The modeling results using a C/O elemental ratio of 1.2 (P2) are shown in Fig.~\ref{proto_1.2}. Decreasing the oxygen elemental abundance has a strongest effect on the predicted gas-phase abundance of HDO in the inner region of the envelope, decreasing its abundance by more than a factor of 9, whereas H$_2$O is decreased by a factor of about 6 and D$_2$O a factor of 5. As a consequence, the HDO/H$_2$O ratio is slightly increased whereas the other abundance ratios (D$_2$O/HDO and D$_2$O/H$_2$O) are decreased. The impact on the abundance ratios is however small: HDO/H$_2$O decreases from 0.029 to {\bf 0.020}, D$_2$O/HDO increases from 0.010 to 0.018 and D$_2$O/H$_2$O from $3.1\times 10^{-4}$ to $3.9\times 10^{-4}$. The main effect of changing the C/O elemental ratio is to decrease the abundance of oxygen available to form those species. The non-linearity of the system makes that the impact on the different species is not the same. In the rest of the envelope, although the change in the abundances is not very strong, it has an impact on the abundance ratios since for radii larger than 200~AU, the H$_2$O abundance is decreased whereas the D$_2$O abundance is increased and the HDO abundance is almost unchanged. As a consequence, all abundance ratios (HDO/H$_2$O, D$_2$O/HDO and D$_2$O/H$_2$O) are increased. The increase of D$_2$O abundance at this radius when the oxygen elemental abundance is smaller is a surprising result. This behavior is due to the increase of the HDCO abundance in the ices, which is a precursor of D$_2$O while evaporated in the gas-phase at this radius (see section~\ref{proto_results}). The HDCO molecule is formed on the grains by the dissociation of deuterated methanol CH$_2$DOH, which is increased by a larger C/O elemental ratio since more carbon is available to form precursors.
The HDO/H$_2$O abundance ratio observed by \citet{2013A&A...560A..39C} in the center is still reproduced by the model but the D$_2$O/HDO and D$_2$O/H$_2$O predicted ratios get larger than the observed upper limits. In the outer region, the abundance ratios also show a worse agreement with the observations using a 1.2 elemental C/O ratio, except for D$_2$O/HDO. 

\begin{figure}
\includegraphics[width=1\linewidth]{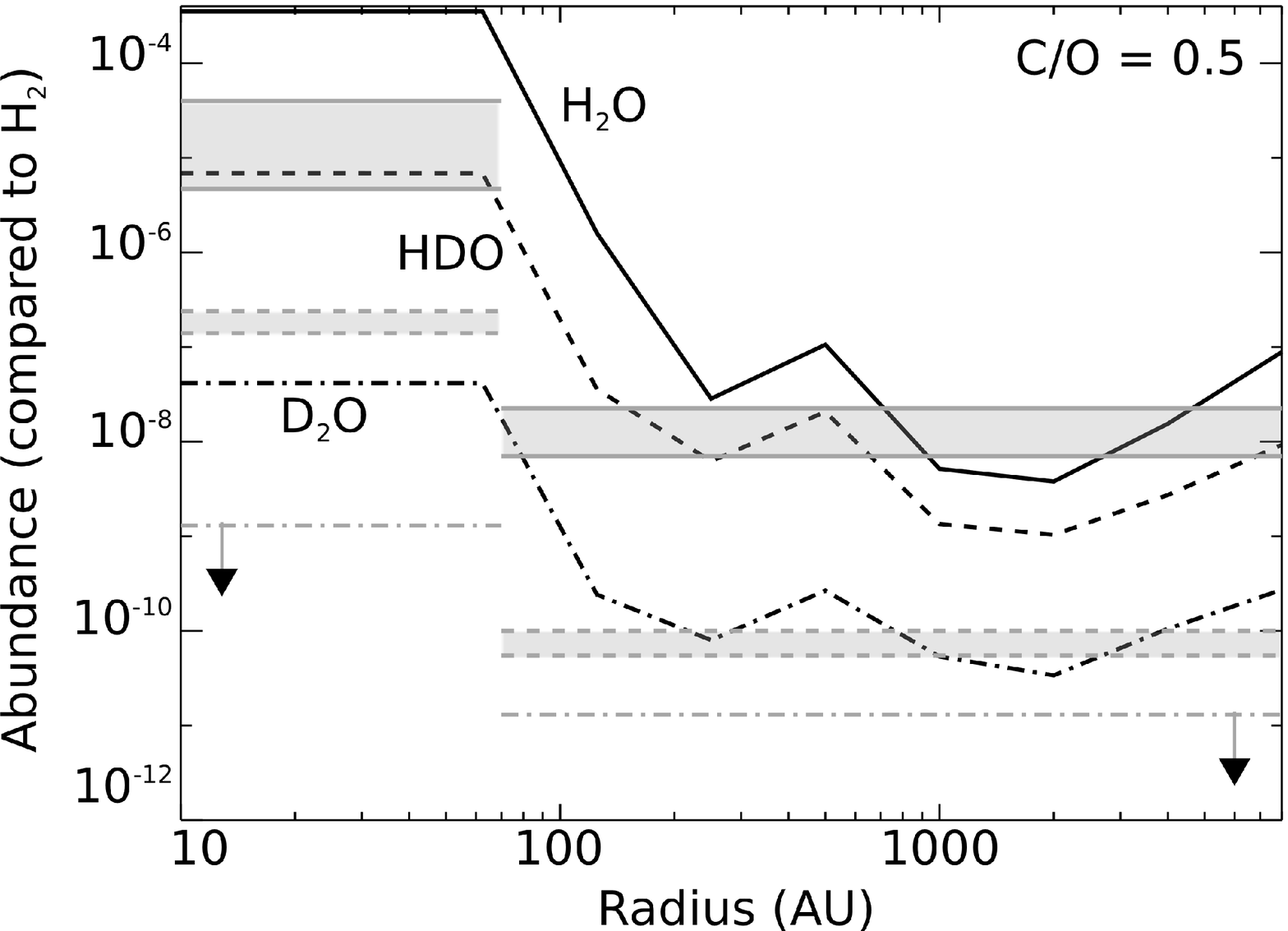}
\includegraphics[width=1\linewidth]{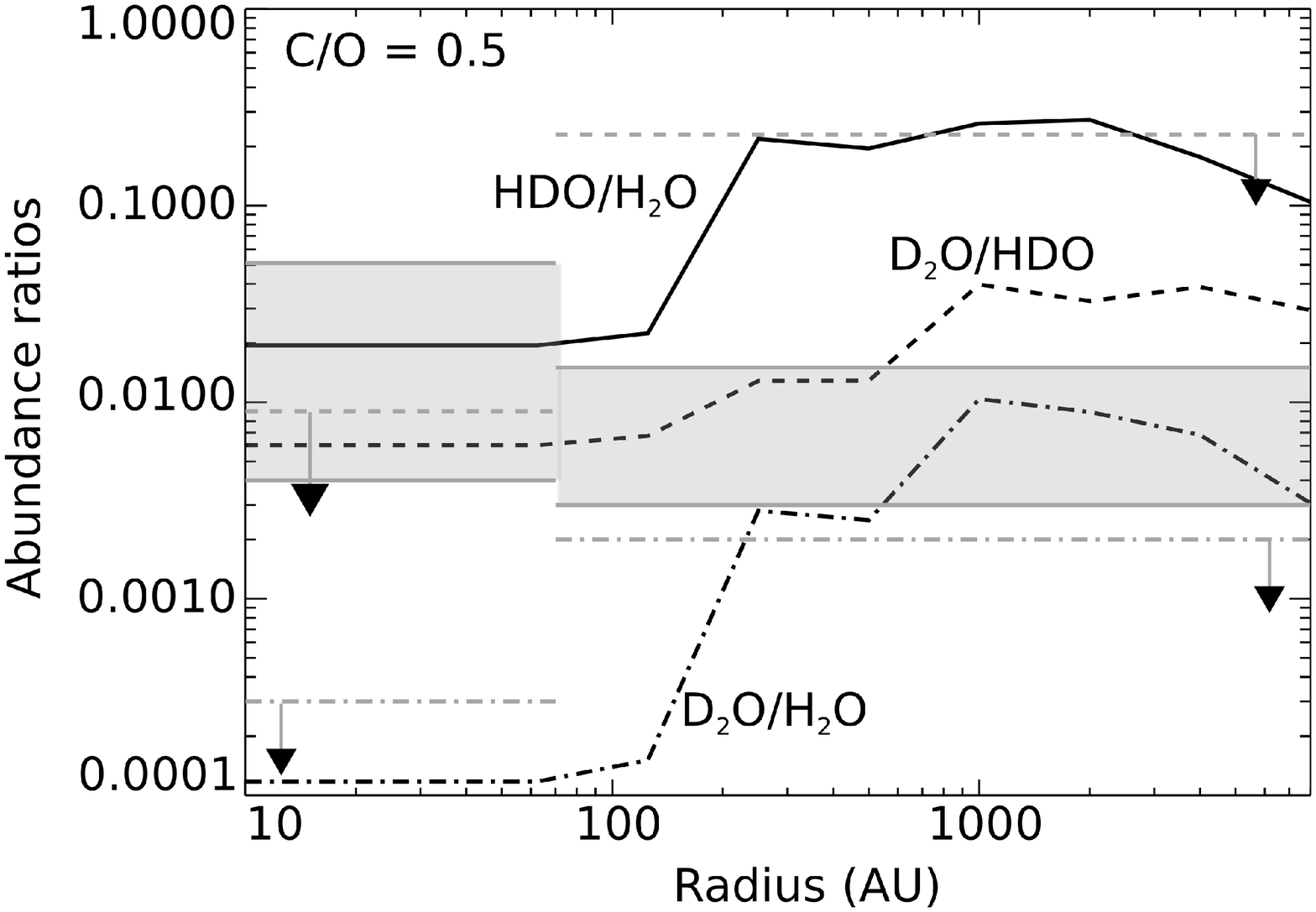}
\caption{Same as Figs.~\ref{ab_proto_standard} and \ref{ratio_proto_standard} but with a cosmic-ray ionization rate of $10^{-16}$~s$^{-1}$.   \label{proto_0.5_zeta}}
\end{figure}

Figure~\ref{proto_0.5_zeta} shows the modeling results when the standard model is run with a cosmic-ray ionization rate of $10^{-16}$~s$^{-1}$ (P3). Abundances of HDO and D$_2$O in the inner regions are slightly decreased (H$_2$O stays constant) whereas all three molecules are strongly increased for radius larger than approximately 100~AU. In the inner region, the gas-phase abundances reflects the ice composition. A larger $\zeta$ produces more dissociations of HDO and D$_2$O at the surface of the grains, producing smaller gas-phase abundances while the ices evaporate. In the colder parts of the protostellar envelope, the increase of the three molecules is produced by the increase of the OH and OD abundances (which react with H$_2$CO and HDCO evaporated from the grains). The increase of those species is due to the dissociation of CH$_3$OH and CH$_2$DOH by cosmic-ray induced UV photons. {\bf For the deuterated network used in those simulations, we do not discriminate the isomers of deuterated CH$_3$OH so that the products of the photodissociation of CH$_2$DOH and CH$_3$OD are the same with statistical branching ratios. Since CH$_3$OD is much less abundant than CH$_2$DOH, we found that OD is mostly produced by the photodissociation of CH$_2$DOH.}

\section{The foreground cloud} \label{cloud}

\subsection{Chemical and physical model description}

To study the chemistry of the foreground cloud, called "photodesorption layer" in \citet{2012A&A...539A.132C,2013A&A...553A..75C}, several models were performed and photo evaporation processes have been added to the model. For this work, a grid of models was performed for the following parameters:  temperatures (gas and dust)  of 15 and 30 K, total H densities of $2\times 10^4$ and $2\times 10^5$~cm$^{-3}$, visual extinctions A$_V$ of 2, 3 and 4, cosmic-ray ionization rates of $10^{-17}$ and $10^{-16}$~s$^{-1}$, and C/O elemental ratios of 0.5 and 1.2 (see section~\ref{proto_parameters}). This grid was defined based on the few observational constraints from \citet{2012A&A...539A.132C}. Species are initially in the atomic form with the same elemental abundances as for the protostellar envelope (see section~\ref{proto_parameters}). The photo-evaporation processes of ices by direct and indirect UV photons have been included following the formalism from \citet{2011A&A...534A.132V}. For that we have used the same yields of photo-evaporation for deuterated and non-deuterated species as suggested by \citeauthor{2009ApJ...693.1209O} (2009a). Experimental results from \citeauthor{2009ApJ...693.1209O} (2009a) and \citeauthor{2009A&A...496..281O} (2009b) have found yields of photo-evaporation about 10$^{-3}$ molecules per grain per incident UV photon, except for N$_2$. They also showed that this yield may depend on the temperature and the thickness of the ices. We have included the temperature dependence as suggested by the authors but we will show that the efficiency of this process against exothermicity of surface reactions is small under the conditions we have used. All the parameters, which impact on the model predictions have been studied, are summarized in Table~\ref{explore_param}.

\subsection{Gas-phase and ice composition}\label{abundances}

\begin{figure*}
\includegraphics[width=0.4\linewidth]{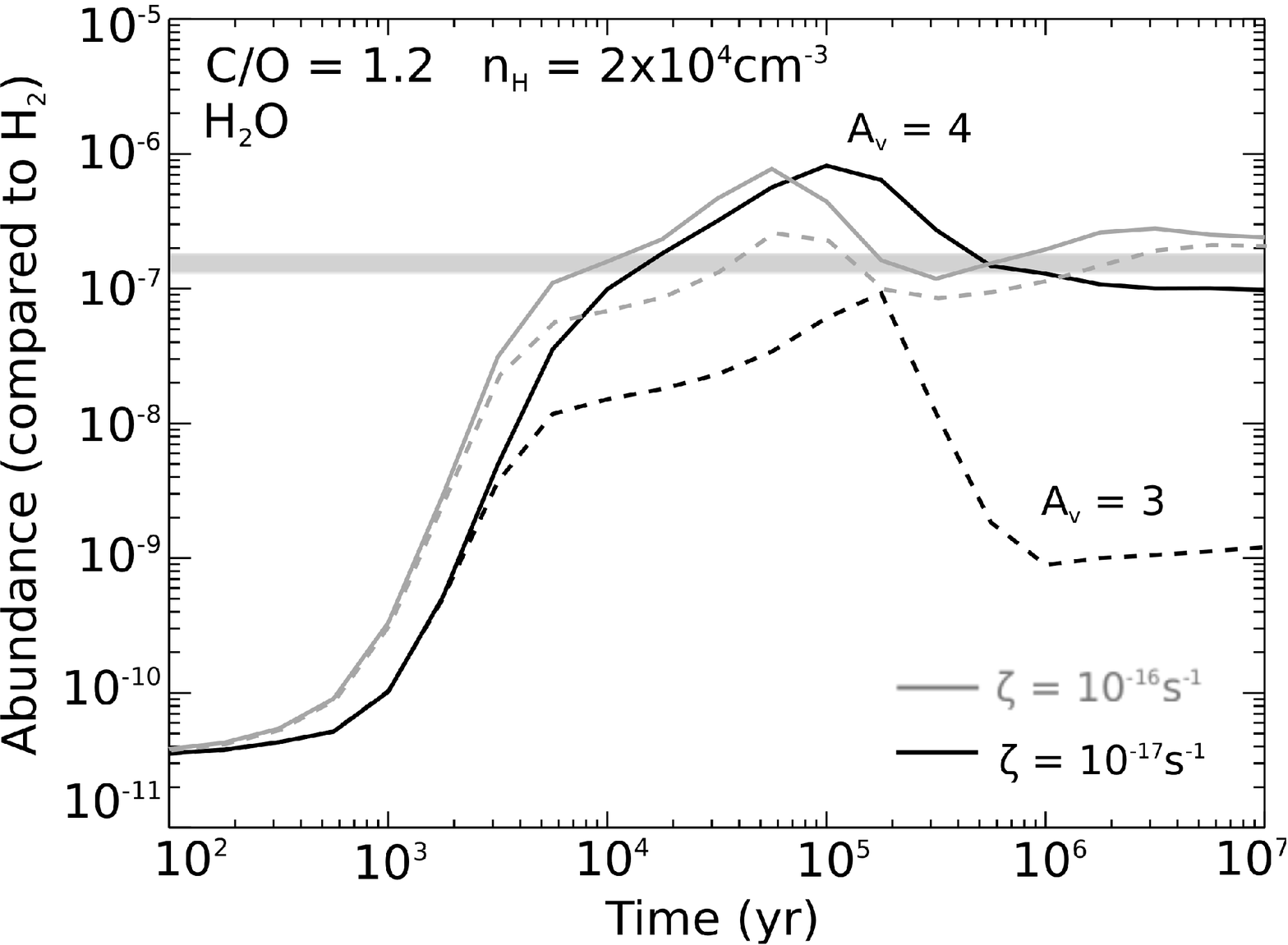}
\includegraphics[width=0.4\linewidth]{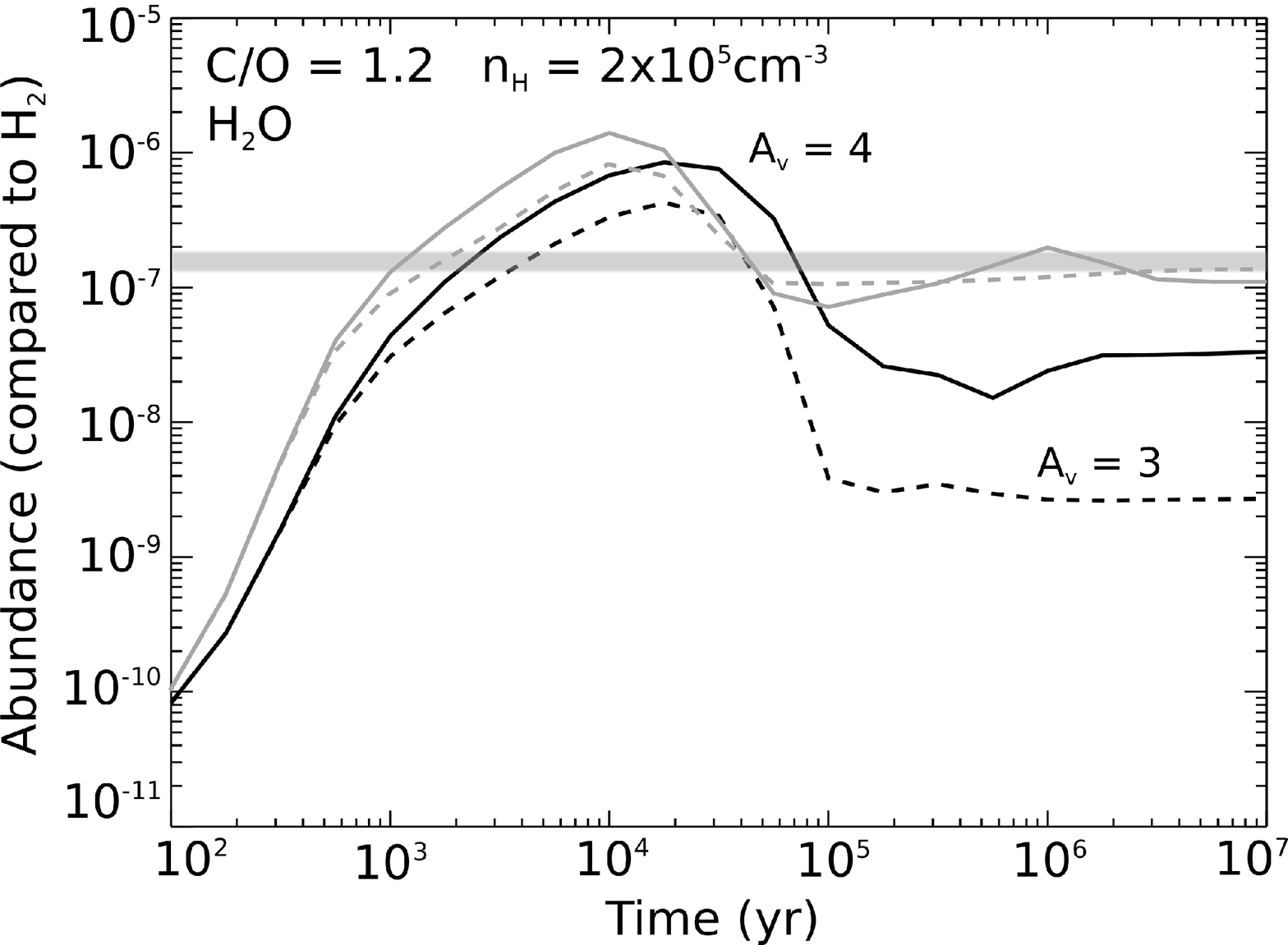}
\includegraphics[width=0.4\linewidth]{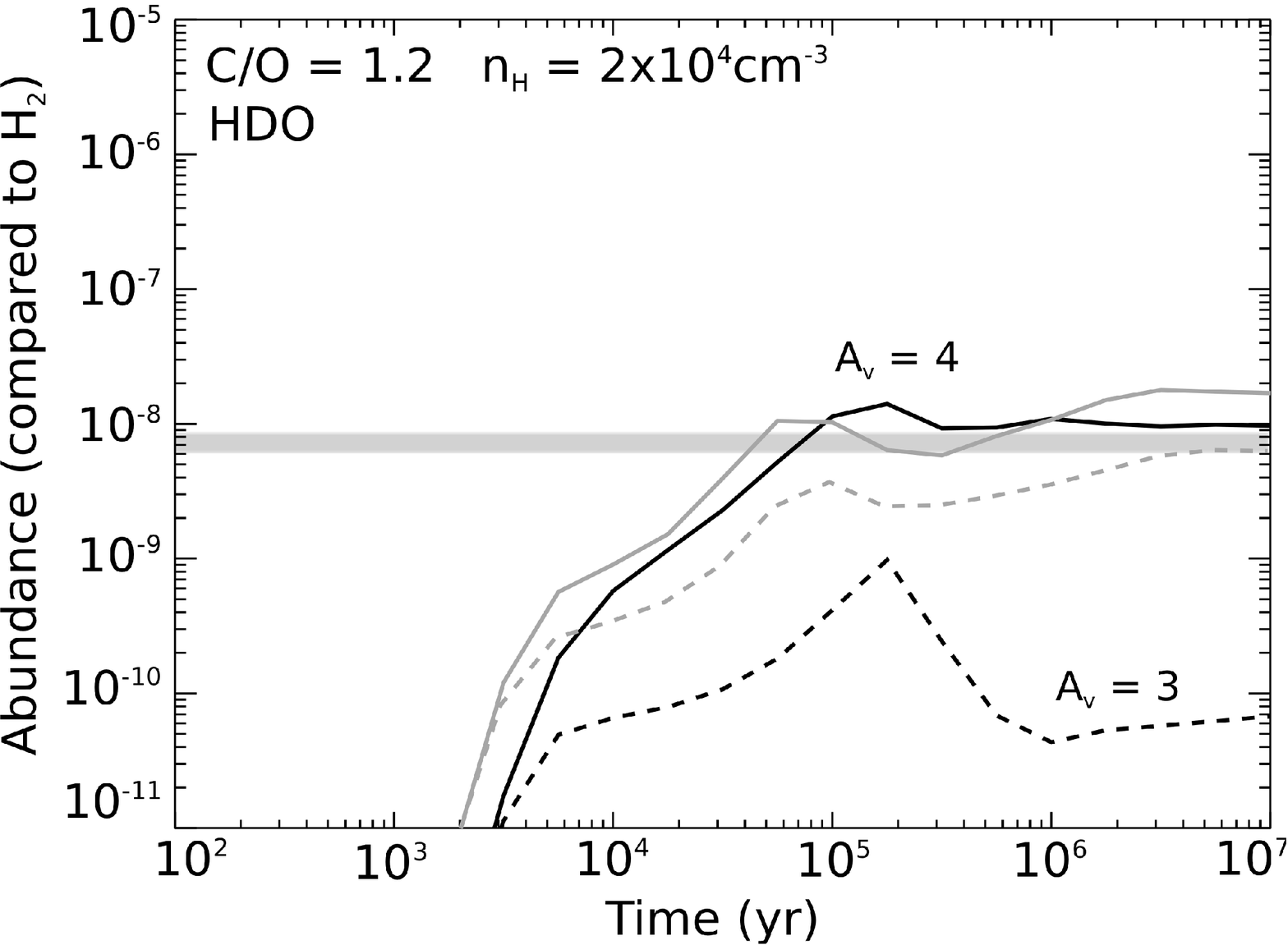}
\includegraphics[width=0.4\linewidth]{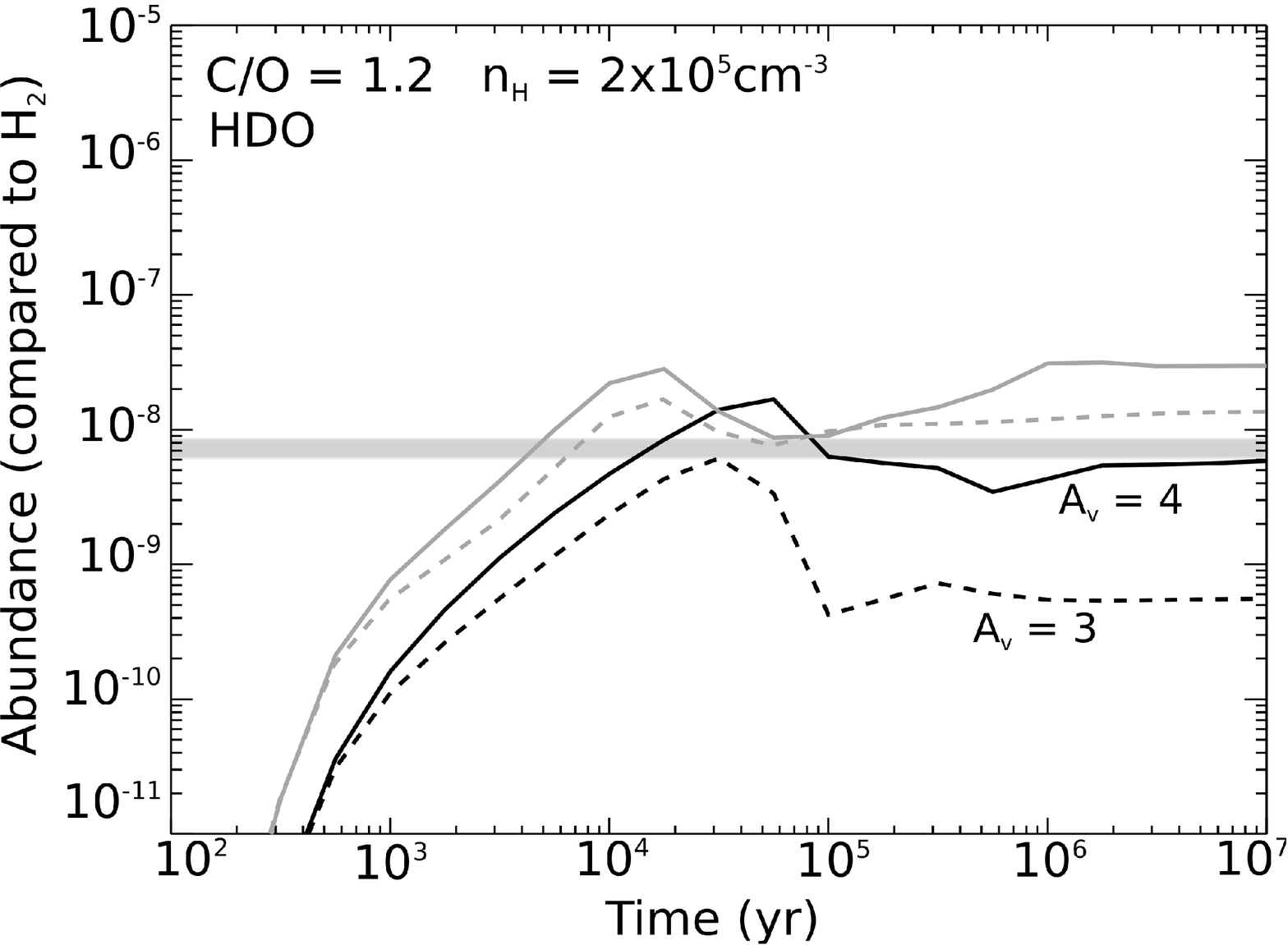}
\includegraphics[width=0.4\linewidth]{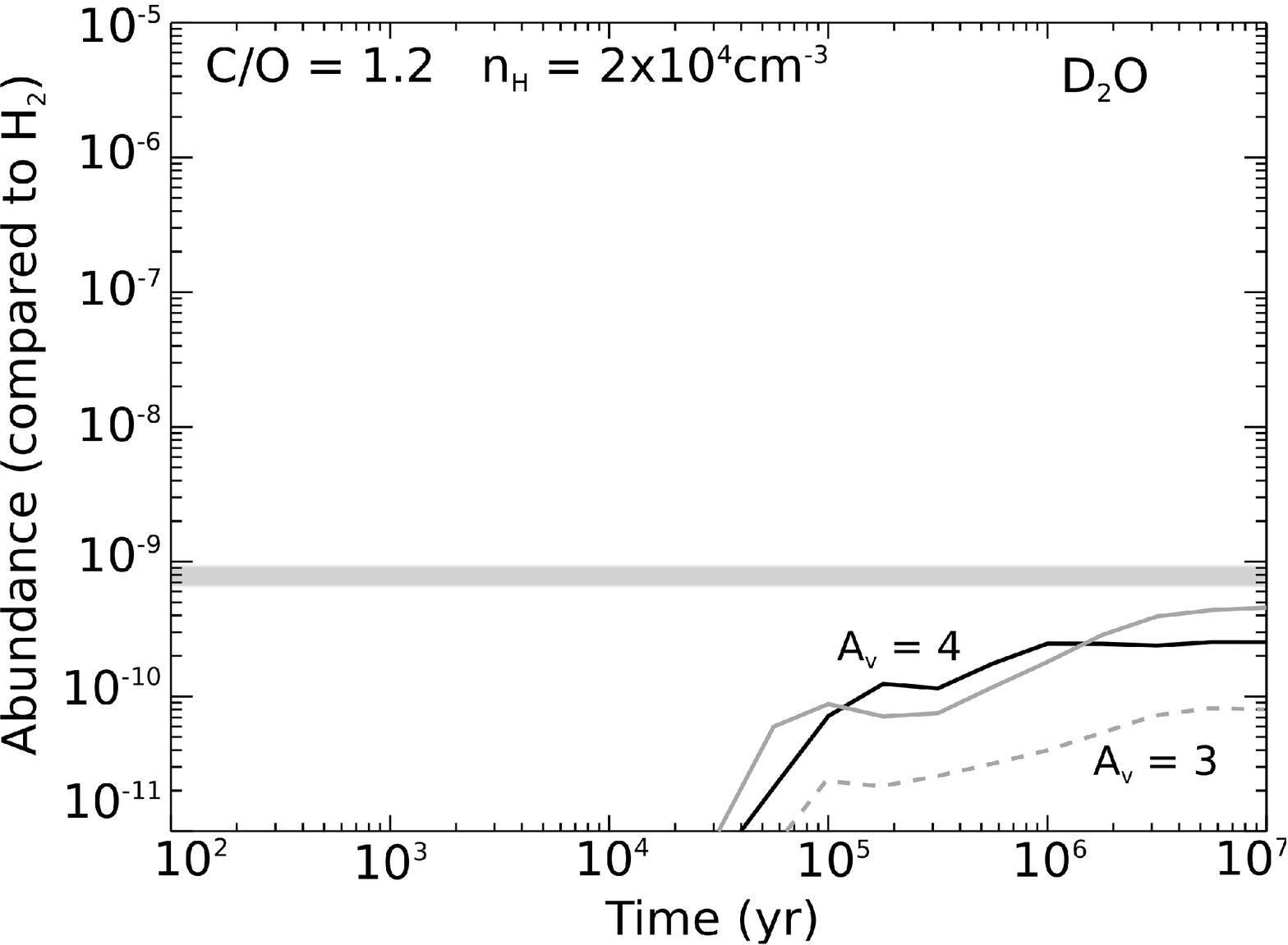}
\includegraphics[width=0.4\linewidth]{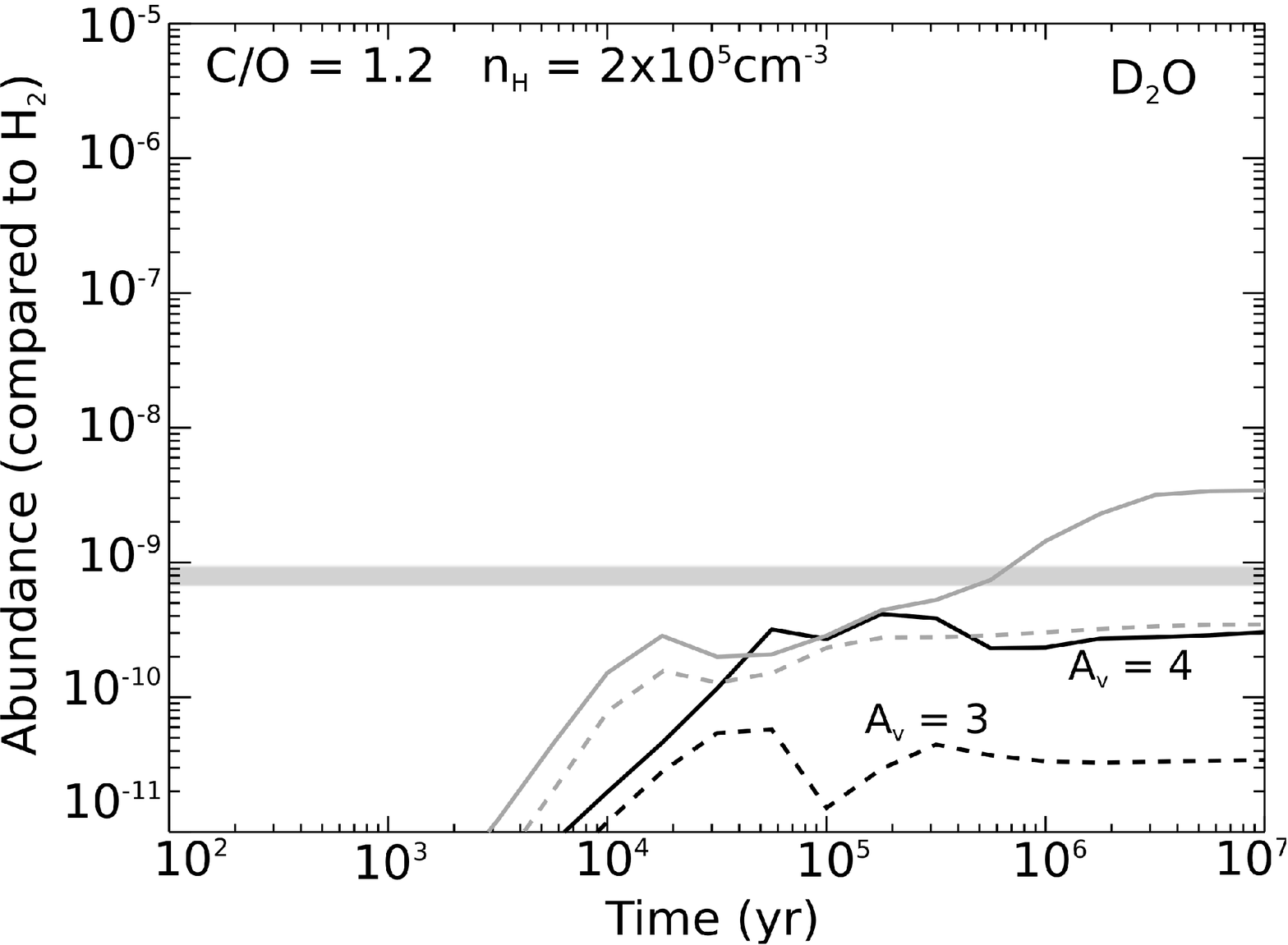}
\caption{Gas-phase abundances of H$_2$O, HDO and D$_2$O as a function of time predicted by the model for the foreground cloud.  Gas and dust temperature is the same for all models: 15~K. The elemental C/O abundance ratio is 1.2. The left side of the model has been obtained for a total H density of $2\times 10^4$~cm$^{-3}$ and the right side for a total H density of $2\times 10^5$~cm$^{-3}$. Black curves have been obtained for a cosmic-ray ionization rate $\zeta$ of $10^{-17}$~s$^{-1}$ whereas grey curves have been obtained for a $\zeta$ of  $10^{-16}$~s$^{-1}$. Solid lines represent an A$_V$ of 4 and dashed lines an A$_V$ of 3. Horizontal grey areas represent the observed values in the foreground cloud for A$_V$ between 3 and 4 (see Table~\label{obs}).
 \label{cloud_C_O_1.2}}
\end{figure*}

\begin{figure*}
\includegraphics[width=0.4\linewidth]{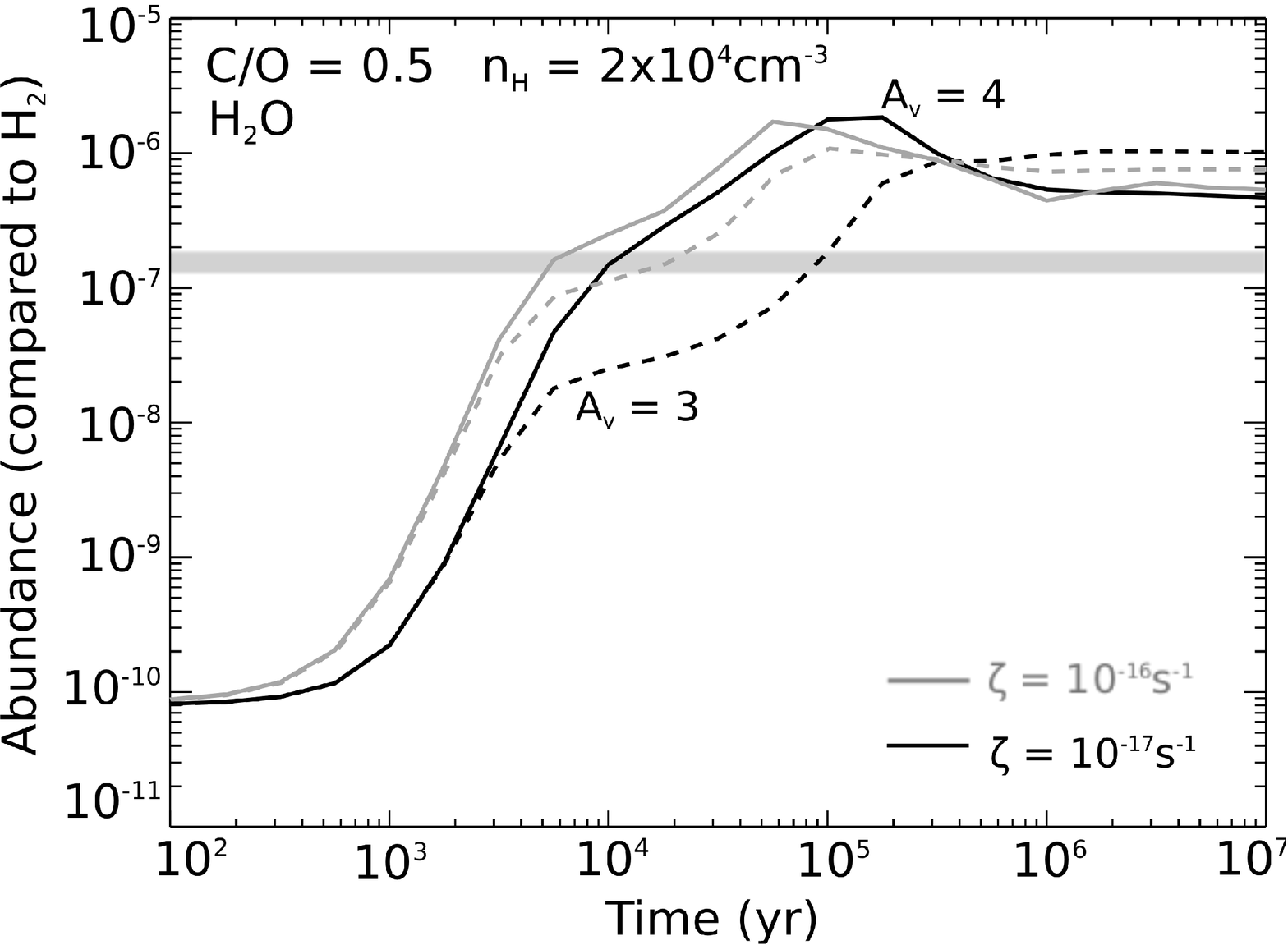}
\includegraphics[width=0.4\linewidth]{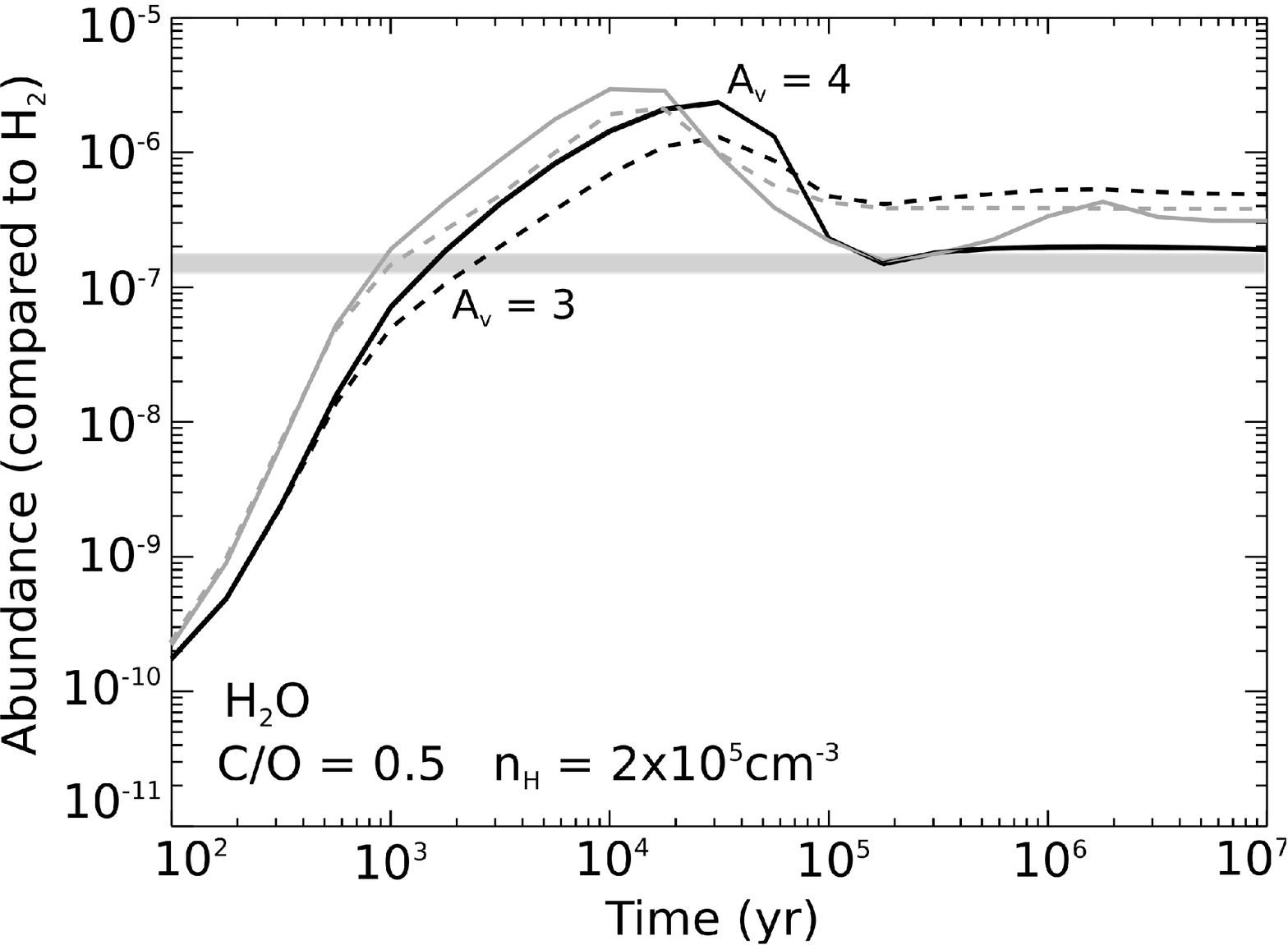}
\includegraphics[width=0.4\linewidth]{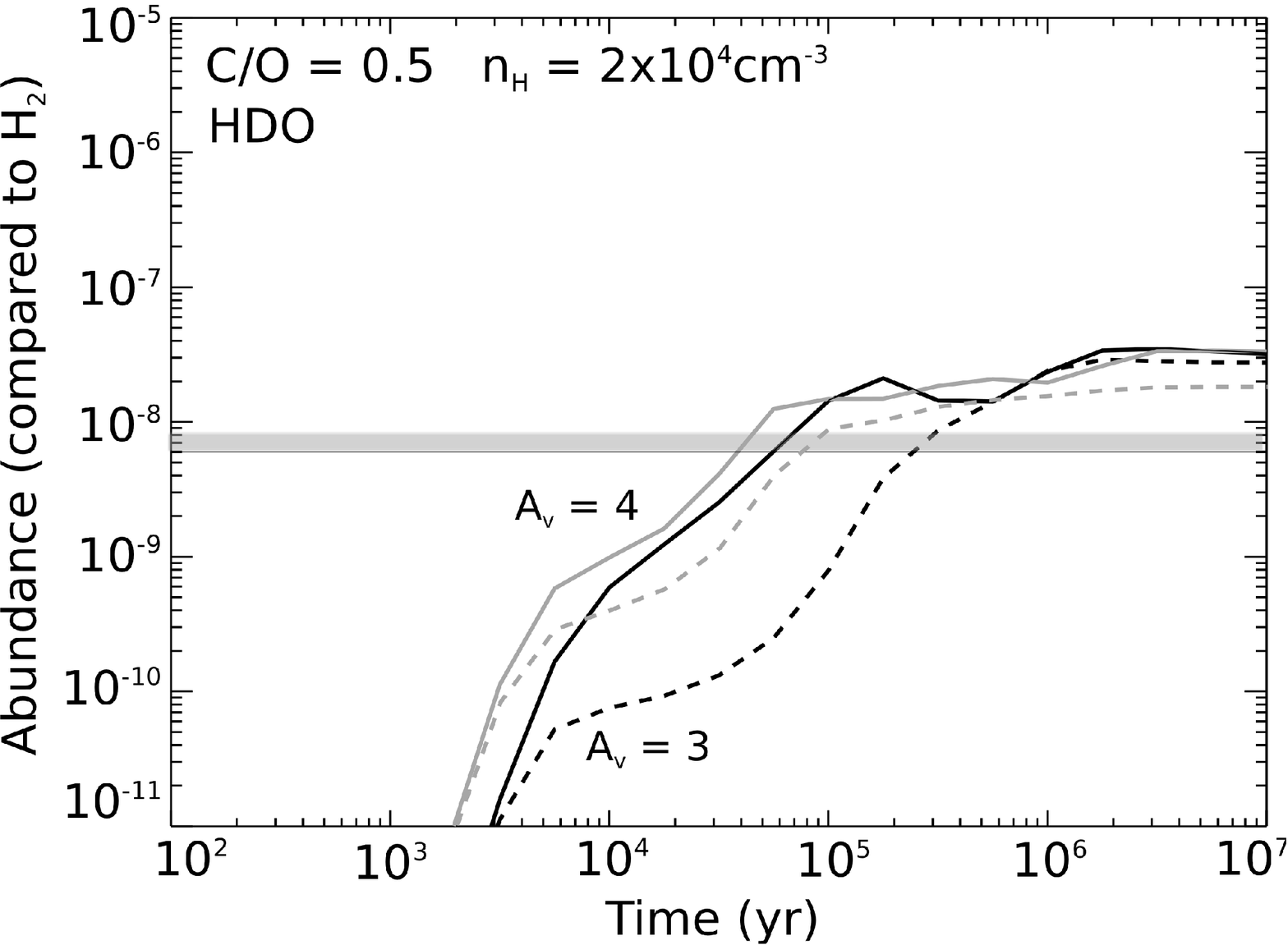}
\includegraphics[width=0.4\linewidth]{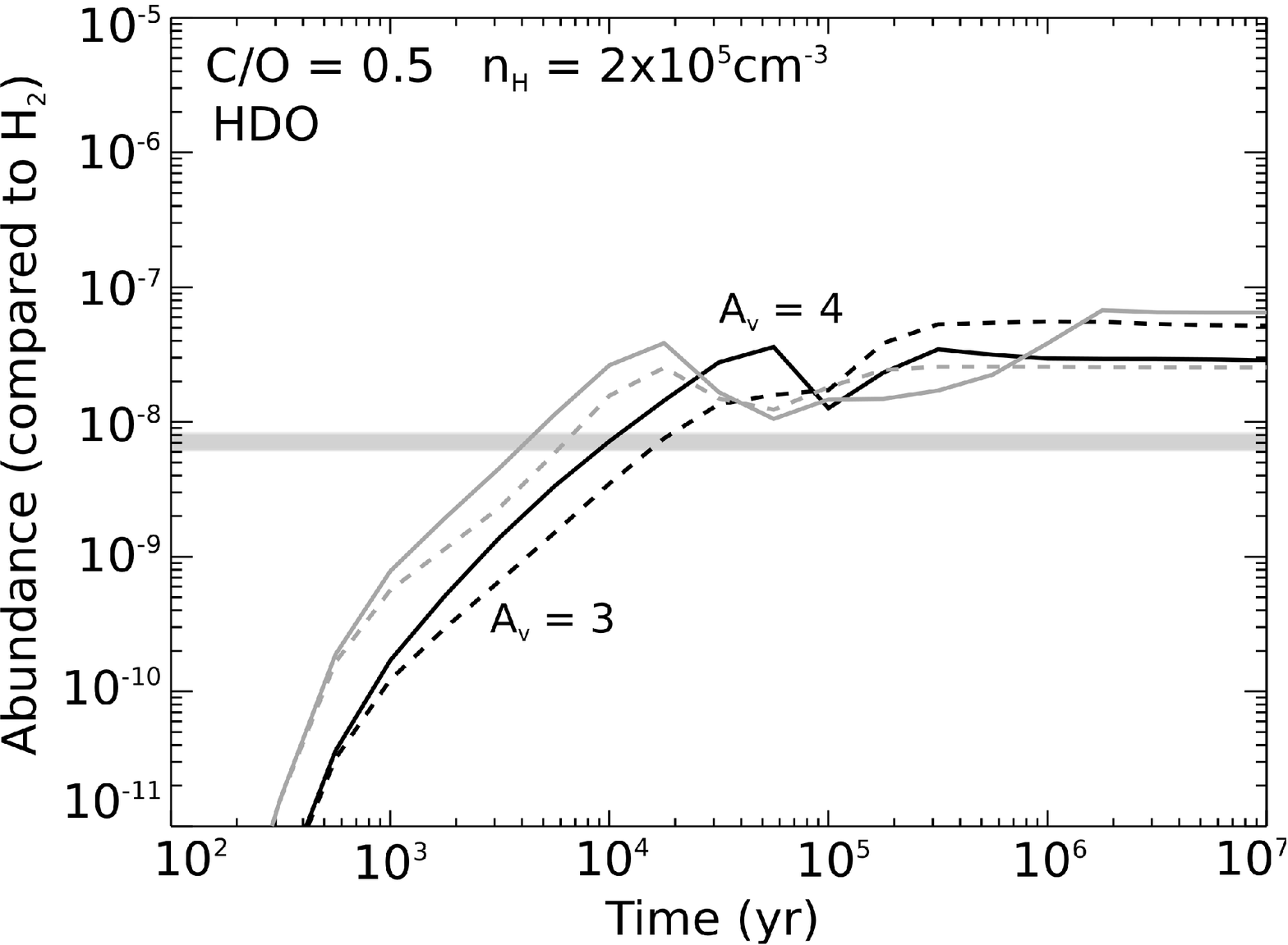}
\includegraphics[width=0.4\linewidth]{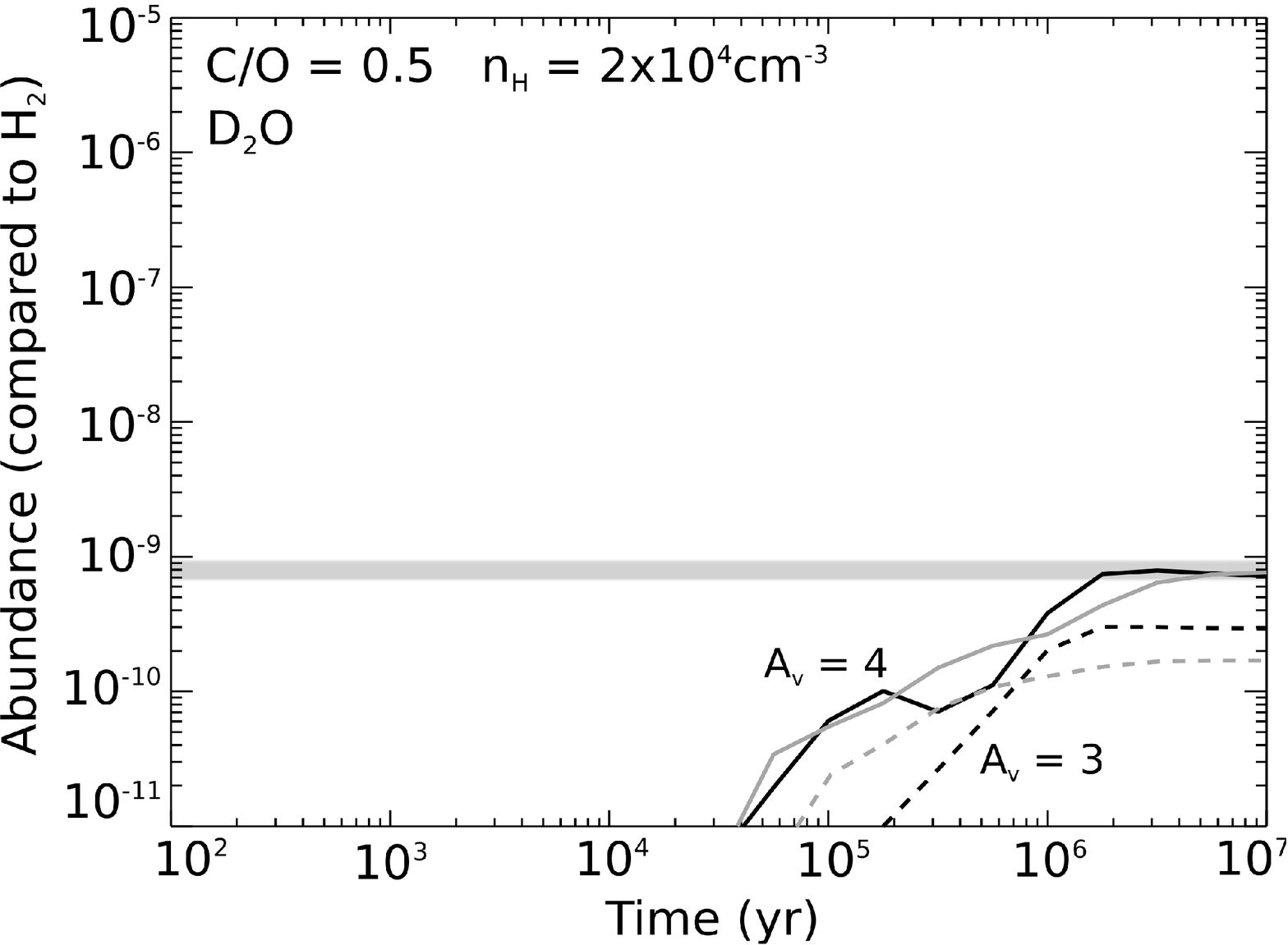}
\includegraphics[width=0.4\linewidth]{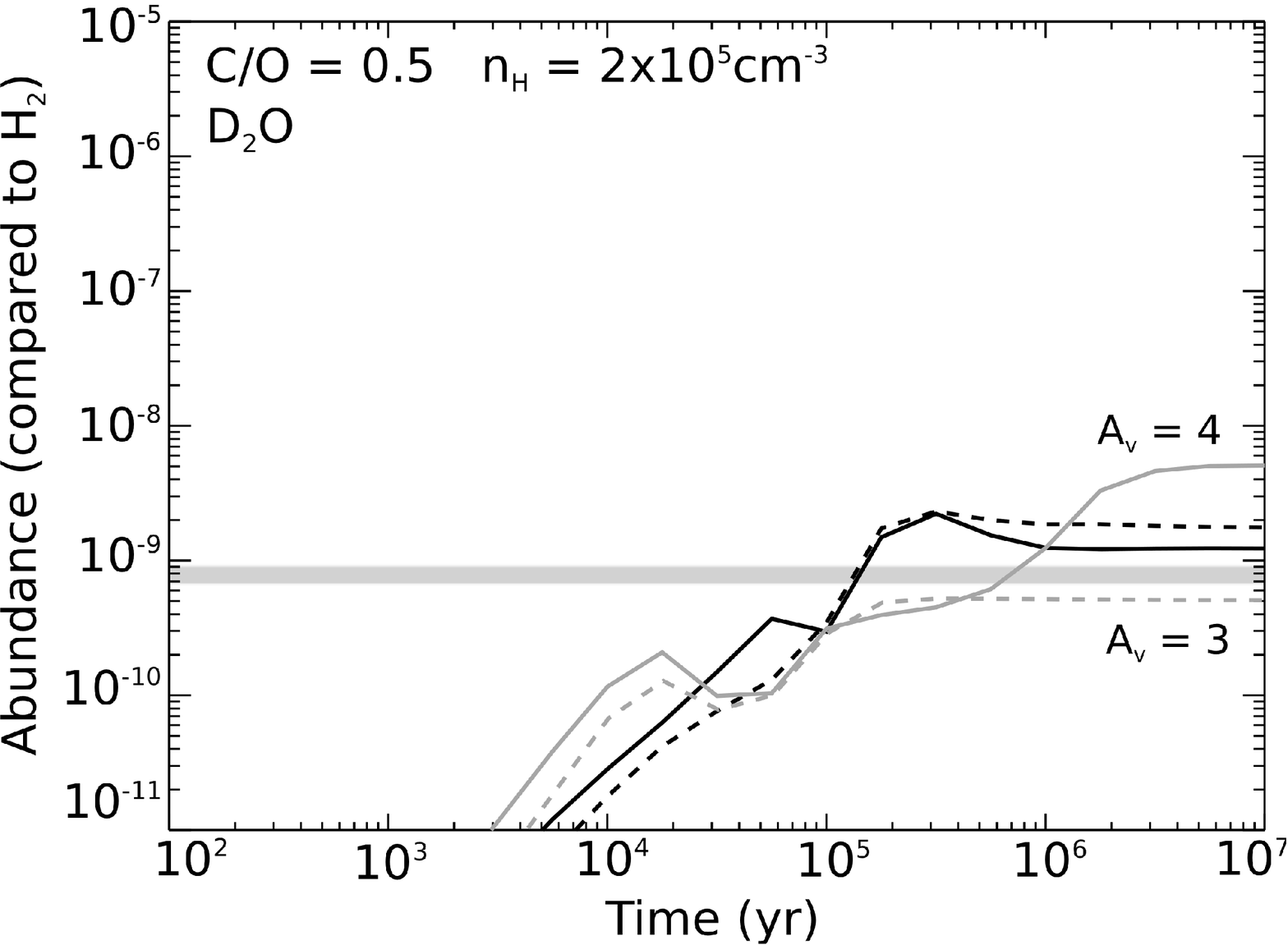}
\caption{ Same as Fig.~\ref{cloud_C_O_0.5} but a C/O of 0.5.  \label{cloud_C_O_0.5}}
\end{figure*}

The first parameter we can exclude is a temperature larger than 15~K. For a temperature of 30~K, the singly and doubly deuterated water abundances are strongly underestimated by the model at all times. Similarly, a visual extinction smaller than 3 produces small amounts of D$_2$O, much below the observed ones. We then show in Figs.~\ref{cloud_C_O_1.2}, \ref{cloud_C_O_0.5}, \ref{cloud_C_O_1.2_ice} and \ref{cloud_C_O_0.5_ice} the  modeling results (in the gas and in the ices) for a temperature of 15~K, A$_V$ of 3 and 4, C/O ratios of 0.5 and 1.2, and cosmic-ray ionization rates of $10^{-17}$ and $10^{-16}$~s$^{-1}$. The gas-phase and ice predicted abundances are more sensitive to the parameters when the carbon abundance is larger than the oxygen one so that H$_2$O and CO are oxygen reservoir competitors. When the C/O elemental ratio is smaller than one, water is the reservoir of oxygen whereas, when C/O is larger than 1 most of the oxygen is stored in CO. 

In the cases C/O = 1.2 and A$_V$ smaller than 4, the photodissociations in the gas-phase and at the surface of the grains break the H$_2$O molecules so that the abundance of this species (and its deuterated forms) is small and most of the oxygen ends up in the gas-phase and icy carbon monoxide CO. When the A$_V$ is large, more H$_2$O, HDO and D$_2$O can be formed and remain abundant in the gas-phase. Note that for all three molecules and all models, the abundances on the grain surfaces are larger than the ones in the gas-phase, except in the case of A$_V$ = 3 and $\zeta = 10^{-17}$~s$^{-1}$.  

The larger abundances of water obtained for a $\zeta$ of $10^{-16}$~s$^{-1}$ are due to more efficient ion-molecule reaction path than with $\zeta = 10^{-17}$~s$^{-1}$, which then deplete on the grains. The HDO and D$_2$O abundances are decreased by the higher $\zeta$ on the grains and increased in the gas-phase, except in the case A$_V$ = 3. The deuterated water HDO on grains is mainly formed by the surface reactions H$_{\rm ice}$ + OD$_{\rm ice}$ and D$_{\rm ice}$ + OH$_{\rm ice}$, and the sticking of gas-phase HDO onto the surfaces, whereas it is destroyed by photo-dissociations with direct UV photons and UV photons induced by H$_2$/cosmic-ray interactions. The larger cosmic-ray ionization rate does not increase the production of HDO on the surface but speeds up strongly its dissociation. In addition, since H atoms are more abundant than D atoms on the surfaces, the photodissociation of HDO $\rightarrow$ OH + D is more likely followed by the surface reaction OH + H. In the gas-phase, HDO is formed by the dissociative recombination of H$_2$DO$^+$, which abundance is increased by the larger $\zeta$. The gas-phase HDO is mostly removed by sticking onto the grain surfaces. Similar conclusions can be drawn for D$_2$O. 

In the case of the smaller C/O ratio (0.5), the predicted abundances are less sensitive to the parameters. This is particularly true for the ice abundances, except for the case of a larger $\zeta$, for which the H$_2$O$_{\rm ice}$ abundance is increased, and the HDO$_{\rm ice}$ and D$_2$O$_{\rm ice}$ abundances are decreased. It is interesting to notice that the late time decrease of the gas-phase abundances with the A$_V$ is not as significant as in the case of C/O = 1.2. In the case of an oxygen elemental abundance larger than the carbon one, there is always enough oxygen to reform these molecules even when the photodissociations are efficient.

\subsection{Abundance ratios and comparison with the observations}\label{ratios_comp}

\begin{figure*}
\includegraphics[width=0.4\linewidth]{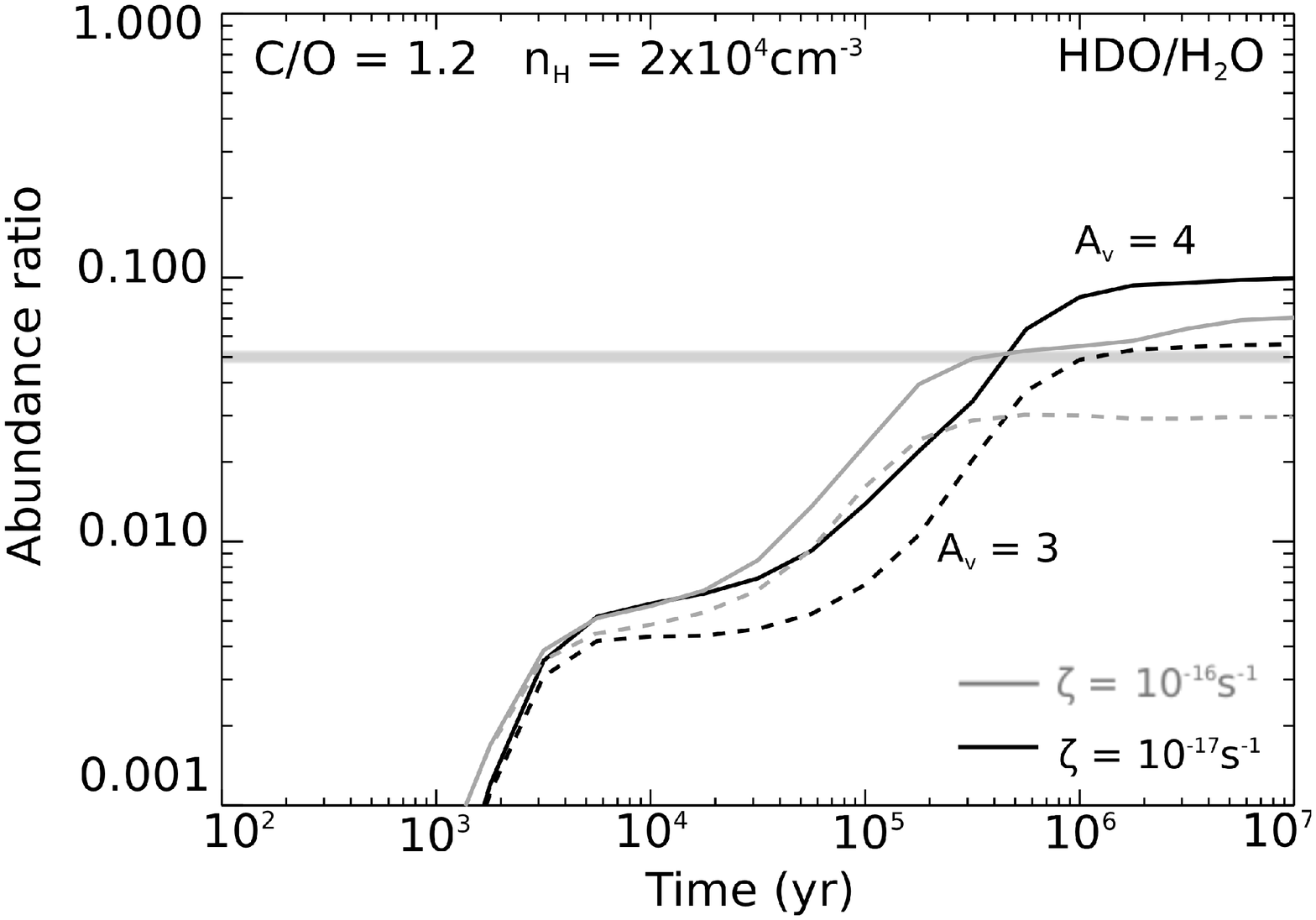}
\includegraphics[width=0.4\linewidth]{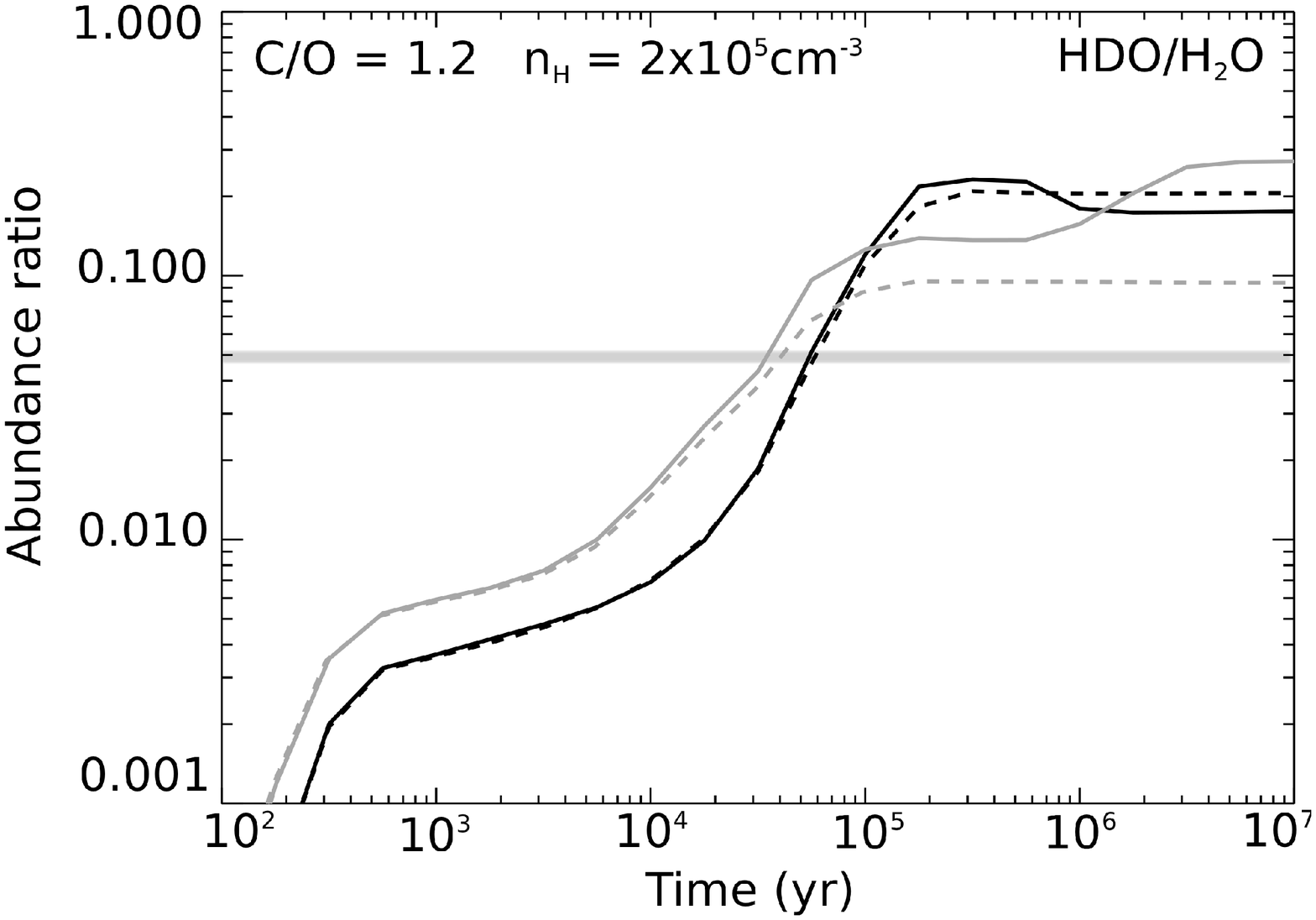}
\includegraphics[width=0.4\linewidth]{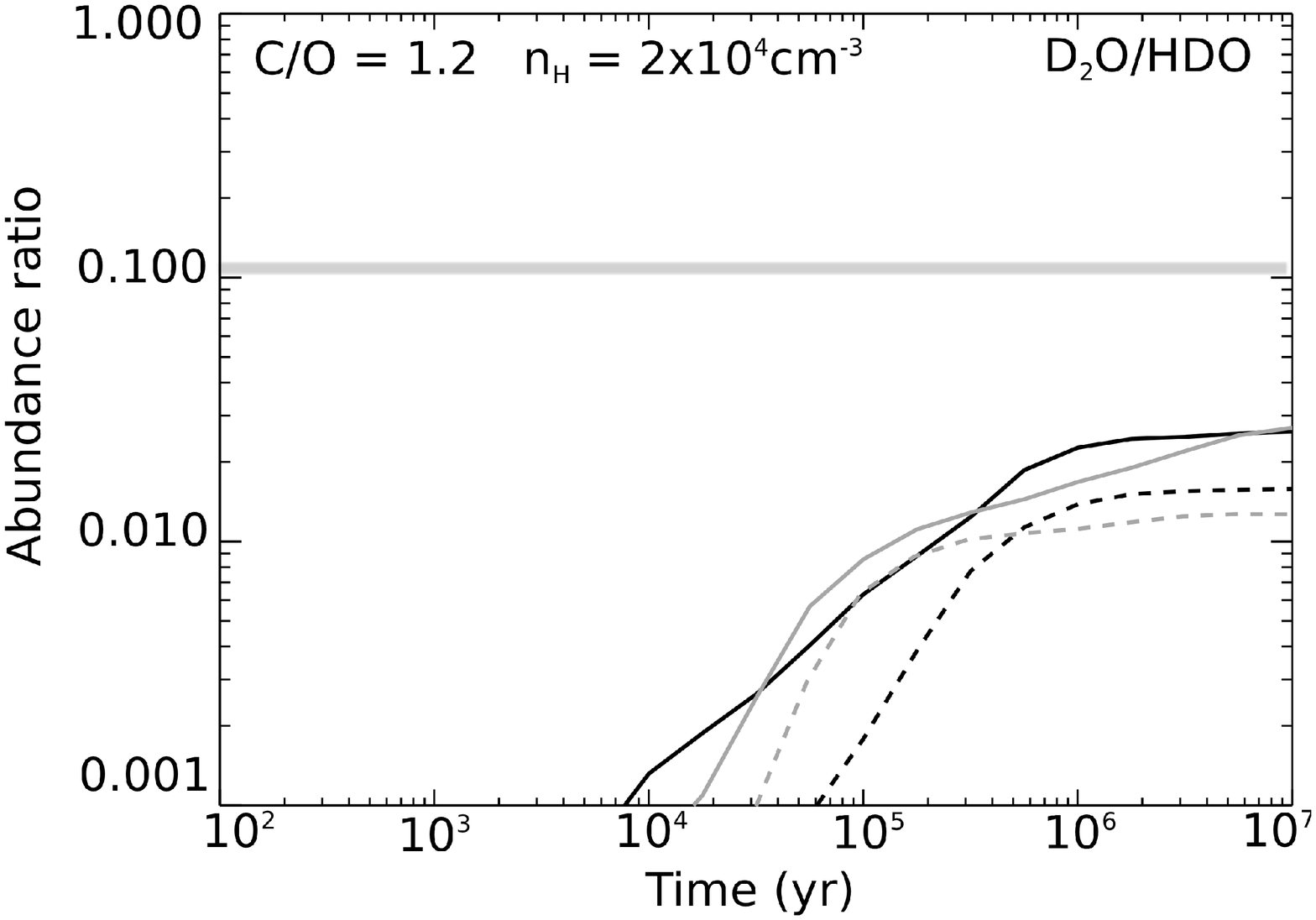}
\includegraphics[width=0.4\linewidth]{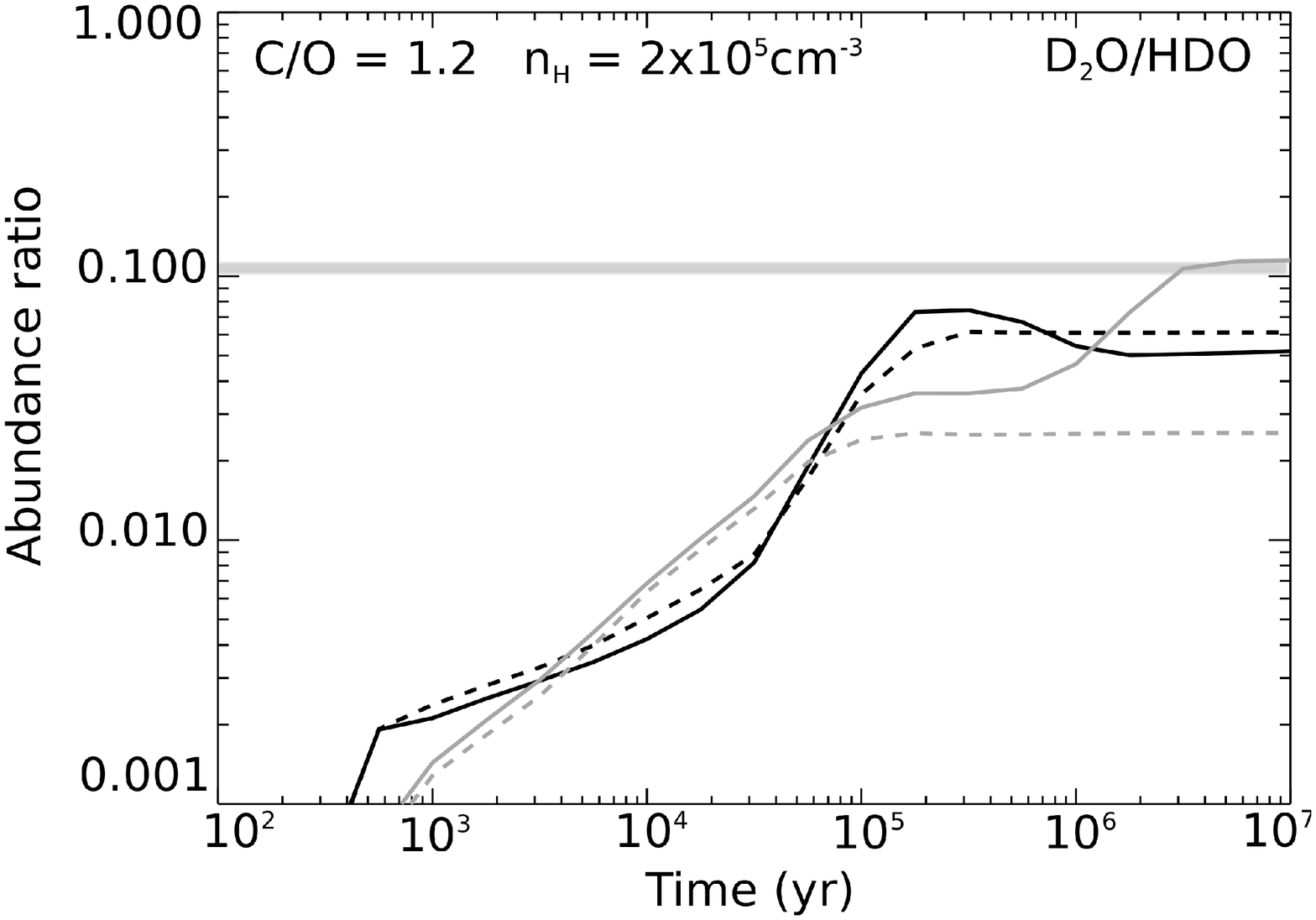}
\includegraphics[width=0.4\linewidth]{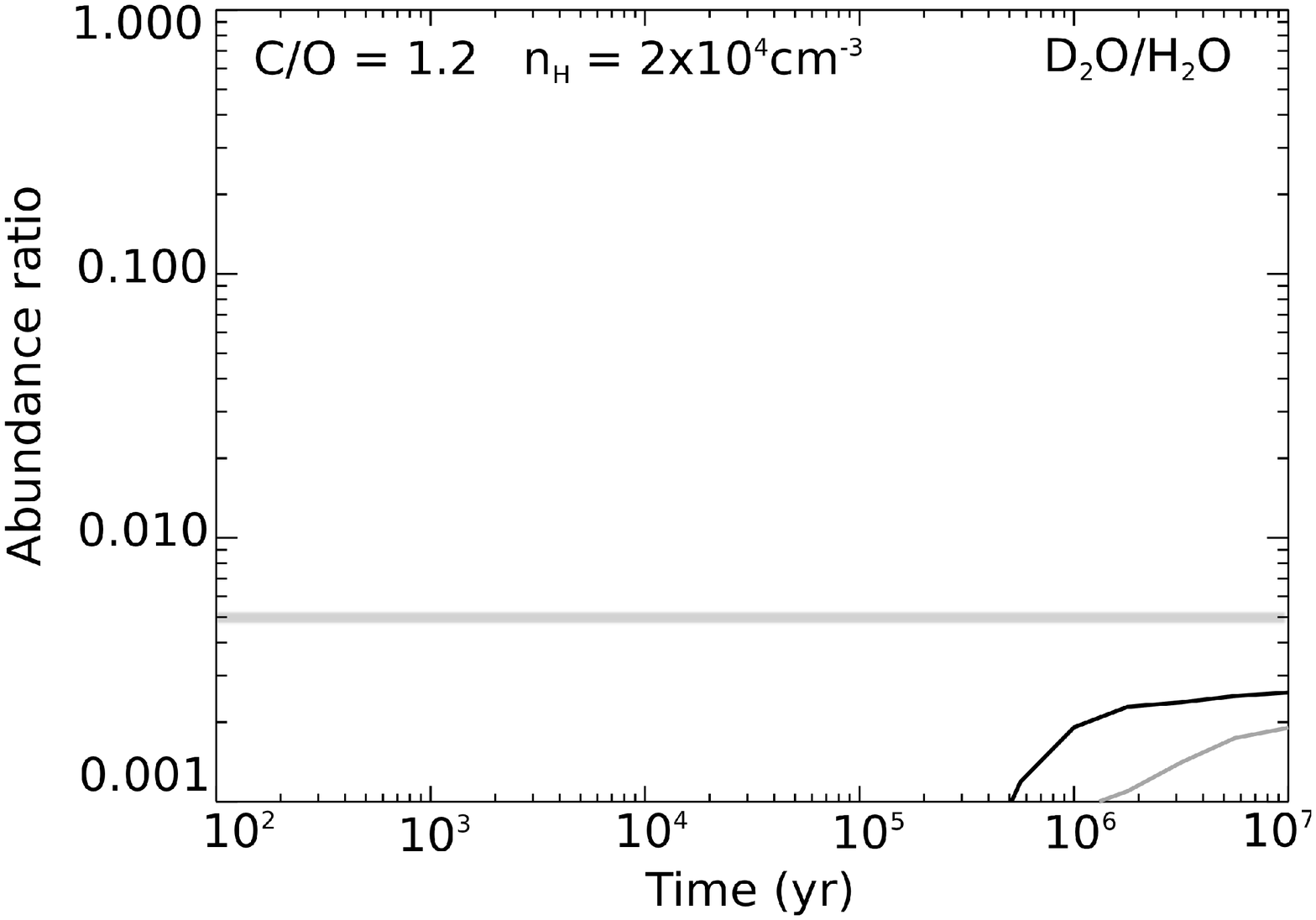}
\includegraphics[width=0.4\linewidth]{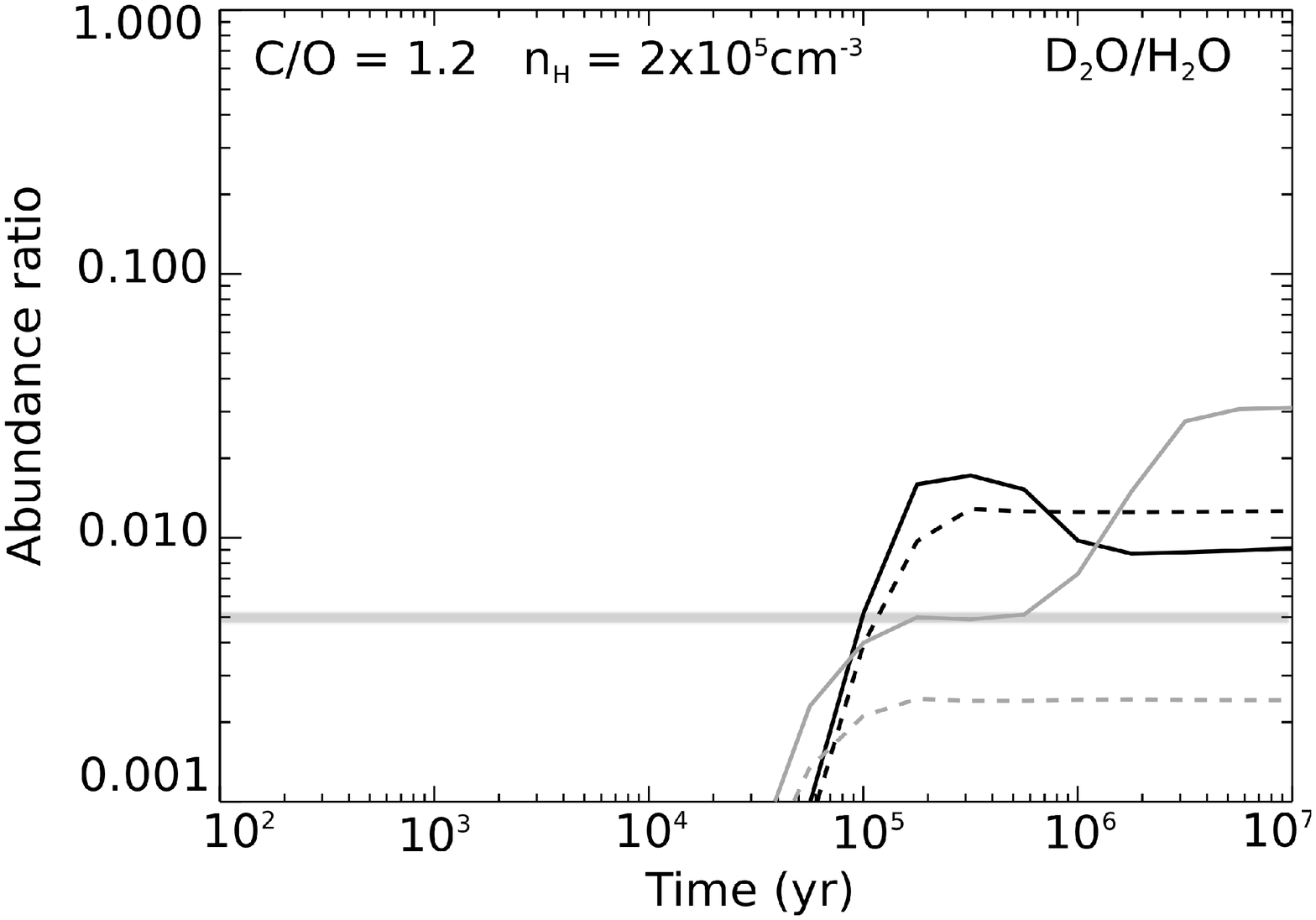}
\caption{ Same as Fig.~\ref{cloud_C_O_0.5_ratio} but a C/O of 1.2.  \label{cloud_C_O_1.2_ratio}}
\end{figure*}

\begin{figure*}
\includegraphics[width=0.4\linewidth]{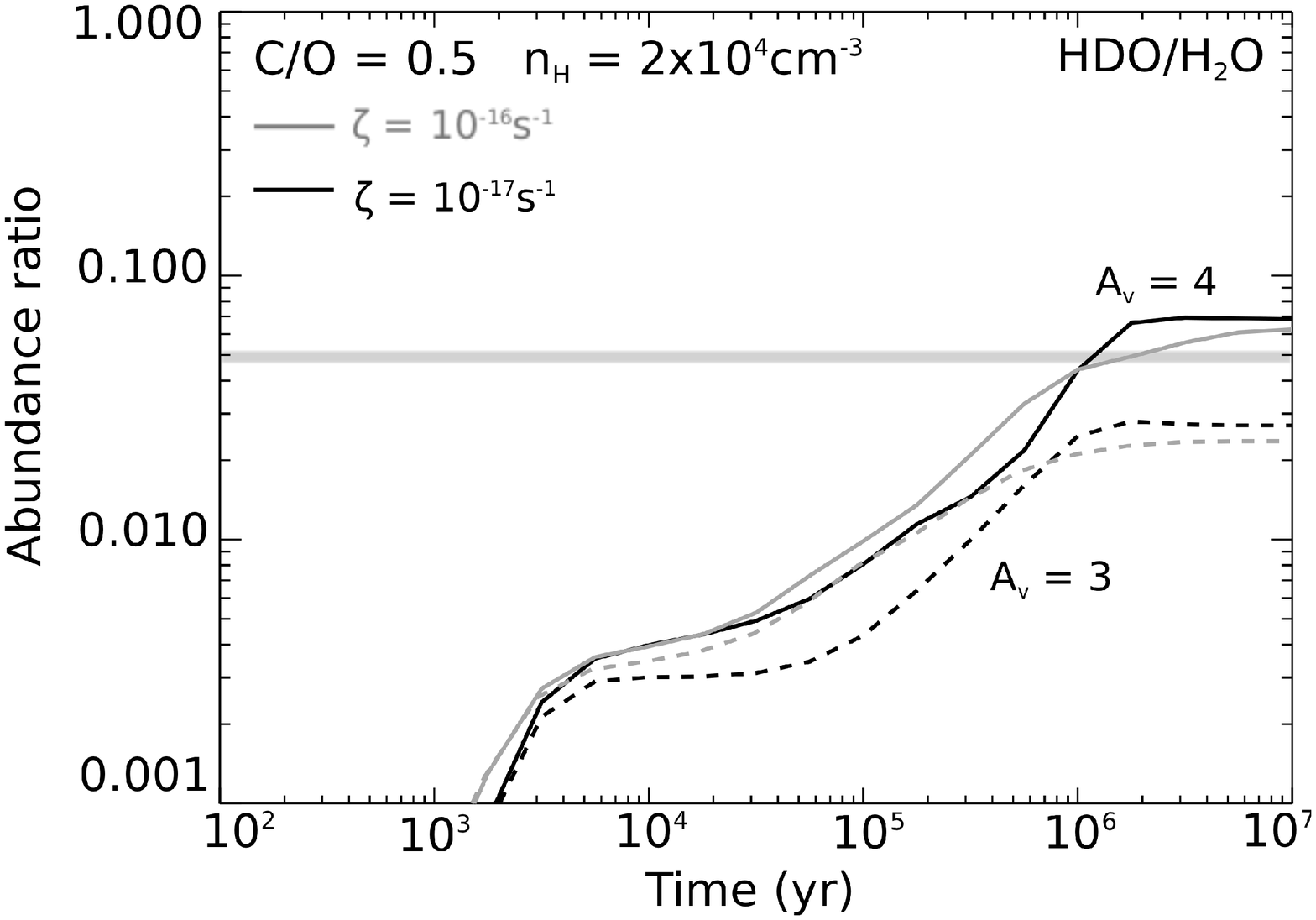}
\includegraphics[width=0.4\linewidth]{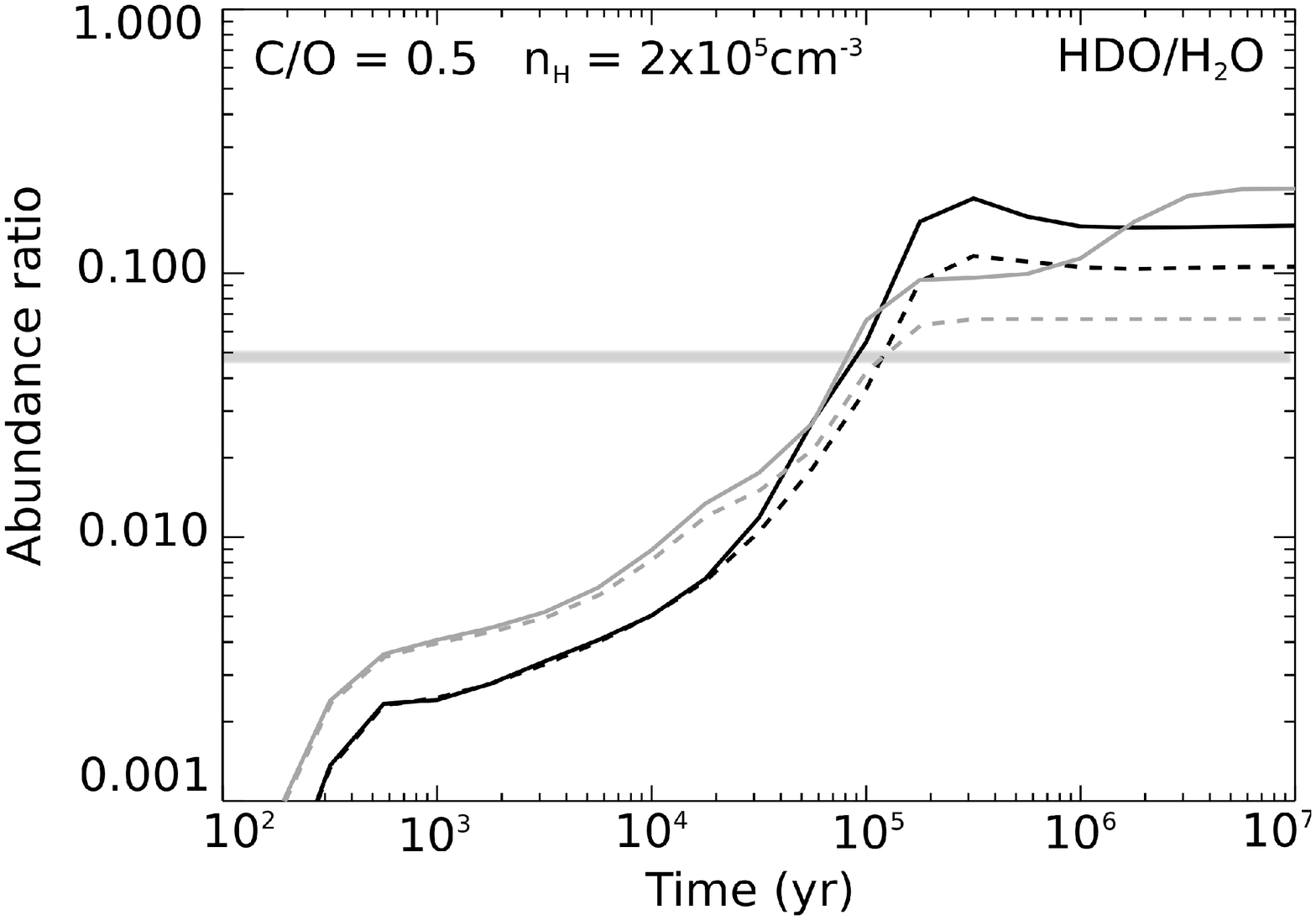}
\includegraphics[width=0.4\linewidth]{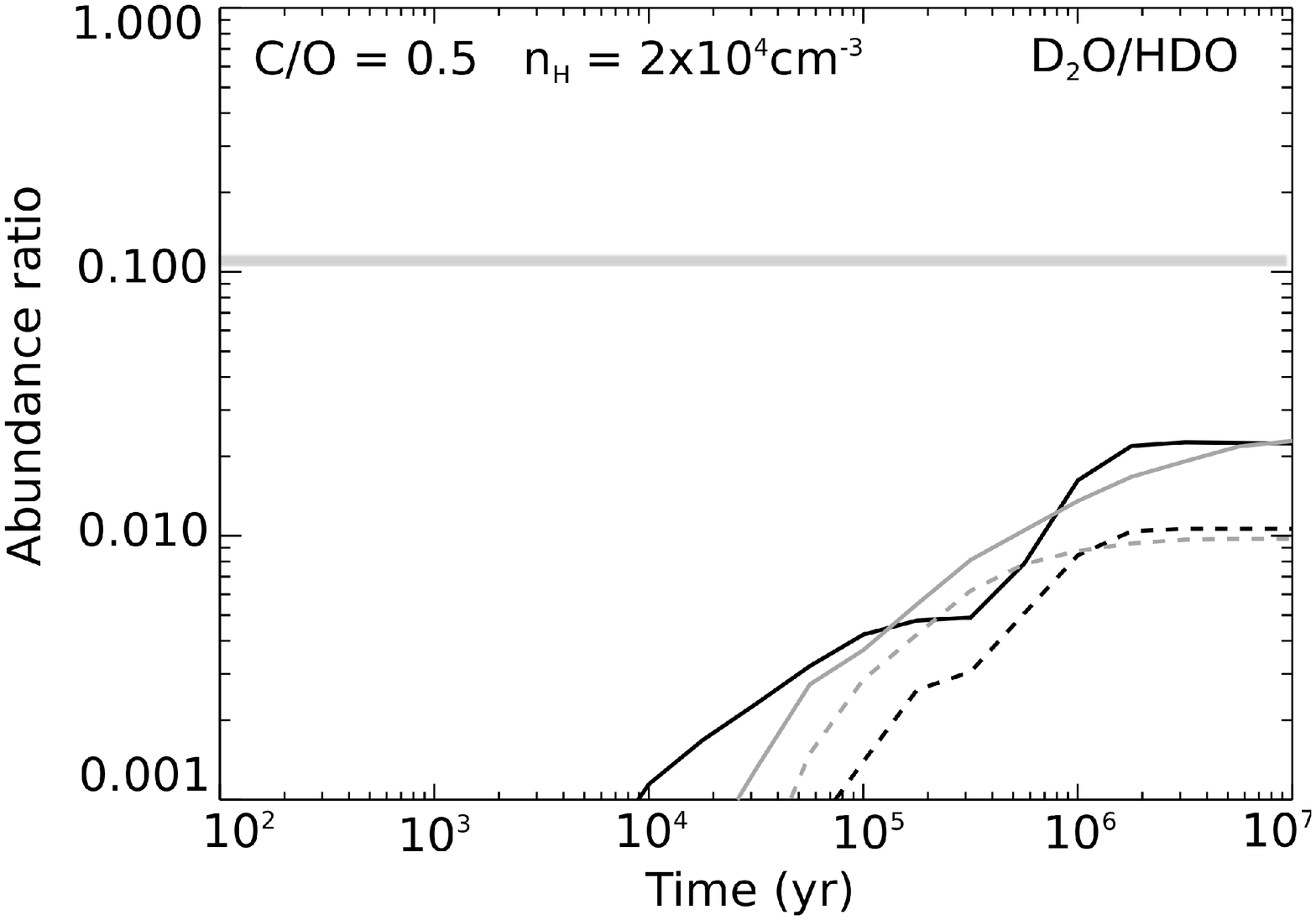}
\includegraphics[width=0.4\linewidth]{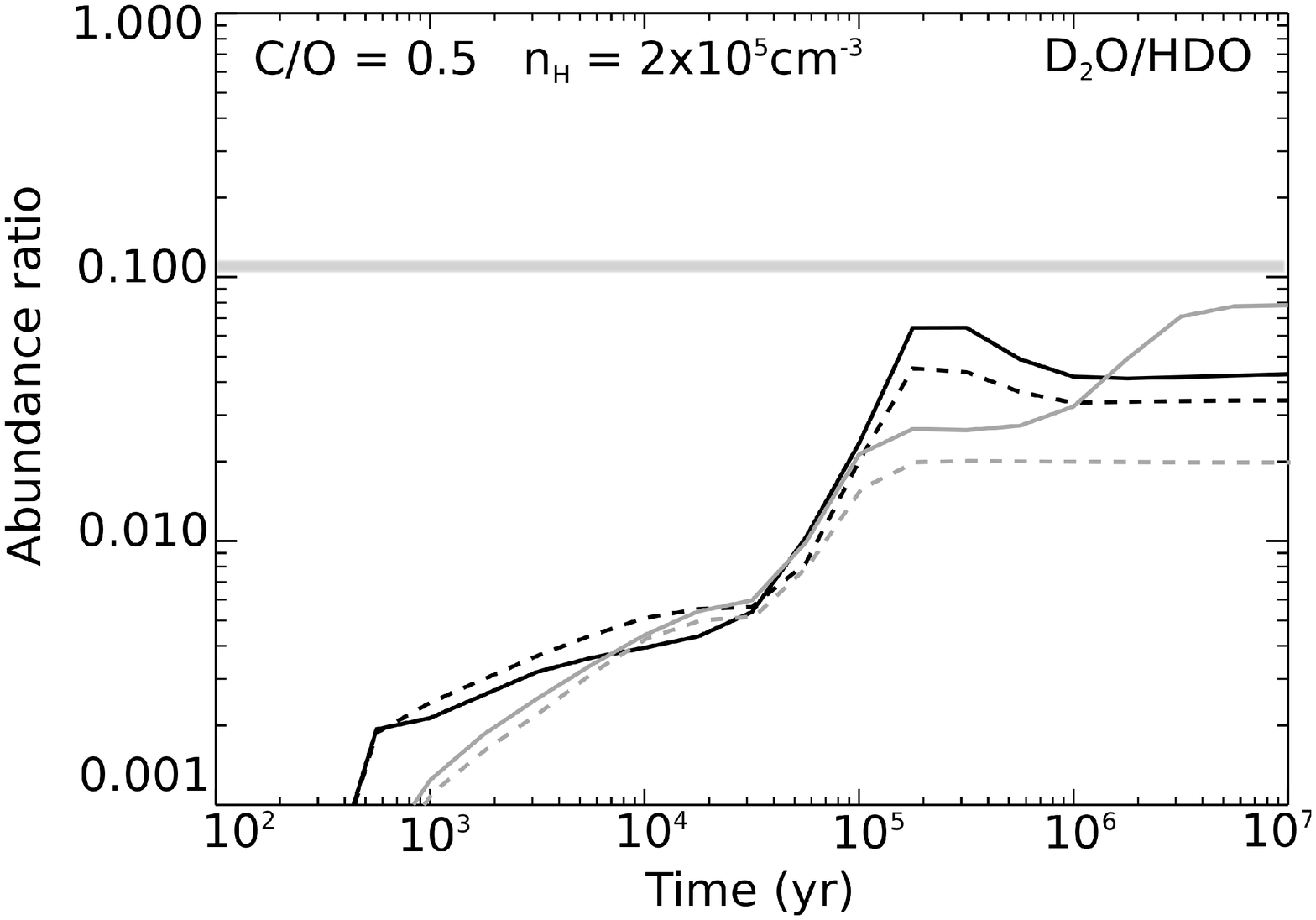}
\includegraphics[width=0.4\linewidth]{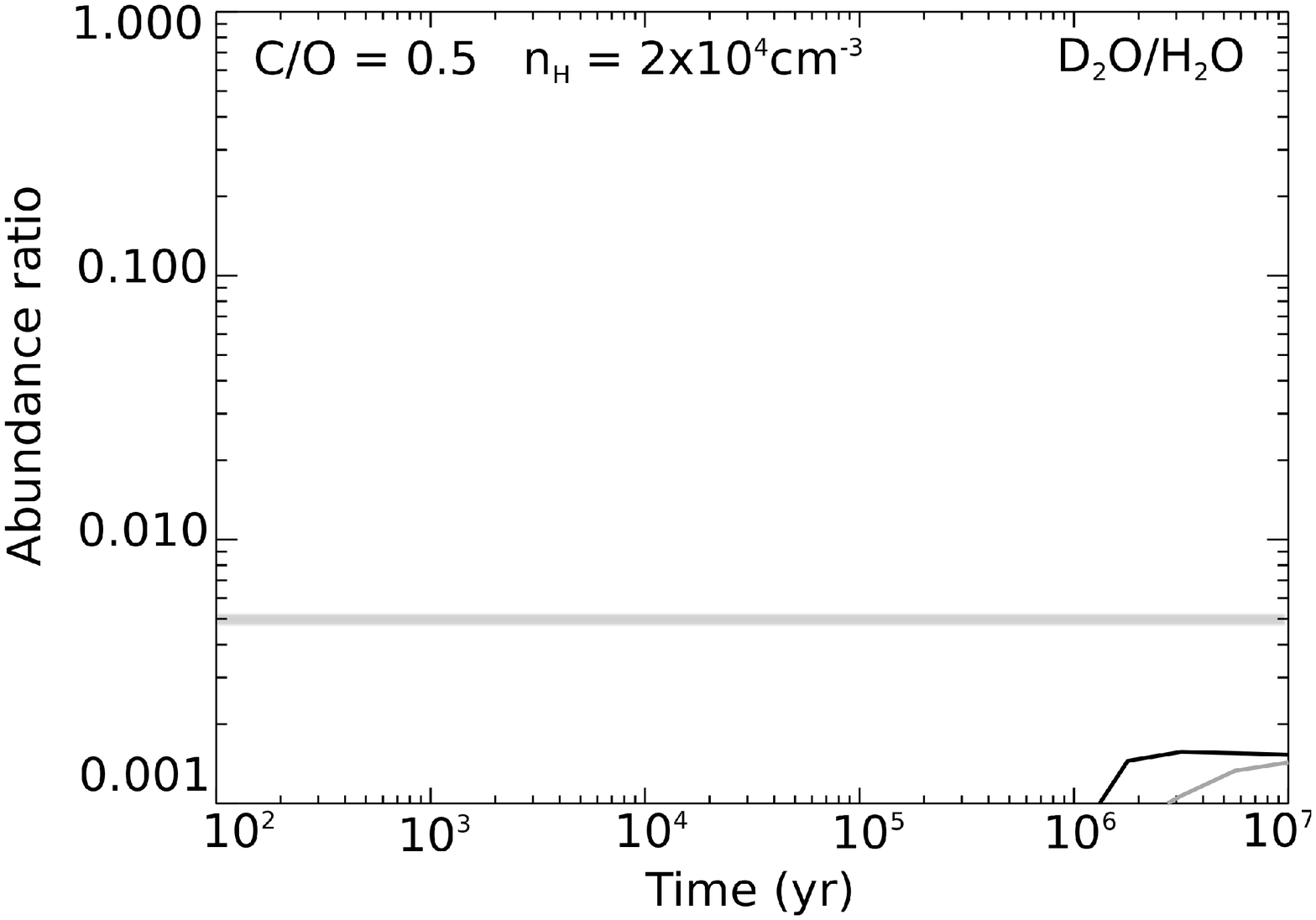}
\includegraphics[width=0.4\linewidth]{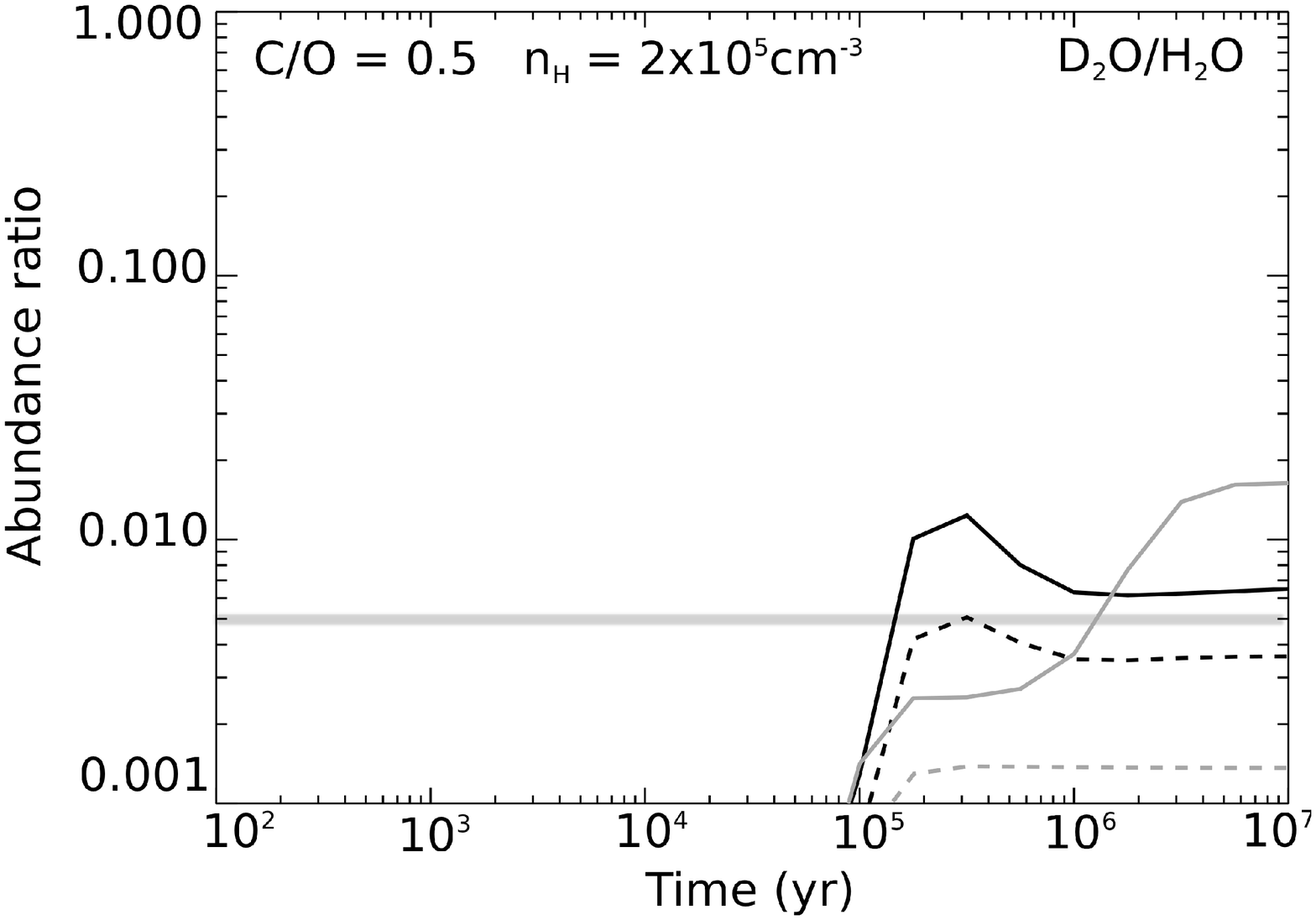}
\caption{Gas-phase abundance ratios HDO/H$_2$O, D$_2$O/HDO and D$_2$O/H$_2$O as a function of time predicted by the model for the foreground cloud.  Gas and dust temperature is the same for all models: 15~K. The elemental C/O abundance ratio is 0.5. The left side of the model has been obtained for a total H density of $2\times 10^4$~cm$^{-3}$ and the right side for a total H density of $2\times 10^5$~cm$^{-3}$. Black curves have been obtained for a cosmic-ray ionization rate $\zeta$ of $10^{-17}$~s$^{-1}$ whereas grey curves have been obtained for a $\zeta$ of  $10^{-16}$~s$^{-1}$. Solid lines represent an A$_V$ of 4  and dashed lines an A$_V$ of 3.  \label{cloud_C_O_0.5_ratio}}
\end{figure*}

The difficulty to compare those models with the observations comes from the small constraints on the physical structure of the source and so the large number of free parameters in our model. The observed abundances in the foreground cloud have been superimposed on Figs.~\ref{cloud_C_O_1.2} and \ref{cloud_C_O_0.5} to the model results. D$_2$O gives the strongest constraints on the model. For a C/O ratio of 1.2 and if we consider a factor of 2 uncertainty in the observed abundances, only the model with a visual extinction of 4 and a $\zeta$ of $10^{-16}$~s$^{-1}$ can reproduce the observations at late times ($> 3\times 10^6$~yr) for a density of $2\times 10^4$~cm$^{-3}$ and between $2\times 10^5$ and $10^6$~yr for a larger density of $2\times 10^5$~cm$^{-3}$. Under those conditions, the observed abundances of HDO and H$_2$O are also reproduced. Using the C/O elemental ratio of 0.5, less constraints can be brought on the other parameters. D$_2$O for instance is sufficiently produced whatever A$_V$ and $\zeta$ for a density of $2\times 10^5$~cm$^{-3}$. For a ten times less dense cloud, a large A$_V$ of 4 is required. From a general point of view, the constraints set by the HDO and D$_2$O abundances are small in that case. The predicted gas-phase abundance ratios HDO/H$_2$O, D$_2$O/HDO and D$_2$O/H$_2$O as a function of time are shown in Figs.~\ref{cloud_C_O_0.5_ratio} and \ref{cloud_C_O_1.2_ratio} with the observational constraints. Whatever the C/O elemental ratio, the models with the smallest density ($2\times 10^4$~cm$^{-3}$) seem to underestimate the D$_2$O/HDO and D$_2$O/H$_2$O ratios. 

\begin{figure*}
\includegraphics[width=0.4\linewidth]{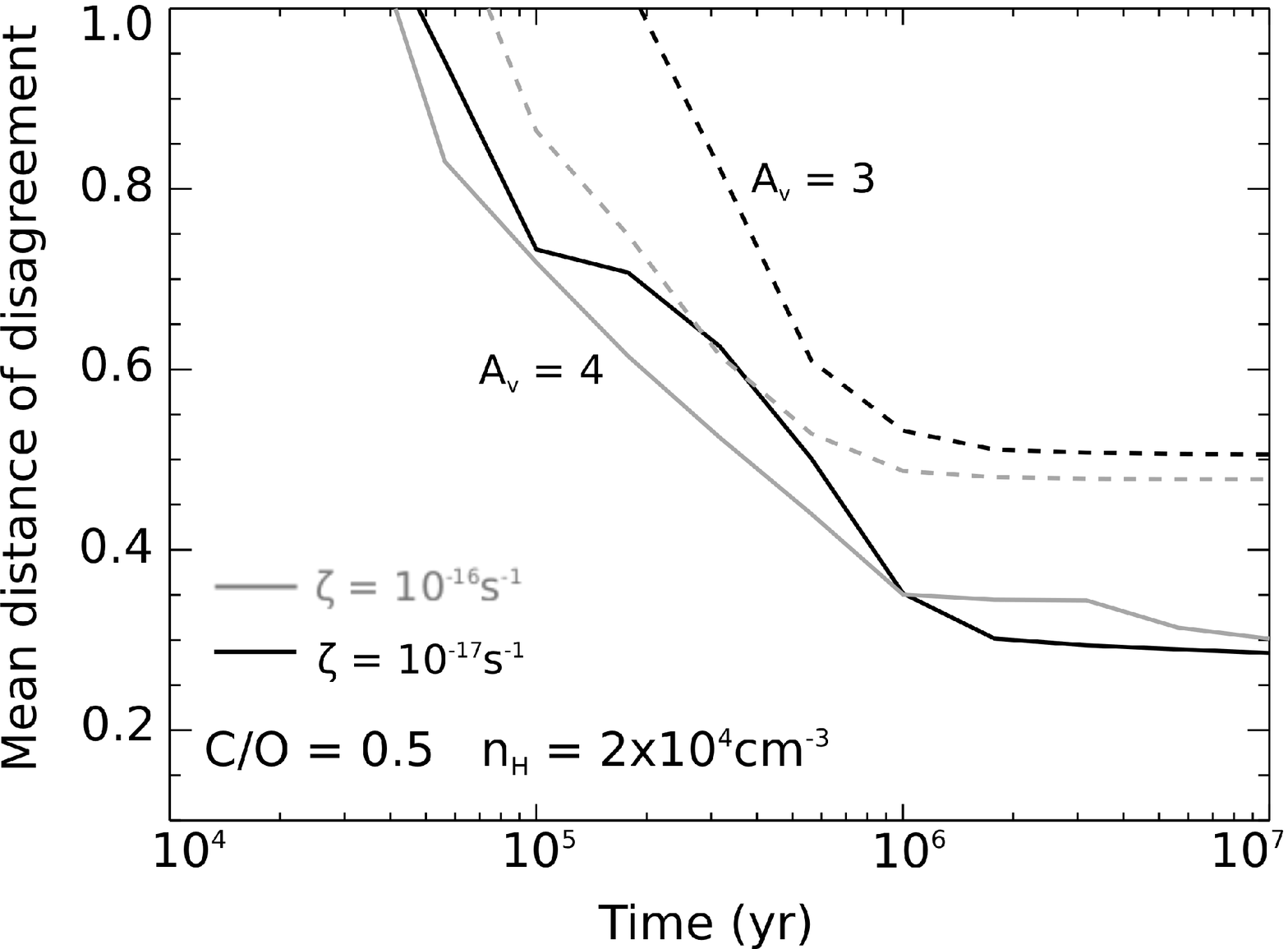}
\includegraphics[width=0.4\linewidth]{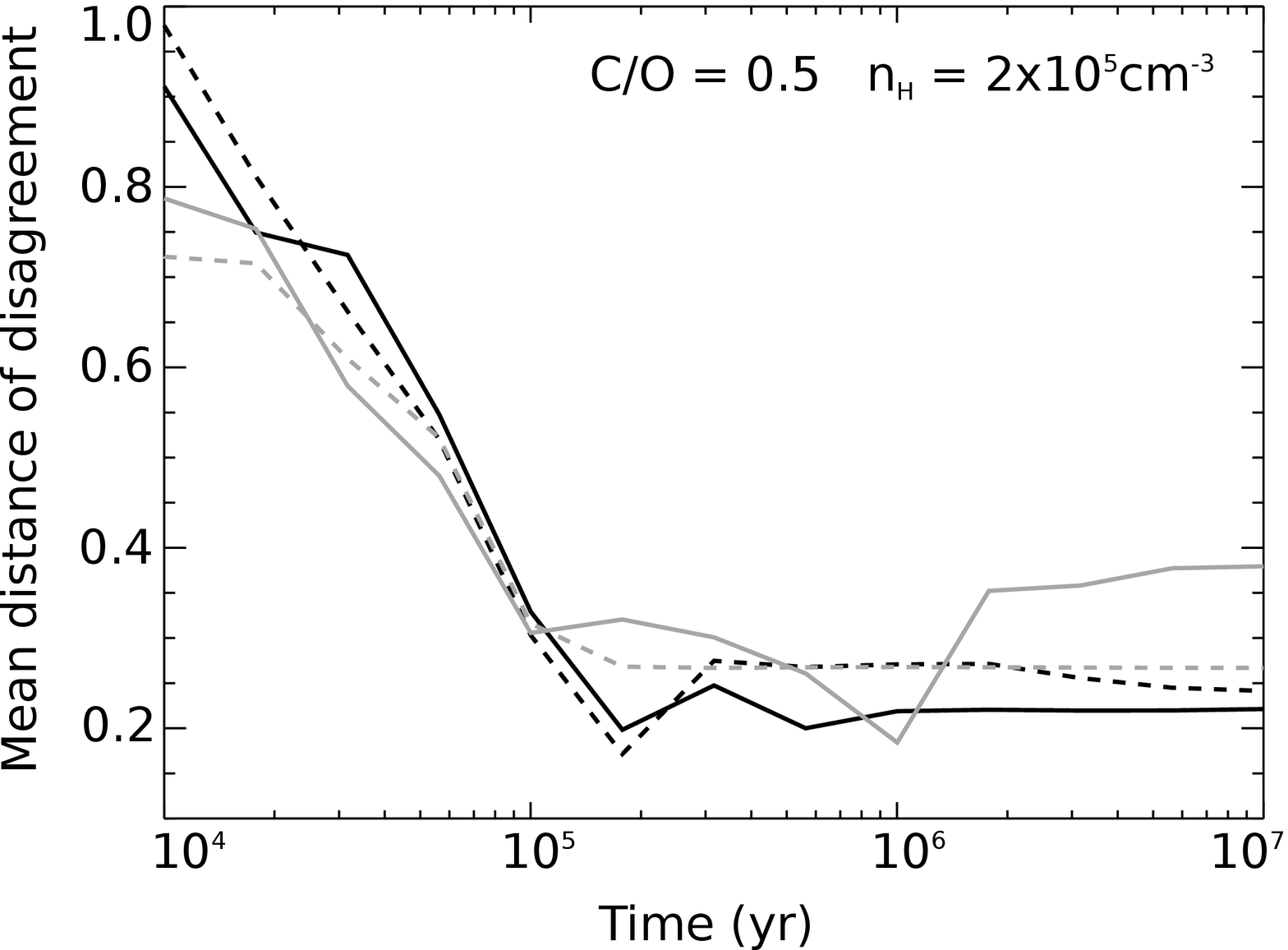}
\includegraphics[width=0.4\linewidth]{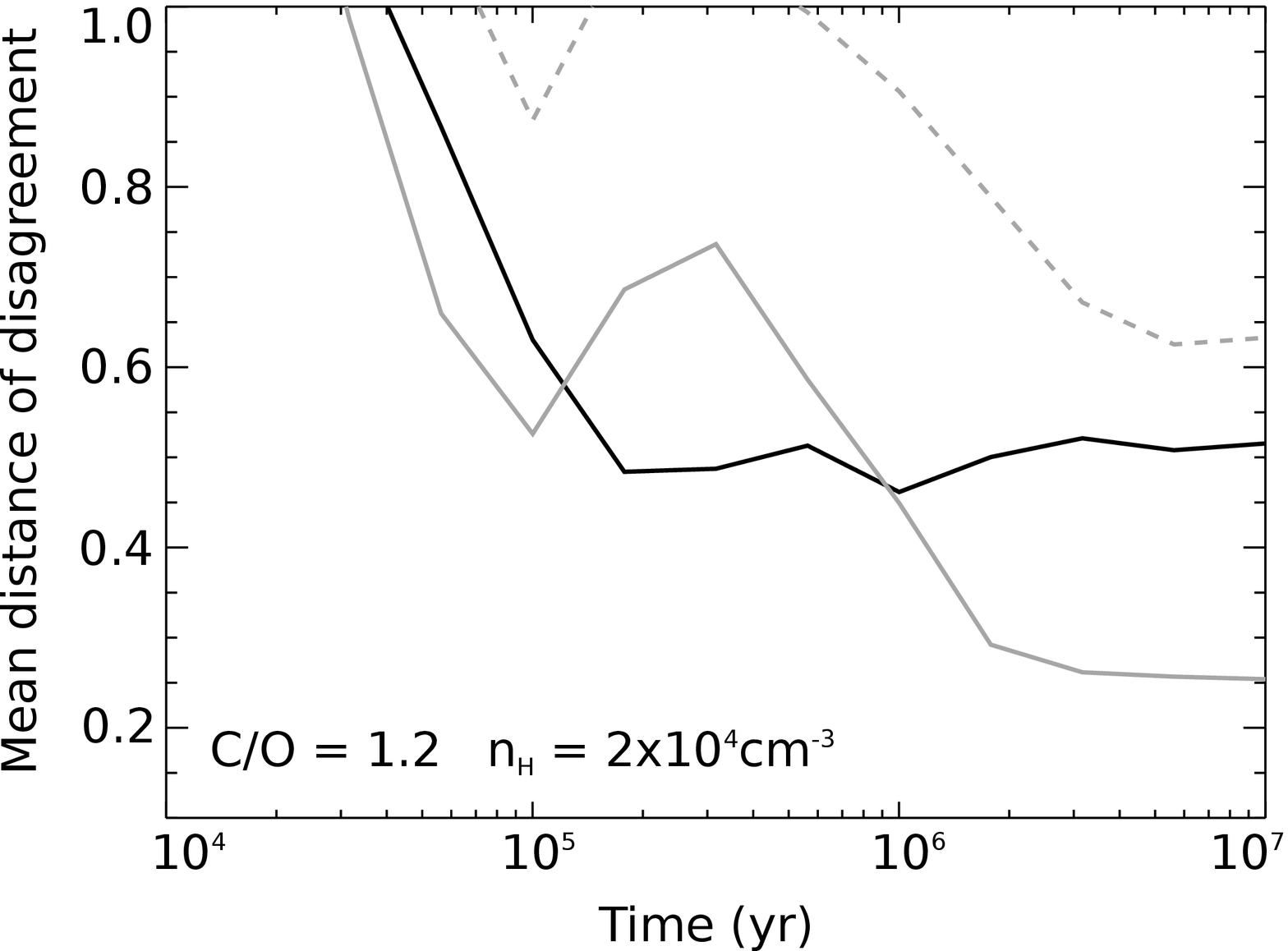}
\includegraphics[width=0.4\linewidth]{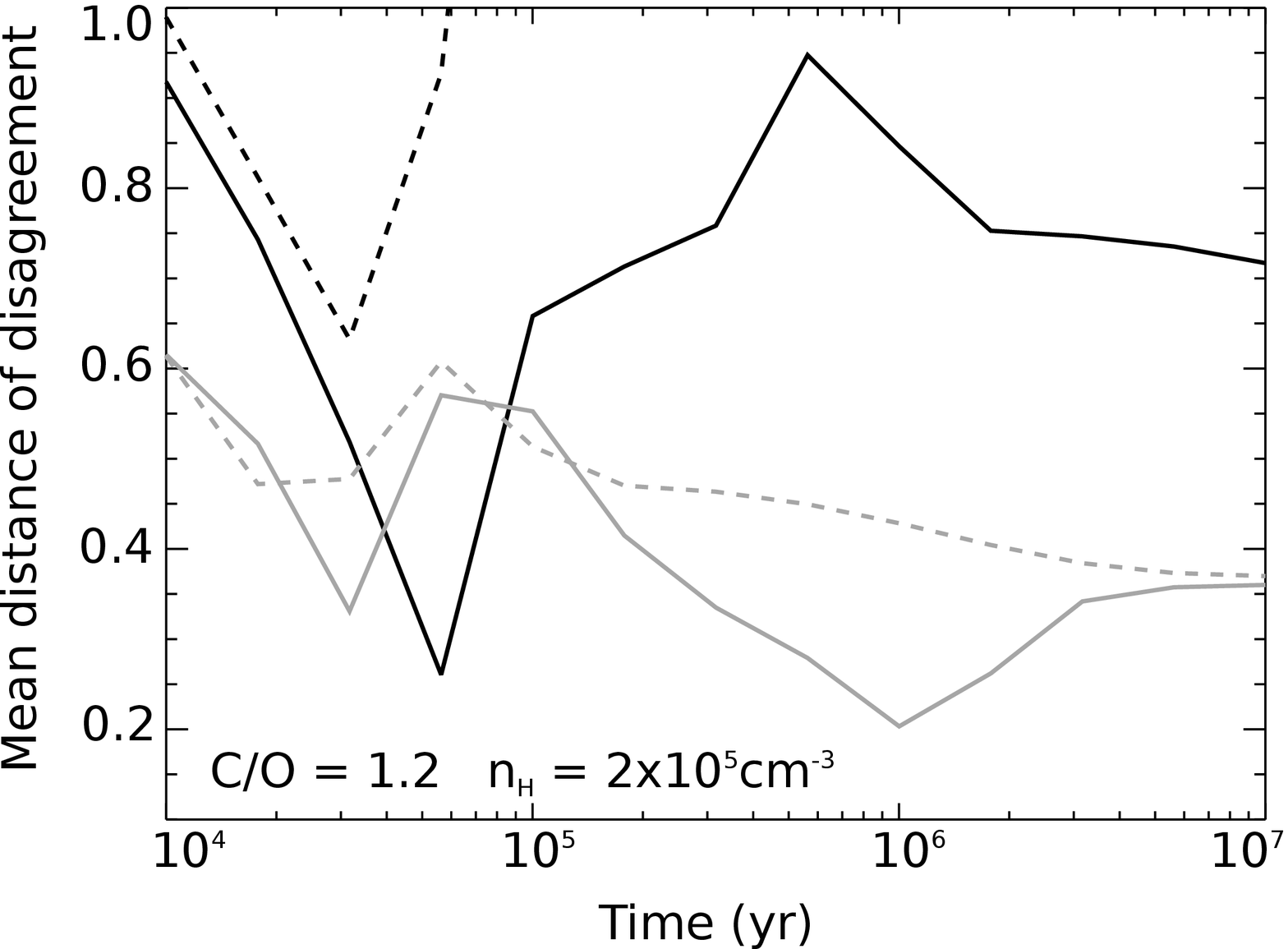}
\caption{Mean distance of disagreement as a function of time for the foreground cloud. Gas and dust temperature is the same for all models: 15~K. The elemental C/O abundance ratio is 0.5 for the upper panels and 1.2 for the lower panels. The left side of the model has been obtained for a total H density of $2\times 10^4$~cm$^{-3}$ and the right side for a total H density of $2\times 10^5$~cm$^{-3}$. Black curves have been obtained for a cosmic-ray ionization rate $\zeta$ of $10^{-17}$~s$^{-1}$ whereas grey curves have been obtained for a $\zeta$ of  $10^{-16}$~s$^{-1}$. Solid lines represent an A$_V$ of 4  and dashed lines an A$_V$ of 3.  \label{agreement}}
\end{figure*}

To compare the models to the observations in a more quantitative way, we have used the method introduced in \citet{2014MNRAS.437..930L}. For the three molecules, we have compared the observed and modeled abundances by calculating the mean distance of disagreement as a function of time : $\Sigma \vert \rm log(X_i)-log(X_{i,obs})\vert /3$, with $\rm X_i$ the abundance of species i (H$_2$O, HDO or D$_2$O) computed by the model and $\rm X_{i,obs}$ the observed abundance. Such calculation has been done for all models and the results are shown in Fig.~\ref{agreement}. Considering the models with a C/O elemental ratio of 0.5, the higher density ($2\times 10^5$~cm$^{-3}$) gives the best agreement whatever the A$_V$ and $\zeta$. For these later cases, the mean distance of disagreement is between 0.2 and 0.3 (in log), which means that the observed and modeled abundances show a difference between a factor of 1.6 and 2 in average.
For a larger C/O ratio, only the case with a large $\zeta$ and A$_V$ gives a satisfactory agreement for times equal to or larger than $10^6$~yr. 

\section{Discussion}\label{discussion}

\subsection{Effect of the photo-evaporation}

In the simulations that we have done for the protostellar outer envelope and the foreground cloud, the abundances of water and its deuterated forms are larger on the grain surfaces than in the gas-phase by several orders of magnitude in many cases. Without any non-thermal evaporation processes of the icy grain mantles, the predicted gas-phase abundances of these species would be much smaller. \citet{2009ApJ...690.1497H} found that using their model, they would produce large abundances of gas-phase water, similar to ours, at the border of dense clouds by photo-evaporation of the ices at A$_V$ between 3 and 8. Although we have included photo-evaporation of grain mantles by direct and indirect UV photons, it is not this process that drives the evaporation of the mantles and produce  large gas-phase H$_2$O abundances. In our case, the exothermicity of the surface reactions, that allow for partial evaporation of the products, releases H$_2$O, HDO and D$_2$O into the gas-phase by the following reactions: 
\begin{equation}
\rm H_{ice} +  OH_{ice} \rightarrow H_2O_{gas},
\end{equation}
\begin{equation}
\rm H_{ice} +  OD_{ice} \rightarrow HDO_{gas}, 
\end{equation}
\begin{equation}
\rm D_{ice} +  OH_{ice} \rightarrow HDO_{gas}, 
\end{equation}
\begin{equation}
\rm D_{gas} +  OD_{ice} \rightarrow D_2O_{gas}. 
\end{equation}
Removing this process decreases the gas-phase abundances of water and deuterated water by a factor of 10 in some cases. Following \citet{2007A&A...467.1103G}, the fraction of products allowed to evaporate is 0.7\%.  This value is, however, far from being constrained.  Previous experiments from \citet{Kroes2006} have set a 1\% limit for the efficiency of this process whereas \citet{2013NatSR...3E1338D} have recently found that the chemi-desorption could be more efficient in particular during the formation of water. Increasing the fraction of desorption in our model would increase the abundances of these molecules in the gas-phase. 

The direct photo-evaporation of icy mantles by UV photons depends on the intensity of the UV irradiation field but does not increase linearly with it. Although the rate of photo-evaporation is proportional to the intensity of the UV field, the rate is also multiplied by a dimensionless factor that indirectly depends on the UV field: $\rm f(X) = n_s(X)/n_{ice}$ where $\rm n_s(X)$ is the number density of species X in the ices and  $n_{ice}$ is the total number density of all ice species. This factor aims at taking into account the 0 order behavior of this reaction when more than one monolayer of molecules is present on the surfaces \citep[see][]{2011A&A...534A.132V}. When the intensity of the UV field is changed, the number of species on the surfaces is also changed. All simulations presented in this paper have been obtained for a standard interstellar UV field (1~G$_0$). For an A$_V$ of 4, if the UV field is 10 times larger, the HDO and D$_2$O abundances are decreased due to more photodissociations. Even in that case, the photo-evaporation processes remain unimportant compared to the exothermicity of the surface reactions, i.e. models with and without photo evaporation will produce the same results.

\subsection{Effect of C/O and cosmic-ray ionization rate on D/H}

\begin{figure*}
\includegraphics[width=0.4\linewidth]{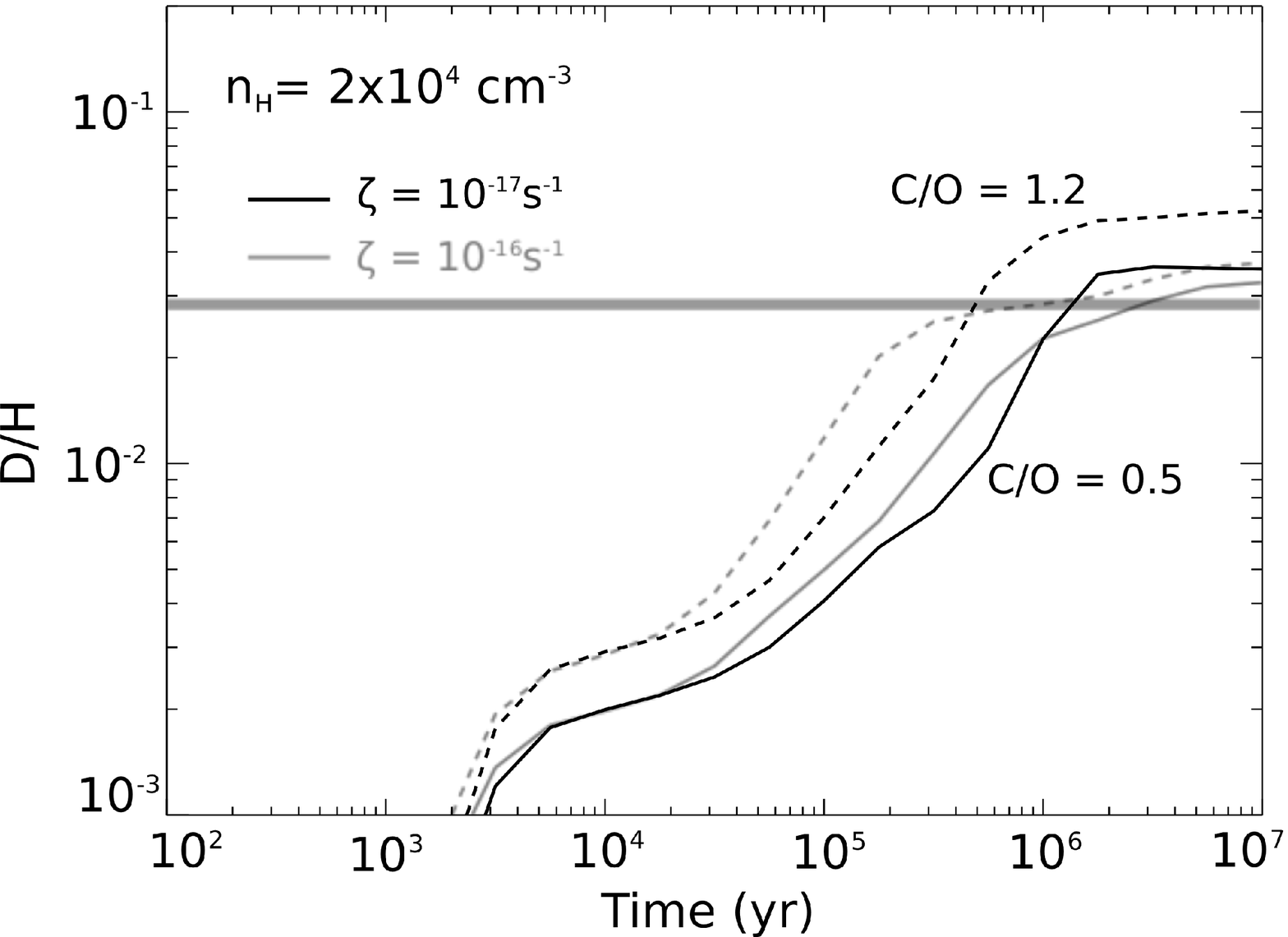}
\includegraphics[width=0.4\linewidth]{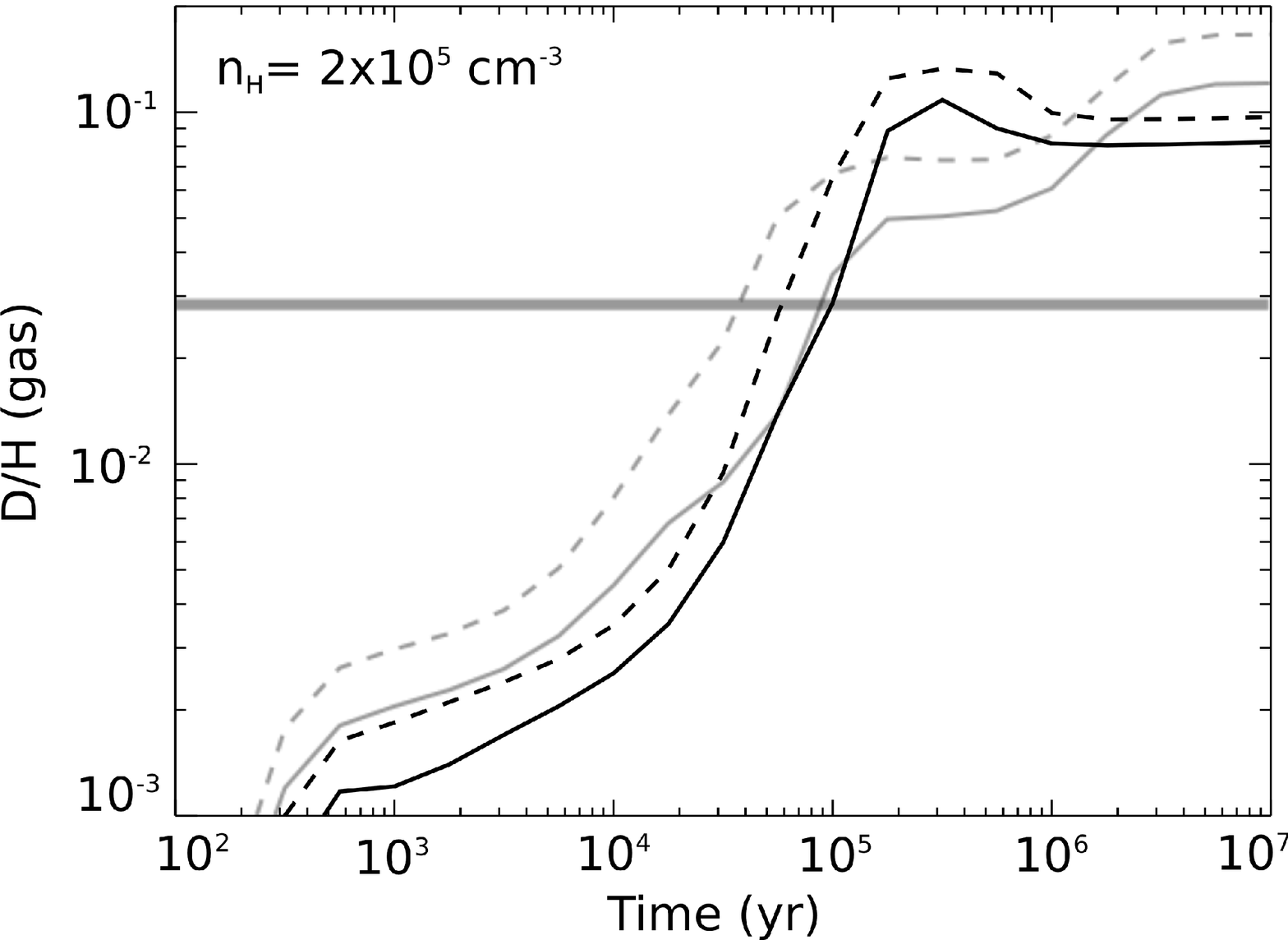}
\includegraphics[width=0.4\linewidth]{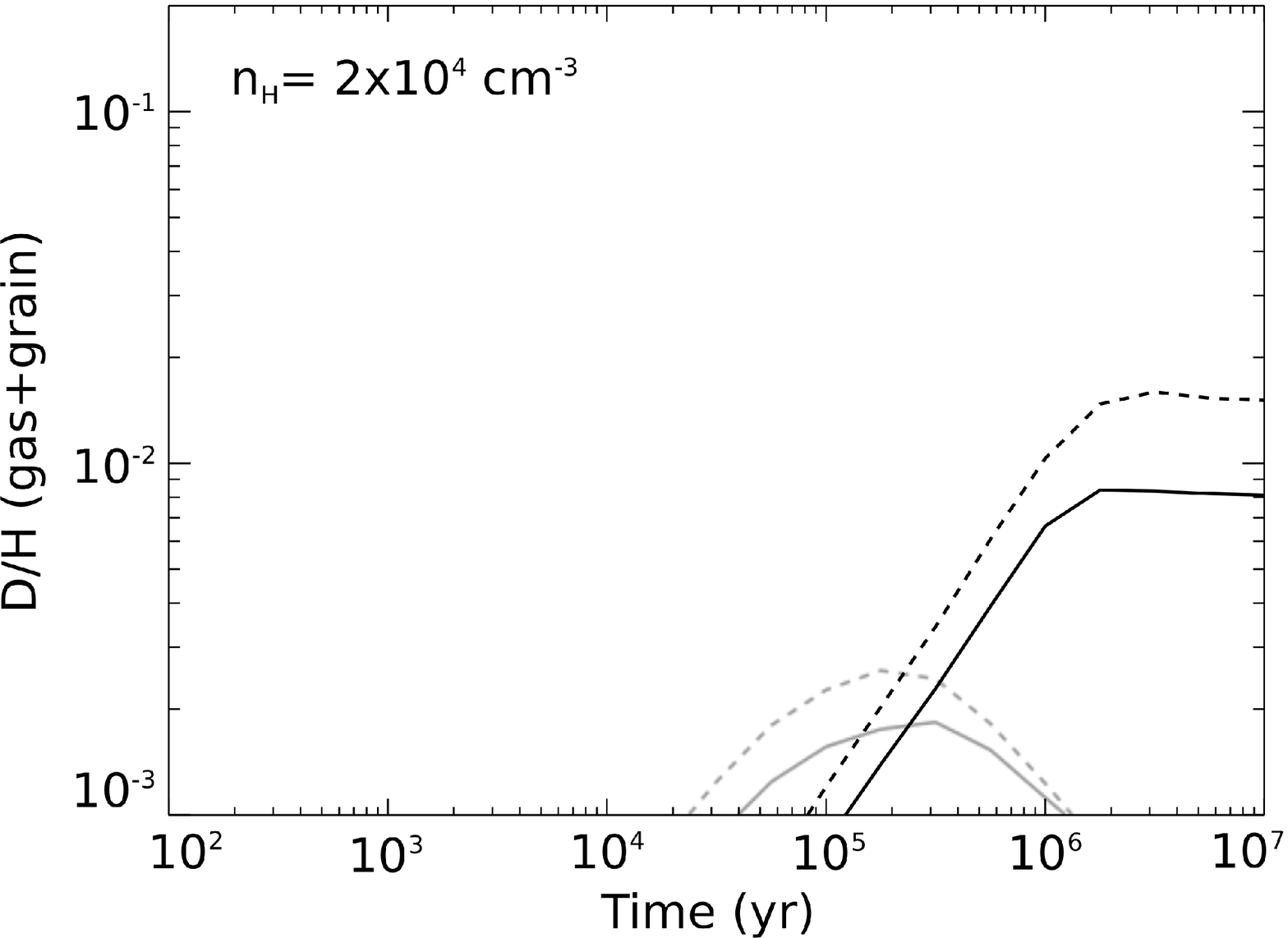}
\includegraphics[width=0.4\linewidth]{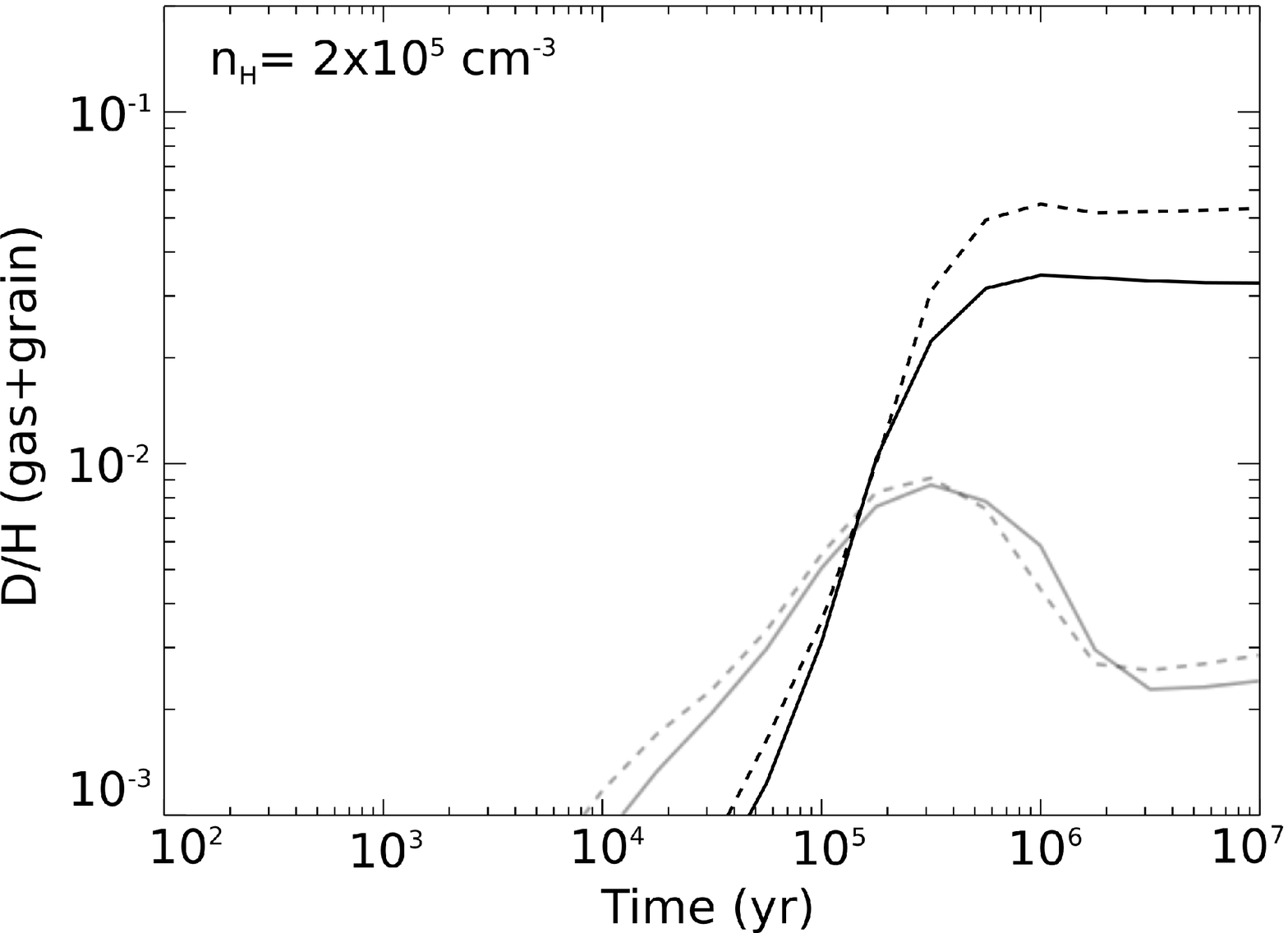}
\caption{D/H ratio predicted by the model in the gas-phase deuterated water (top row) and considering both the gas-phase and grain surface deuterated water (bottom row). Two densities are considered: $2\times 10^4$~cm$^{-3}$ on the left and $2\times 10^5$~cm$^{-3}$ on the right, two C/O elemental ratios (C/O = 0.5 for the solid curves and C/O = 1.2 for dashed curves) and two cosmic-ray ionization rates ($10^{-17}$~s$^{-1}$ for black curves and $10^{-16}$~s$^{-1}$ for grey curves).  Grey area represents the D/H ratio observed in the foreground cloud. \label{D_H}}
\end{figure*}

To look at the effect of the C/O elemental ratio and the cosmic-ray ionization rate on the D/H ratio measured in deuterated water, we show in Fig.~\ref{D_H} the following abundance ratio $\rm (0.5X_{HDO} + X_{D_2O})/X_{H_2O}$ as a function of time for two densities ($2\times 10^4$ and $2\times 10^5$~cm$^{-3}$), two C/O elemental ratios (C/O = 0.5 and 1.2) and two cosmic-ray ionization rates ($10^{-17}$ and $10^{-16}$~s$^{-1}$). The temperature is 15~K and the A$_V$ is 4. We present the D/H ratio predicted 1) in the gas-phase only and 2) considering both the gas-phase and grain surface abundances.
The D/H ratio increases with time in all cases as the molecules are built up, except for the case of a larger $\zeta$ where the species at the surface of the grains are destroyed. Considering only water in the gas-phase, the D/H ratio is less sensitive to the C/O elemental ratio and $\zeta$ than to the density of the medium. Larger densities produce larger D/H ratios. Considering the full budget of water (both in the gas-phase and at the surface of the grains), D/H is strongly sensitive to the cosmic-ray ionization rate, even more than to the density. The larger $\zeta$ decreases strongly D/H at times larger than $10^5$~yr. In general, large C/O seems to produce slightly larger D/H.

\subsection{Elemental abundance of deuterium}

The deuterium elemental abundance is set to $1.5\times 10^{-5}$ compared to H, which is the average value found by absorption spectroscopy measurements towards stars in the vicinity of the Sun \citep[e.g.][]{2006ApJ...647.1106L}. \citet{2006ApJ...647.1106L} argued that these measurements inside the so-called "Local Bubble" may not be representative of the deuterium abundance in the galactic disk beyond. Indeed, spatial variations in the depletion of deuterium from the gas phase onto dust grains (Jura 1982; Draine 2004, 2006) can explain these local variations in the observed gas-phase D/ H ratios. 
In our standard model, we used the value of $1.5\times 10^{-5}$ and envision the influence of its uncertainty on our modeling.\\
When the deuterium elemental abundance is multiplied or divided by a factor of two compared to its nominal value given in section~\ref{proto_parameters}, the HDO gas-phase and ice abundances are also multiplied and divided but by less than a factor of two. Similarly, D$_2$O is modified by less than a factor of four. The abundance ratios are also modified by an amount depending on the model but, from a general point of view, an increase of D elemental abundance increases the three ratios HDO/H$_2$O, D$_2$O/HDO and D$_2$O/H$_2$O. 

\subsection{Importance of the ortho/para ratio of H$_2$ on the formation of deuterated water}\label{dis_opH2}

As mentioned in the presentation of the deuterated network (section~\ref{deut_network}), we did not include the ortho H$_2$. Ortho and para H$_2$ will react differently for some reactions. One reaction is of particular importance for the deuteration: H$_2$D$^+$ + H$_2$ $\rightarrow$ H$_3^+$ + HD. Para H$_2$ in this reaction reacts very slowly compared to ortho H$_2$. According to \citet{2009JChPh.130p4302H}, the difference between the two rate coefficients is more than four orders of magnitude at 10~K. A faster rate coefficient decreases the efficiency of the production of deuterated species. We have tested the effect of considering the higher reactivity of ortho H$_2$ by replacing the rate coefficient of this reaction by the one from ortho H$_2$  \citep[$\rm k(T) = 4.67\times 10^{-11} \exp{(0.82/T)}$,][]{2009JChPh.130p4302H}) times the ortho to para H$_2$ ratio. For the ortho to para H$_2$ ratio, we have considered three values that are fixed with time and temperature: $10^{-3}$ which is the value constrained by \citet{2012A&A...537A..20D} in the outer envelope of IRAS16293, 0.1 which is the value proposed by several authors in cold dense clouds \citep{2011ApJ...739L..35P,2012ApJ...759L..37C}, and 3 which is the statistical value of their formation on the grains. 
In all the simulations that we have done, a o/p H$_2$ ratio of $10^{-3}$ does not affect significantly the results presented in the paper. For larger values, it does and we will now discuss the main consequences. 

At the low temperatures considered for the foreground cloud, the Boltzmann value of the o/p H$_2$ ratio is very low ($10^{-4}$). The grains however are expected to form o-H$_2$ and p-H$_2$ in the ratio 3:1 so the main uncertainty should be the time of conversion from 3 to it Boltzmann value. We have shown the results of the tests in Fig.~\ref{opfigs}. Considering an o/p H$_2$ ratio of 0.1 and above decreases the predicted abundances of HDO and D$_2$O in the gas-phase too strongly compared to the observed values listed in Table~\ref{obs_tab}. Using a full gas-grain model including the ortho and para forms of H$_2$, \citet{2013A&A...550A.127T} concluded that the o/p H$_2$ ratio should be small (smaller than  $3\times 10^{-4}$) in IRAS16293 to reproduce the deuterated water in the outer parts of this source. \\
In the protostellar envelope, the gas temperature increases with time as the shells of material move towards the center. Considering fixed o/p H$_2$ ratios is then a strong approximation but allows us to frame the effect. The modeling results are shown in Fig.~\ref{opfigs_proto}. Increasing the o/p H$_2$ ratio decreases the abundances of HDO and D$_2$O in the gas-phase without changing significantly H$_2$O in the entire envelope. In the inner hot corino region, the HDO and D$_2$O abundances get closer to the observational constraints from \citet{2013A&A...560A..39C} but in that case the abundance ratios HDO/H$_2$O and D$_2$O/H$_2$O are not reproduced anymore. In the case of a fixed o/p H$_2$ ratio of 0.01, the HDO/H$_2$O abundance ratio is $2\times 10^{-3}$ and for an o/p H$_2$ ratio of 3, it becomes $5\times 10^{-4}$. In those cases, the HDO/H$_2$O ratio is in agreement the values observed by \citet{2013A&A...549L...3P}. In the intermediate and outer regions of the envelope, the HDO abundance gets closer to the observations but is still over predicted by the model even for an o/p H$_2$ ratio of 3, as well as the HDO/H$_2$O ratio. The D$_2$O/H$_2$O abundance ratio becomes in agreement with the observations for o/p H$_2$ larger or equal to 0.01. The HDO/H$_2$O ratio is still predicted by the model to be larger than D$_2$O/HDO. To maybe be more realistic, we redid those simulations but assumed an o/p H$_2$ ratio temperature dependent using the relation o/p H$_2$ ratio = $9\times \exp{-170.5/T}$ at the Boltzmann equilibrium. {\bf Using this relation, the o/p H$_2$ ratio increases along with the temperature during the collapse of the protostellar envelope. Considering the shell of material that ends up in the inner region of the envelope (inside 75~AU), the initial o/p H$_2$ ratio is only $5\times 10^{-9}$ (T=8~K) whereas it is above 1.6 at the end of the simulations (T$\leq$100~K). However, the timescale of the collapse implies that the material stays at low temperature for a much longer period of time than at high temperature \citep[see][]{2008ApJ...674..984A}. In fact, as previously mentioned, the deuterated water is formed in the ices at low temperature when the o/p H$_2$ ratio is low. Then when the temperature increases, HDO and D$_2$O are evaporated in the gas-phase but not efficiently destroyed. As a consequence the abundance predictions are not changed compared to our nominal model. If the collapse was much faster, with an initial pre-collapse phase shorter than $2\times 10^5$~yr, the formation of deuterated water could be less efficient producing a smaller HDO/H$_2$O ratio. Further investigation in this direction should however be done to quantitatively conclude on this question. }

\subsection{The physical structure of the protostellar envelope} \label{disc_phys}

To analyze the observed lines of H$_2$O, HDO and D$_2$O in the envelope of IRAS16293, \citet{2012A&A...539A.132C,2013A&A...553A..75C} used the physical structure constrained by \citet{2010A&A...519A..65C} from dust emission at several wavelength (see Fig.~\ref{comp_struct}). These temperature and density profiles are similar to the ones we obtain for a protostellar age of $9\times 10^4$~yr using the RHD model from \citet{2000ApJ...531..350M} except that the densities are approximately ten times larger at a given radius. To be more consistent with the observations, we have thus multiplied the densities at each radius and all along the calculation by a factor of ten. Doing that, the temperatures and timescales of collapse are not consistent anymore. \\
The RHD model starts with a hydrostatic sphere of 3.852~M$\odot$ over a radius of $4\times 10^4$~AU. At the final time ($9\times 10^4$~yr after the birth of the protostar), the mass of the protostar (inside 4~R$\odot$) is approximately 1~M$\odot$ and the luminosity 31~L$\odot$. The mass accretion rate is constant with time to a value of $\sim 5.3\times 10^{-6}$~M$\odot$ yr$^{-1}$. \citet{2010A&A...519A..65C} estimated a protostellar mass of 2~M$\odot$ with a luminosity of 22~L$\odot$. The mass is then two times larger than in our case whereas the luminosity is smaller. The mass accretion rate was determined by \citet{2000A&A...355.1129C} to be $3.5\times 10^{-5}$~M$\odot$ yr$^{-1}$, much larger than in our model. Other numbers for the mass of the protostar have been published in the literature ranging from 0.8~M$\odot$ \citep{2000A&A...355.1129C} to 6.1~M$\odot$ \citep{2004ApJ...608..341S}. Using ALMA observations of dust and complex molecules, \citet{2012A&A...544L...7P} derived a mass of 0.21 and 0.26~M$\odot$ for sources A and B. IRAS1293 is indeed a binary (even maybe triple) system. All observations discussed up to now in this paper are single dish and so do not separate the different sources. The interferometric observations show that the two sources present very different kinematics. B presents clear infall signature whereas A seems to be consistent with a rotating disk. Considering all these uncertainties in the observational constraints of IRAS16293 physical structure, it seems difficult to improve our modeling at this point. 

That said, we can discuss the feedback of increasing the densities in our model and the importance of these assumptions for the chemistry. A denser envelope has two opposite effects on the temperature. Since the core is heated on the inside, the higher density would increase the opacity of the core and then decrease the temperature. On the other hand, a denser and so more massive envelope would mean a larger mass accretion rate, which would produce larger temperatures. To understand the feedback of a denser envelope on the temperature, the RHD model would have to be run again using different parameters. This will be done in the future. Another effect of the denser envelope is to slow down the time of free-fall. To test this effect, we have run our model reducing dynamical time of collapse of all cells by a factor of 3. This does not change significantly the results. In addition, we run our model using the original density profile from the RHD model. This on the contrary modified strongly our results for radius larger than 100~AU. Smaller densities produce less depletion and so the H$_2$O, HDO and D$_2$O abundances are larger but not affected the same way. As a consequence, the abundance ratios are very different from the ones shown in this paper. This result points out the crucial need for accurate estimates of the physical conditions in protostellar envelope for chemical studies. 

\subsection{Comparison with other works} \label{other_works}

Deuteration in star forming regions have been extensively studied from the observation and modeling points of view. We will compare our results with a few other recent published works on the modeling of deuterated water. \citet{2011ApJ...741L..34C} studied the formation of deuterated species in ices in translucent clouds evolving towards denser clouds. The model does not include gas-phase chemistry but allows the sticking of gas-phase species onto the surface of the grains and follows the formation of the species at the surface of the grains. The density of the cloud is computed by the equation of free-fall whereas several dust temperatures are considered from 12~K to 17~K. The authors found that the formation of deuterated water on ices strongly depends on the dust temperature within this range: at low temperature the large abundance of H$_2$ on the surface favors the formation of water whereas at 17~K, H$_2$ evaporates and produces larger HDO/H$_2$O ratios. The considered reactions of oxygen with molecular hydrogen are endothermic and in our model, as in \citet{2013A&A...550A.127T}, the formation of water mostly occurs through reactions with atomic hydrogen. As a consequence, we do not expect to find such a strong temperature sensitivity. 

Three other deuterated models have recently been published using a gas-grain chemical model similar to ours: \citet{2013ApJS..207...27A}, \citet{2013A&A...550A.127T} and \citet{2013A&A...554A..92S}.   \citet{2013ApJS..207...27A} studied the deuteration in a large range of temperatures and density but for a fixed visual extinction (of 10), cosmic-ray ionization rate (of $10^{-17}$~s$^{-1}$) and evolution time (of 1~Myr). Similarly to \citet{2012ApJ...760...40A}, they extended a large gas-grain network to include the deuteration, without considering the ortho form of H$_2$. The main point of this work is to identify the key reactions for the deuteration in the various conditions. For physical conditions similar to hot corinos, they found an abundance ratio between $10^{-3}$ and $10^{-2}$ (see their Table 19), which is smaller compared to our predictions. The difference is very likely due to the different physical conditions used in both models. 
\citet{2013A&A...550A.127T} did also a parameter space study of the deuteration of water using a two phase model (a chemically inactive bulk and a surface layer). Their main conclusion is that the HDO/H$_2$O and D$_2$O/H$_2$O abundance ratios are highly dependent in the density and the visual extinction (between 2 and 5) of the medium. We do find similar results. To quantitatively compared their results to ours, we have used the foreground model for a visual extinction of 4, a temperature of 15~K, a cosmic-ray ionization rate of $1.3\times 10^{-17}$~s$^{-1}$, a C/O ratio of 0.5 and a density of $2\times 10^5$~cm$^{-3}$ and compared with the Fig. 8 from \citeauthor{2013A&A...550A.127T} (for a temperature of 10~K and a density of $10^5$~cm$^{-3}$). The HDO/H$_2$O and D$_2$O/H$_2$O abundance ratios in the ices of our model presents the same shape as a function of time and reaches a steady-state value at approximately the same age. In our case, the maximum HDO/H$_2$O value is 0.065 and the one of D$_2$O/H$_2$O is $10^{-3}$. Those values agree quite well with the ones predicted by \citeauthor{2013A&A...550A.127T} We do also predict that the abundance ratios {\bf increase} with the density. The same comparison can be made with the Fig. 11 from \citet{2013A&A...554A..92S}. Despite the fact that they do not reach a steady-state value, we obtain HDO/H$_2$O abundance ratios in the ices that are similar at $10^7$~yr. In their model, \citet{2013A&A...554A..92S} considered ortho and para forms of H$_2$ and a time dependent o/p ratio. This may explain the differences.

\section{Conclusions}

In this paper, we have done a modeling of the chemistry towards the protostellar envelope of the low-mass protostar IRAS16293-2422 using a gas-grain chemical model (Nautilus), a deuterated gas-grain network and a dynamical model of the protostellar collapse. The model has been used to reproduce the observations of H$_2$O, HDO and D$_2$O towards this source using ground based telescopes and the HIFI spectrometer on board the Herschel telescope observed by \citet{2012A&A...539A.132C,2013A&A...553A..75C}. Part of the emission of molecules was identified to come from a foreground cloud in addition to the protostellar envelope. Using a multi-parameter study of the chemistry, we try to put some constraints on the physical parameters of this foreground cloud as well as the protostellar envelope. The following conclusions can be drawn.\\
For the central part of the protostellar envelope:\\
\begin{itemize}
\item[-] In a 1D model, the abundance ratios HDO/H$_2$O, D$_2$O/HDO and D$_2$O/H$_2$O are predicted to be constant in the inner 100~AU even though the full evaporation of the mantle ices only occurs for radii smaller than 60~AU.
\item[-] The H$_2$O, HDO and D$_2$O observed abundances are strongly overproduced by the model in the inner region of the protostellar envelope. The reason may be the presence of a more complex structure within the inner 100~AU. Future investigations with 2D or 3D physical structure, both for the analysis of the observations and the chemical model will be needed. 
\item[-] The observed abundance ratios HDO/H$_2$O, D$_2$O/HDO and D$_2$O/H$_2$O are better reproduced by the model using a cosmic-ray ionization rate of $10^{-16}$~s$^{-1}$ and a C/O elemental ratio of 0.5. 
\end{itemize}

For the outer part of the protostellar envelope:
\begin{itemize}
\item[-] The observed abundances of H$_2$O, HDO and D$_2$O and abundance ratios HDO/H$_2$O, D$_2$O/HDO and D$_2$O/H$_2$O are not well reproduced by our model in the outer part of the protostellar envelope whatever the parameters used. Abundance ratios are systematically overproduced by the models. One has to note the difference between the chaotic abundance profiles profiles predicted by the model, which are very different from the jump model used for the analysis of the observations. In addition, it is possible that part of the emission that was attributed to the foreground cloud indeed come from the outer part of the envelope. In this case, the observed abundances and D/H ratios could be larger than the ones used for this comparison.
\item[-] Both for the inner and outer regions of the protostellar envelope, the age of the cloud prior collapse does not seem to play a major role for the H$_2$O, HDO and D$_2$O abundances.
\end{itemize}

For the foreground cloud:
\begin{itemize}
\item[-] To reproduce the observation of water and deuterated water in this foreground cloud, the temperature needs to be small (about 15~K) and the visual extinction larger or equal to 3. Within these limits, several C/O ratios and cosmic-ray ionization rates can reproduce the observations at a satisfactory level. If the C/O ratio is 0.5, the cloud density may be large (of about $2\times 10^5$~cm$^{-3}$) whereas if the C/O ratio is 1.2, the observations constrain the cosmic-ray ionization rate to be large (of about $10^{-16}$~s$^{-1}$). 
\item[-] As for the CH modeling from \citet{2014MNRAS.441.1964B}, the larger densities shows a better agreement for earlier times. The "best" times are however not consistent with the present constraints and are ten times larger for deuterated water. For a density of $2\times 10^5$~cm$^{-3}$ for instance, CH observed abundances were reproduced around $10^4$~yr whereas H$_2$O, HDO and D$_2$O are reproduced around $2\times 10^5$~yr in the present study. Better constraints on the physical properties of this foreground cloud are clearly needed. This could be achieved by studying other molecules that could trace this layer of gas using the same method.  
\item[-] This foreground cloud was baptized "photodesorption layer" by \citet{2013A&A...553A..75C} arguing that considering the small A$_V$ the photo-evaporation of ices may control the gas-phase abundances of these species. We have shown that, in our model, it is the exothermicity of the reactions at the surface of the grains that would release H$_2$O, HDO and D$_2$O during their formation at the surface rather than the photo-evaporation.
\item[-] Although some of our models show a good agreement with the observations, the model always found a D$_2$O/HDO abundance ratio smaller than the HDO/H$_2$O one, contrary to the observations. Indeed, the observations suggest that the conversion of HDO into D$_2$O is more efficient than the conversion of H$_2$O into HDO or that D$_2$O would be more efficiently evaporated than HDO.
\end{itemize}

Finally, we have neglected the importance of the ortho form of H$_2$ in this model. Further works are needed to check its importance for the formation of water and deuterated water under the conditions considered for the different layers in the line of sight of IRAS16293.

\section*{Acknowledgments}

VW thanks the following funding agencies for their partial support of this work: the French CNRS/INSU program PCMI and the ERC Starting Grant (3DICE, grant agreement 336474). VW is grateful to Prof. Eric Herbst, George Hassel and Franck Hersant for fruitful discussions on photo-evaporation processes. We would like to thank the referee for providing useful comments that helped us improve the content of the paper.

\bibliographystyle{mn2e}
\nocite{*}
\bibliography{bib}

\appendix
\section{Abundances predicted in the ices in the foreground cloud}

\begin{figure*}
\includegraphics[width=0.4\linewidth]{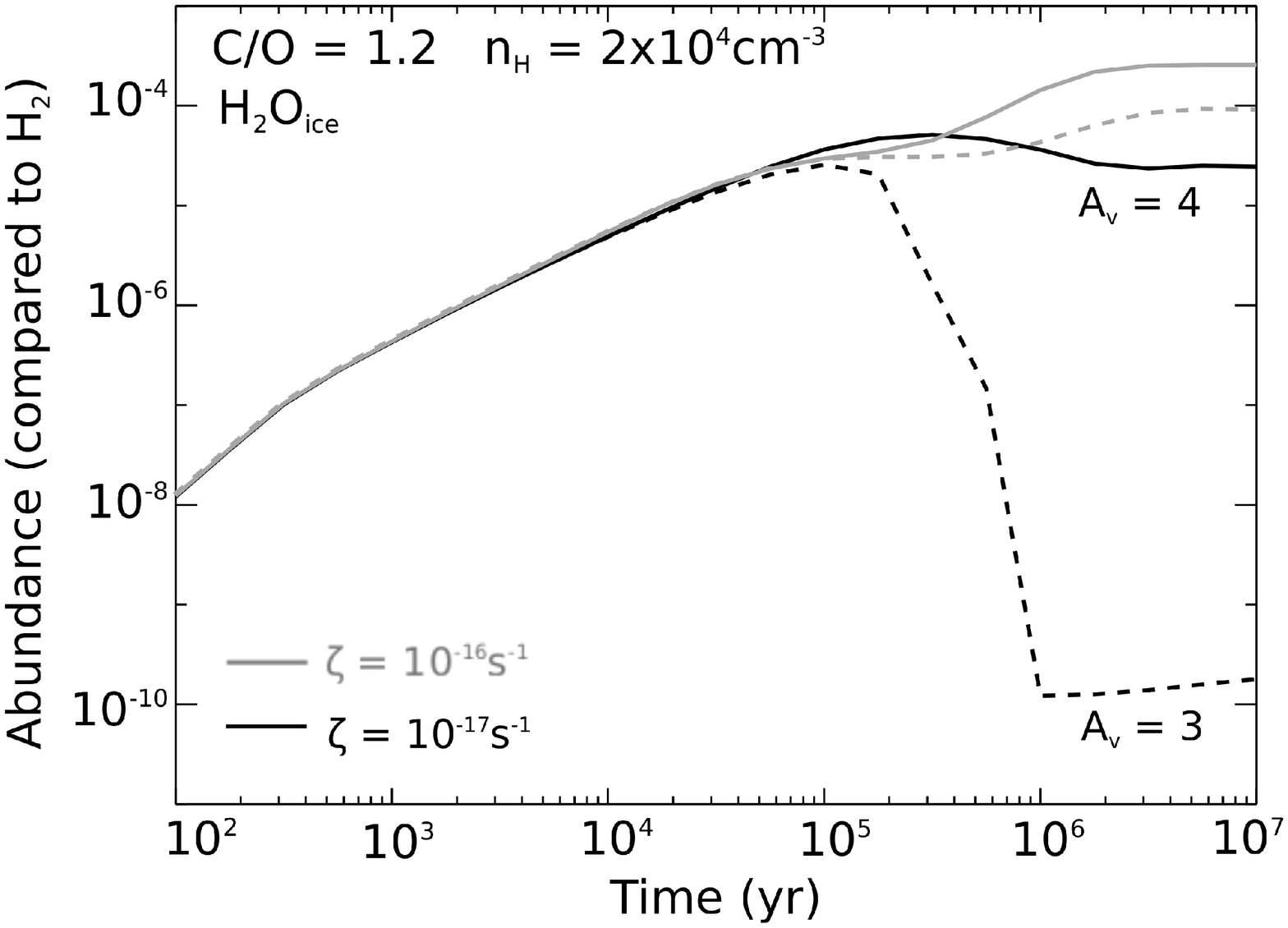}
\includegraphics[width=0.4\linewidth]{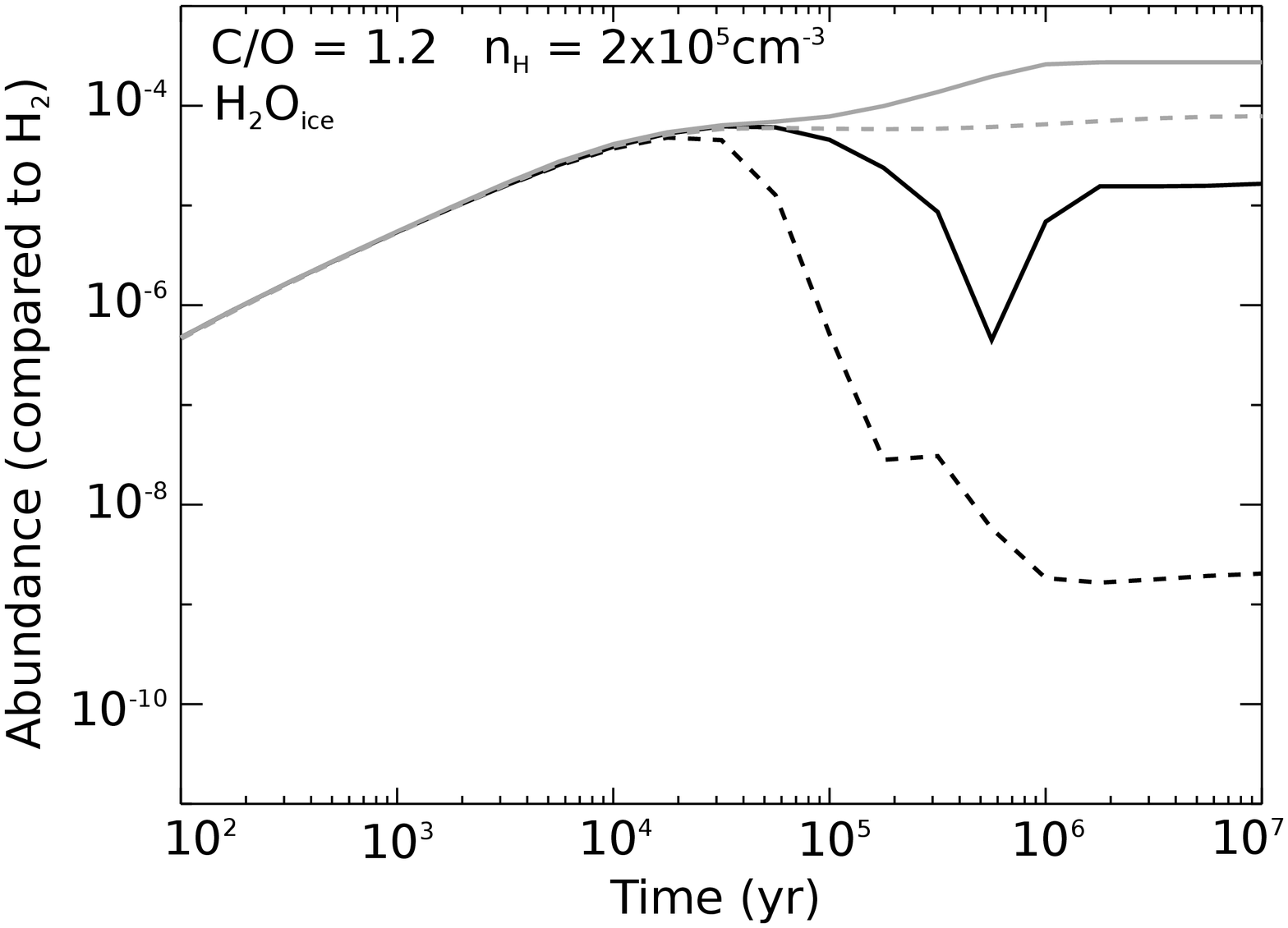}
\includegraphics[width=0.4\linewidth]{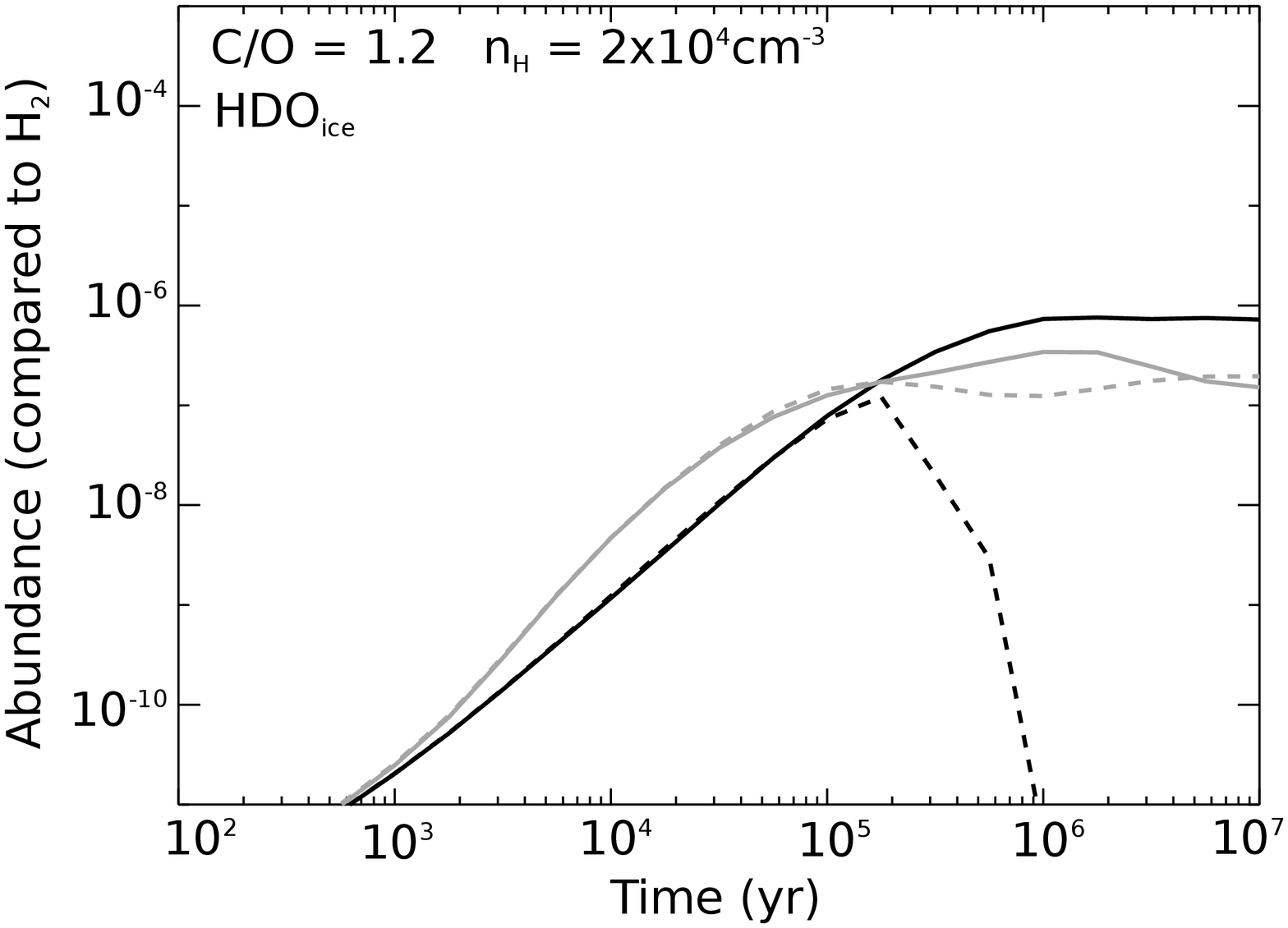}
\includegraphics[width=0.4\linewidth]{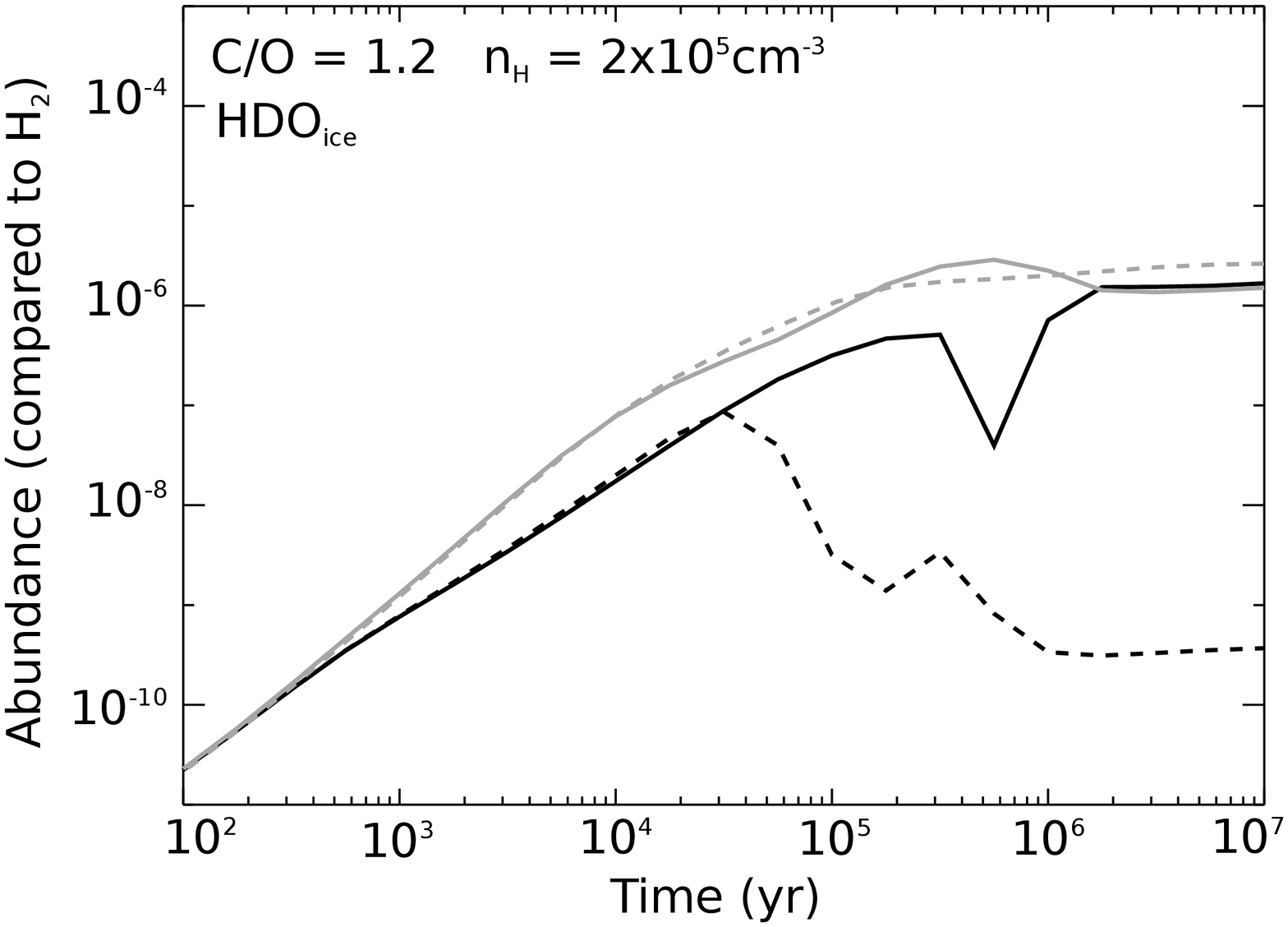}
\includegraphics[width=0.4\linewidth]{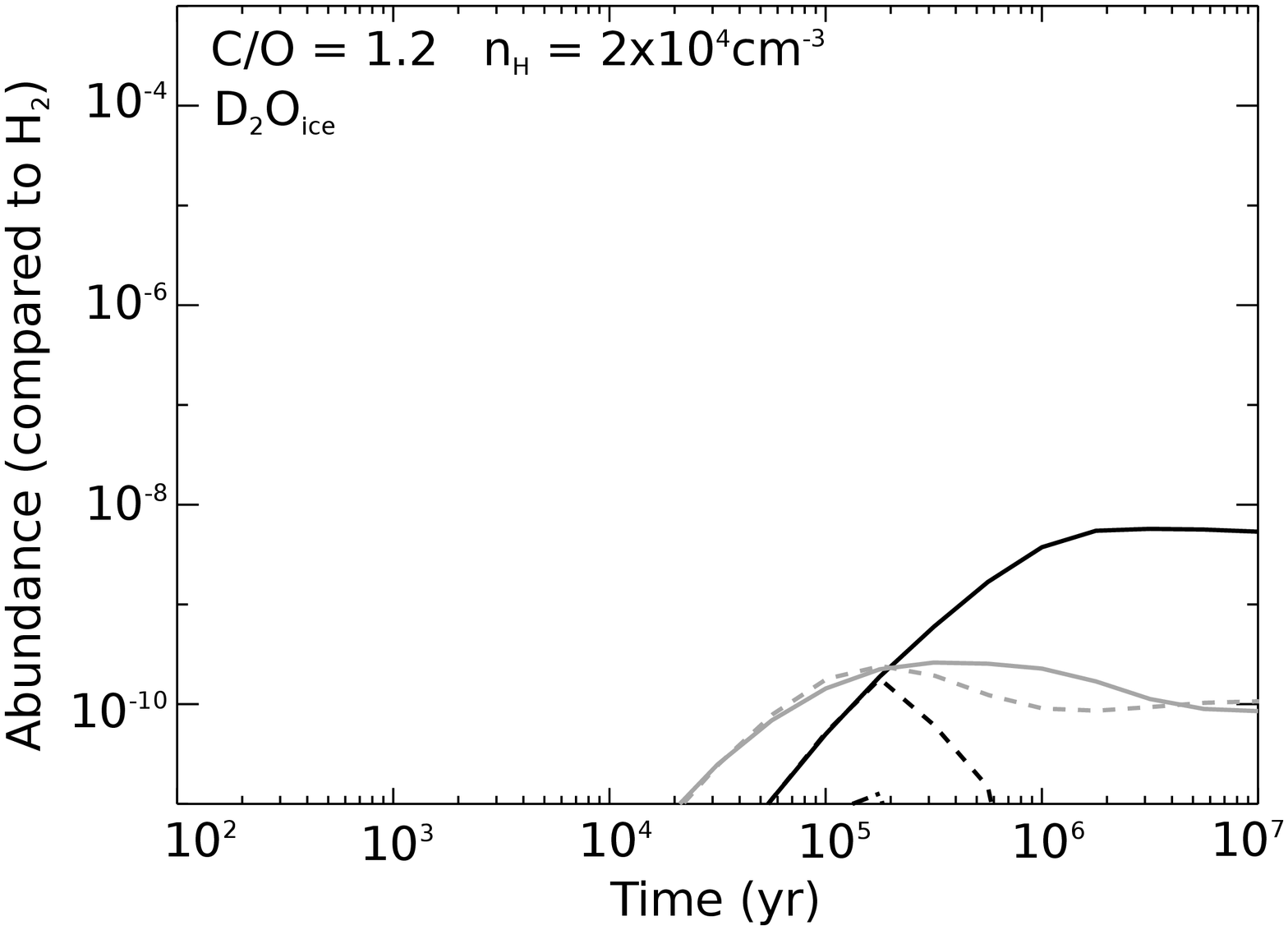}
\includegraphics[width=0.4\linewidth]{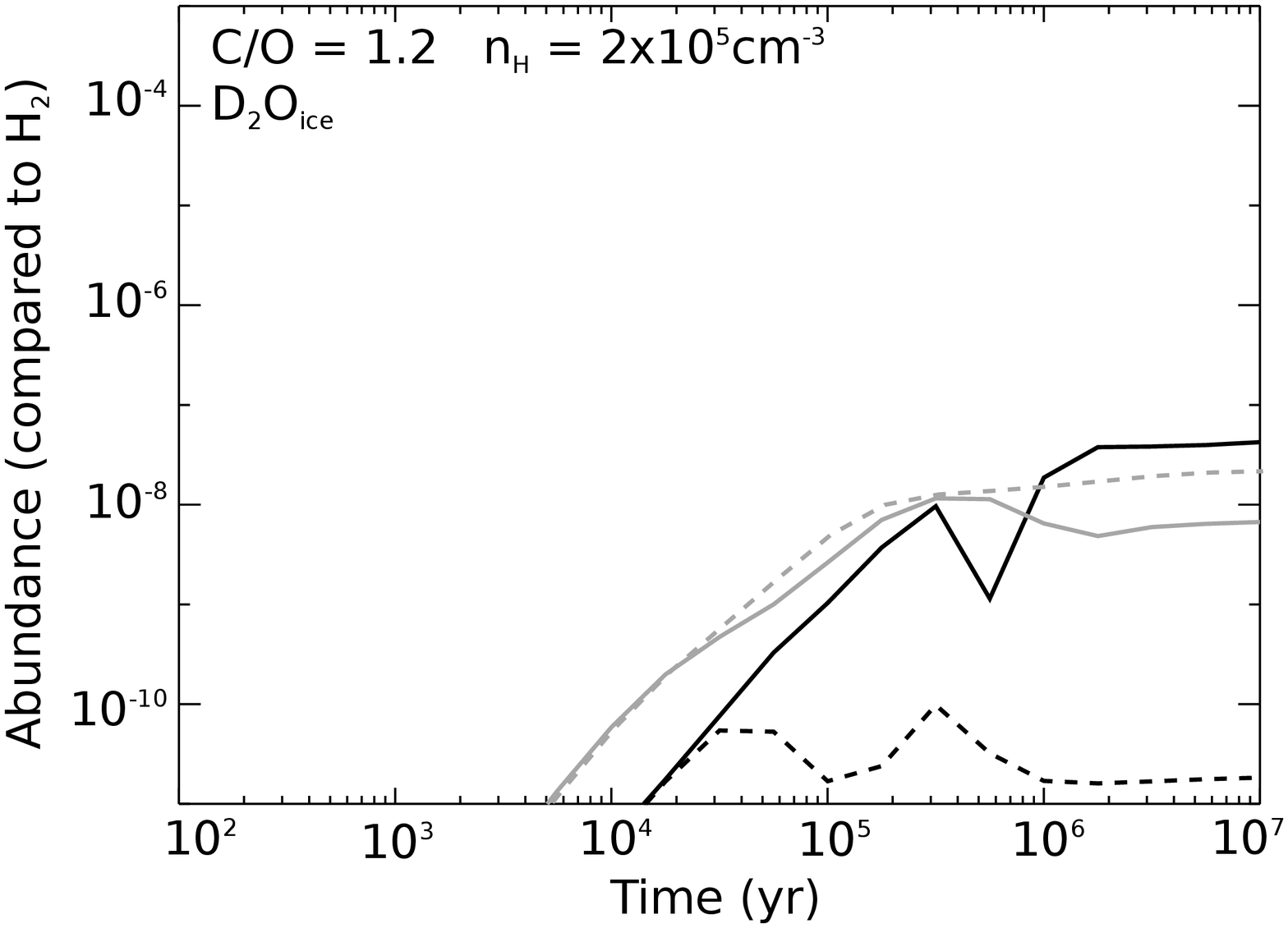}
\caption{Abundances of H$_2$O, HDO and D$_2$O in the ices as a function of time predicted by the model for the foreground cloud.  Gas and dust temperature is the same for all models: 15~K. The elemental C/O abundance ratio is 1.2. The left side of the model has been obtained for a total H density of $2\times 10^4$~cm$^{-3}$ and the right side for a total H density of $2\times 10^5$~cm$^{-3}$.  Black curves have been obtained for a cosmic-ray ionization rate $\zeta$ of $10^{-17}$~s$^{-1}$ whereas grey curves have been obtained for a $\zeta$ of  $10^{-16}$~s$^{-1}$. Solid lines represent an A$_V$ of 4  and dashed lines an A$_V$ of 3.  \label{cloud_C_O_1.2_ice}}
\end{figure*}

\begin{figure*}
\includegraphics[width=0.4\linewidth]{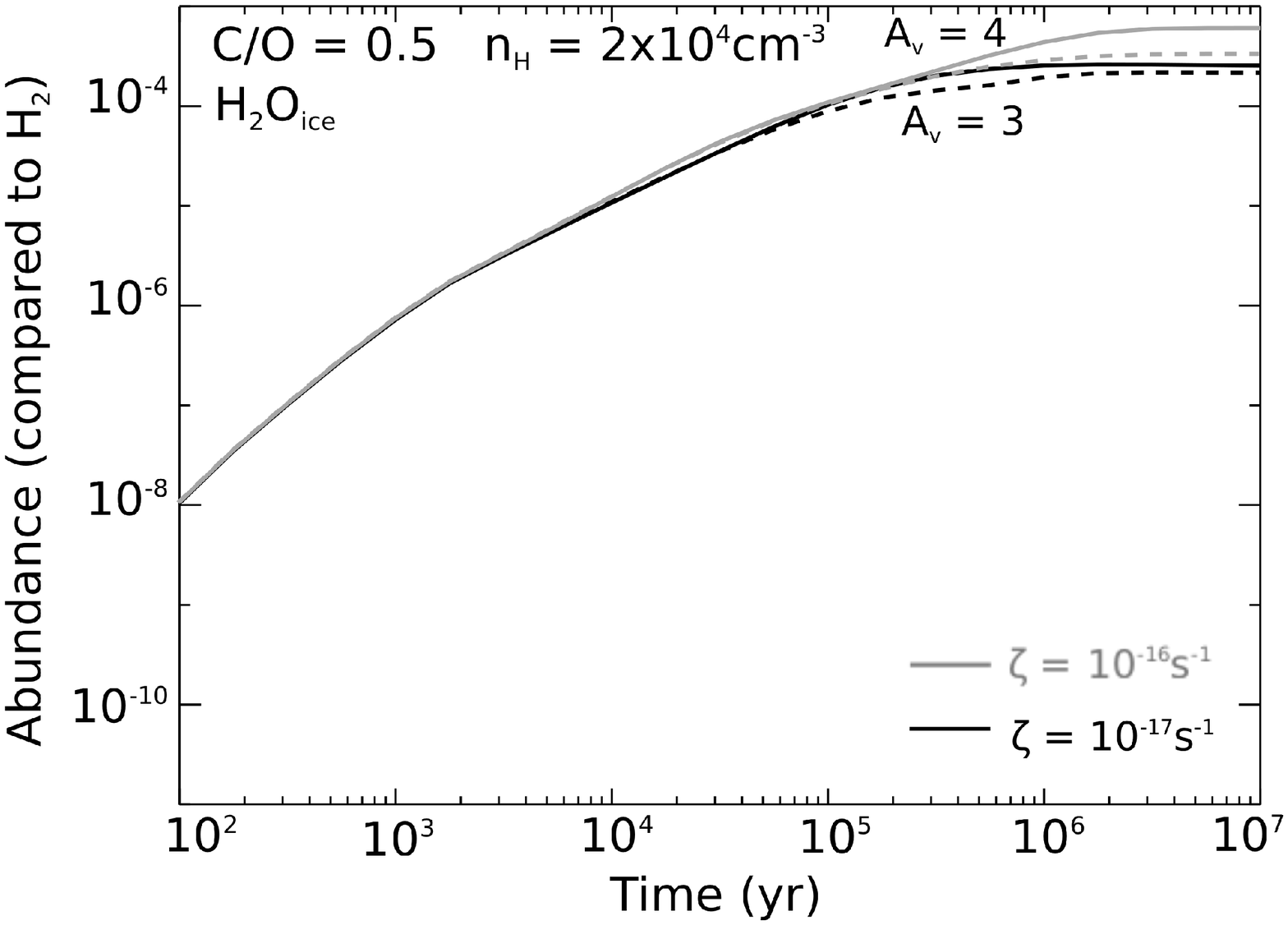}
\includegraphics[width=0.4\linewidth]{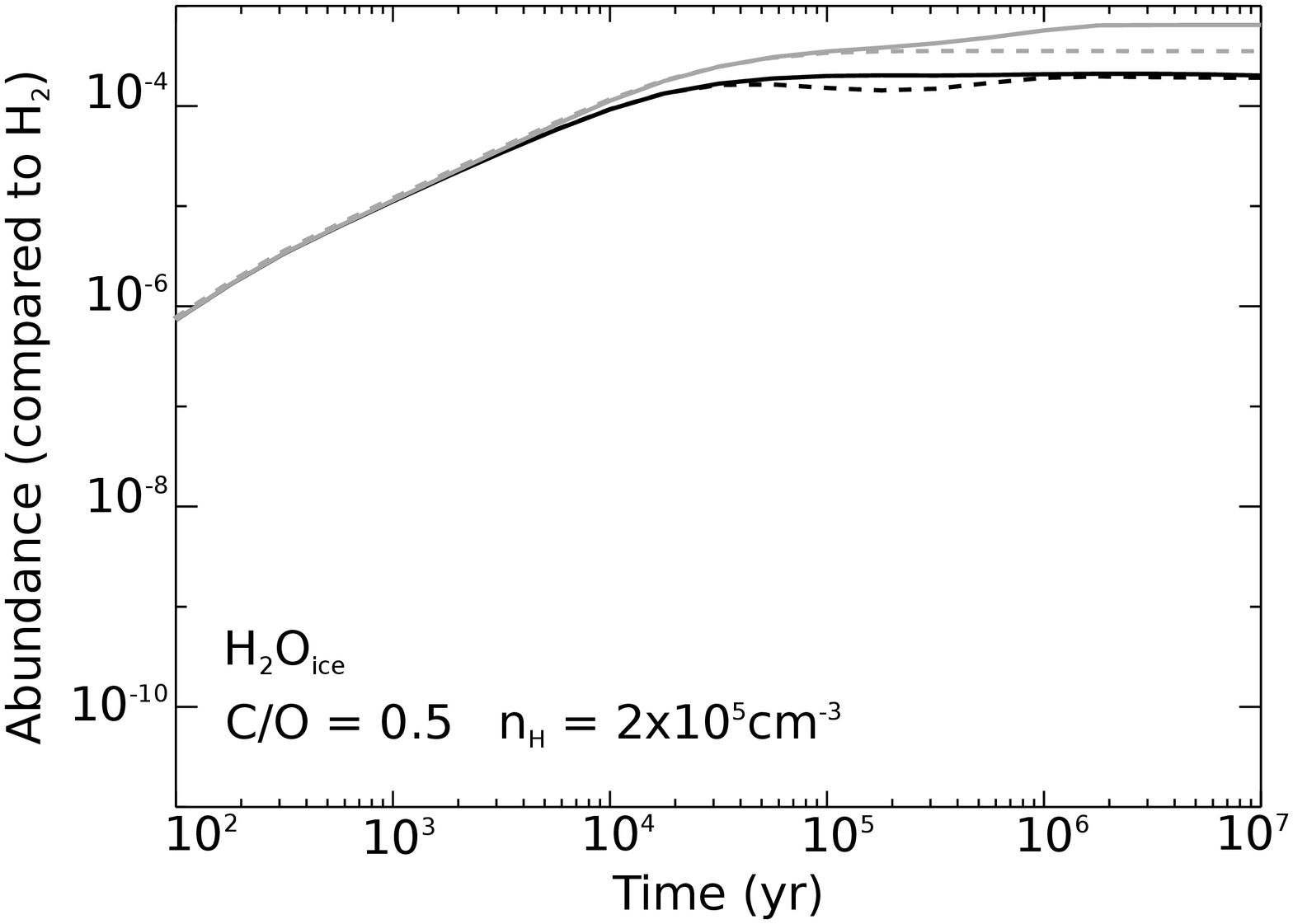}
\includegraphics[width=0.4\linewidth]{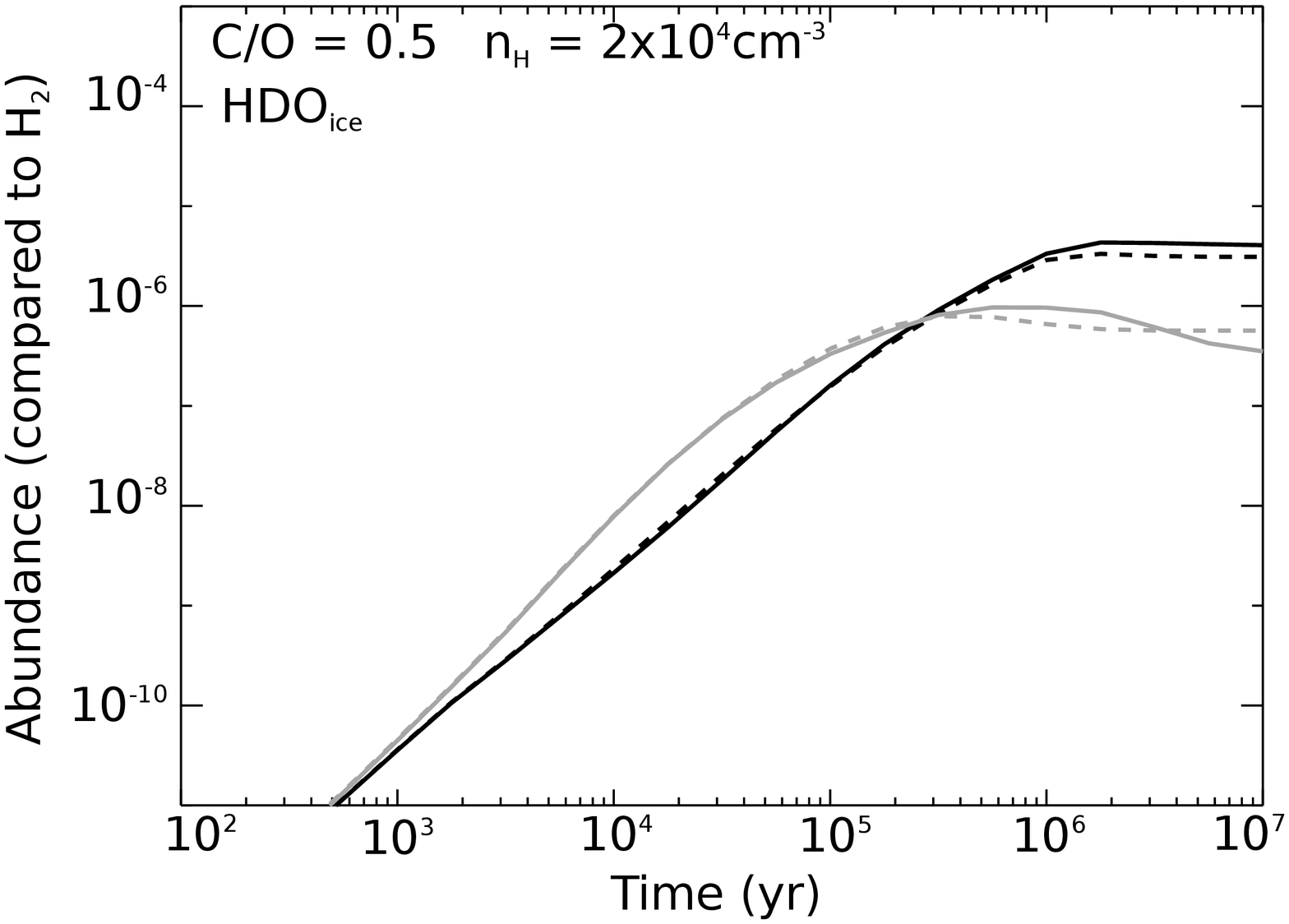}
\includegraphics[width=0.4\linewidth]{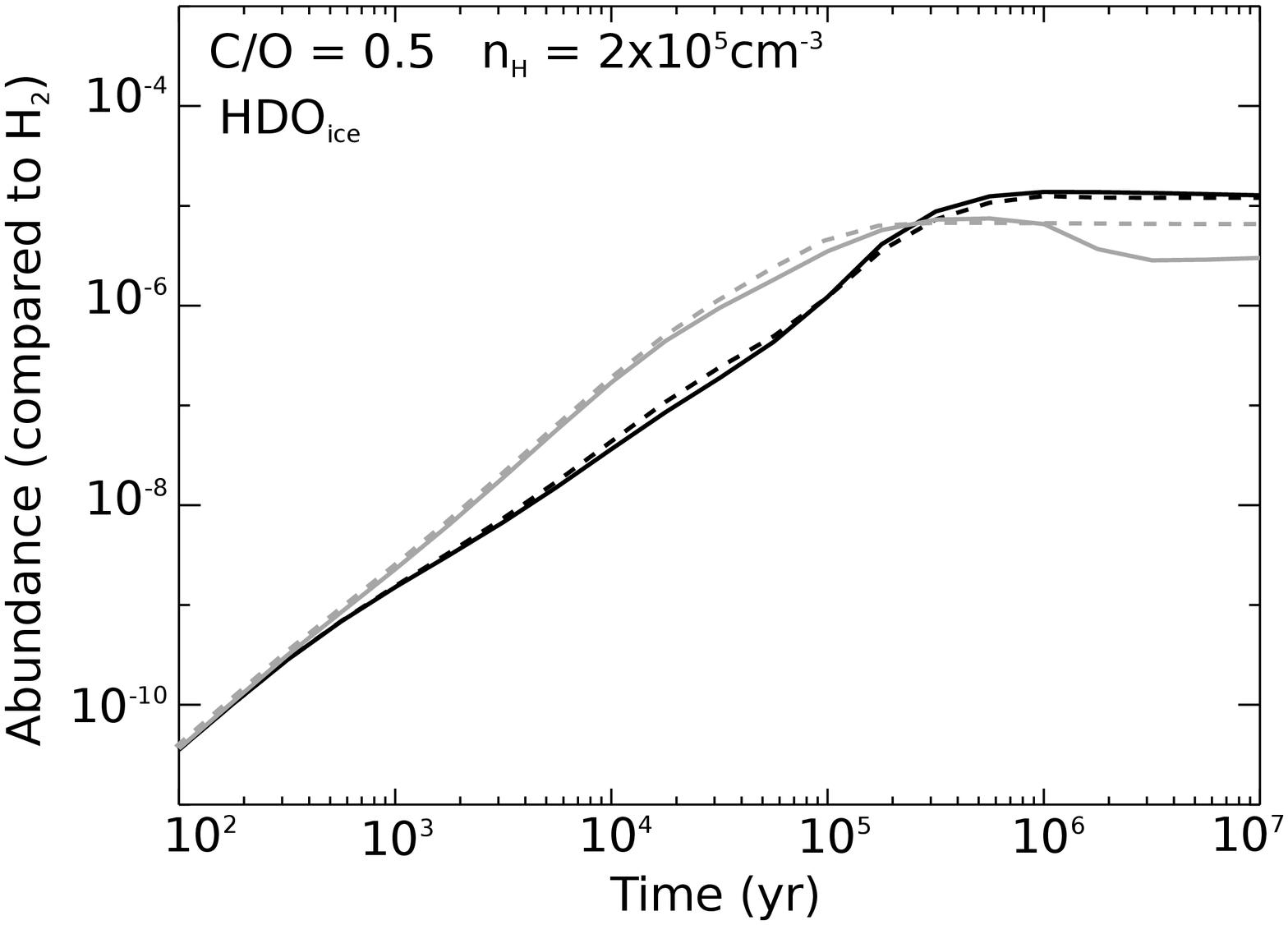}
\includegraphics[width=0.4\linewidth]{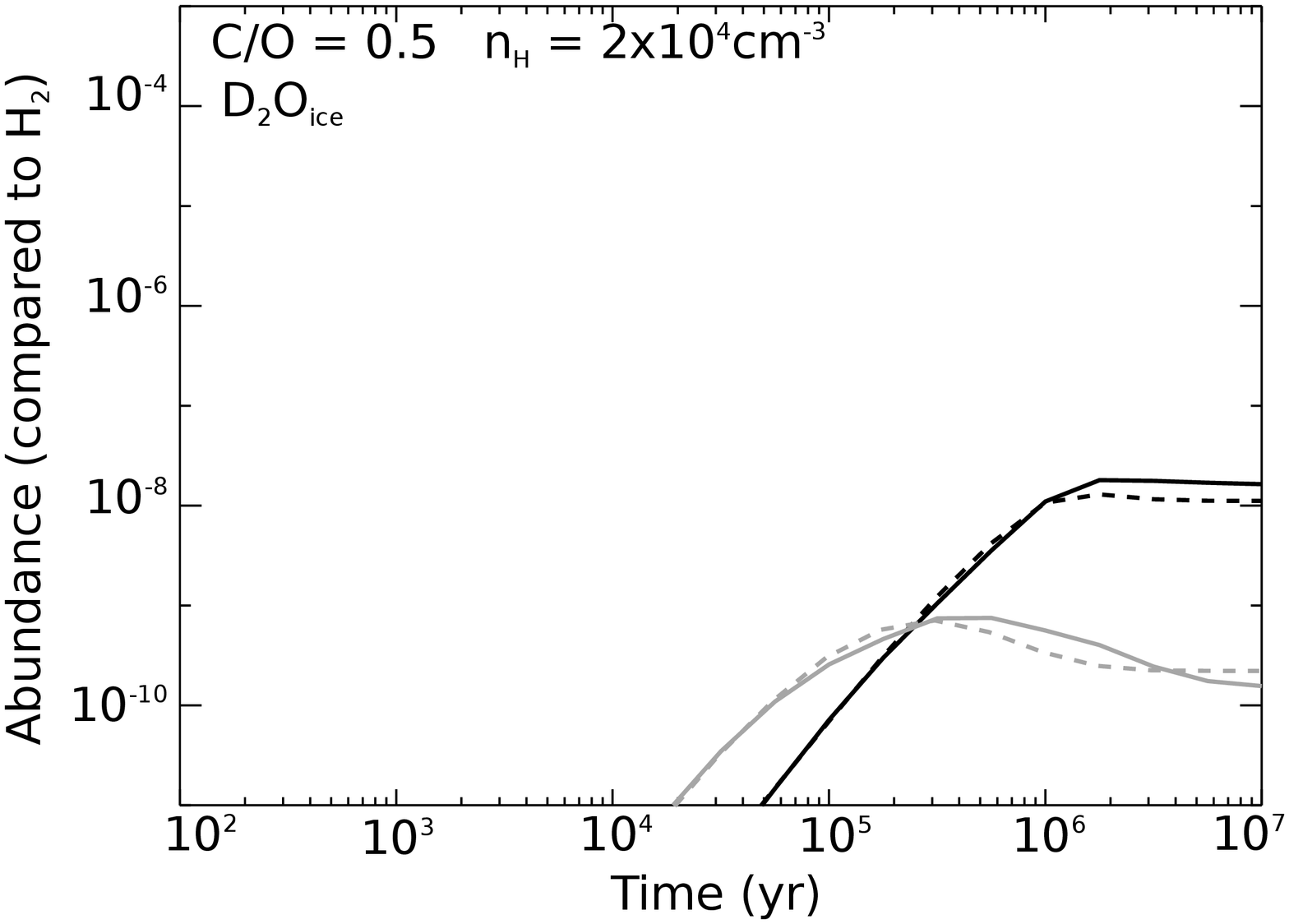}
\includegraphics[width=0.4\linewidth]{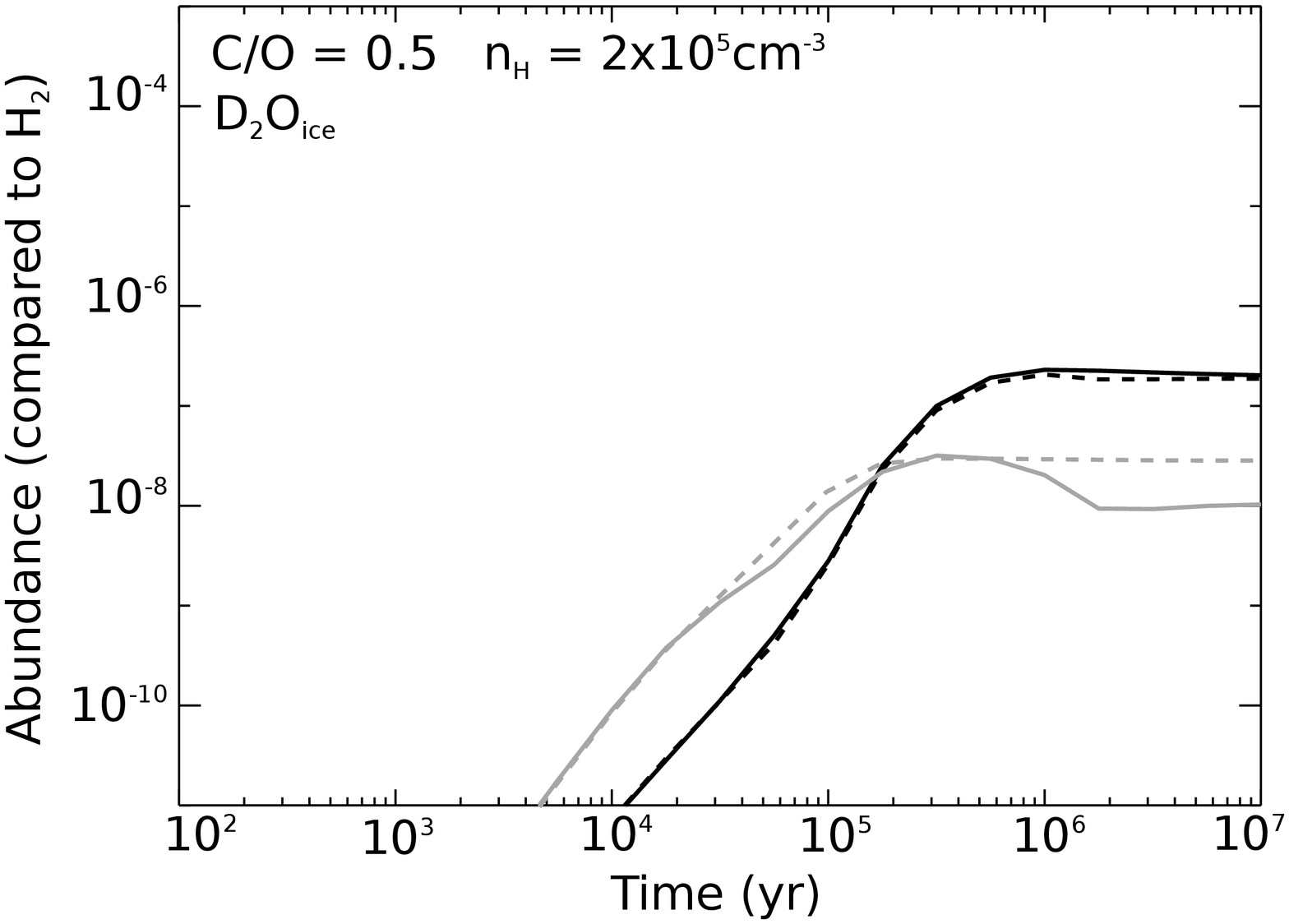}
\caption{Same as Fig.~\ref{cloud_C_O_1.2_ice} but a C/O of 0.5.  \label{cloud_C_O_0.5_ice}}
\end{figure*}

\section{Abundances predicted with different values of the rate coefficient of the reaction H$_2$D$^+$ + H$_2$ $\rightarrow$ H$_3^+$ + HD}

\begin{figure*}
\includegraphics[width=0.5\linewidth]{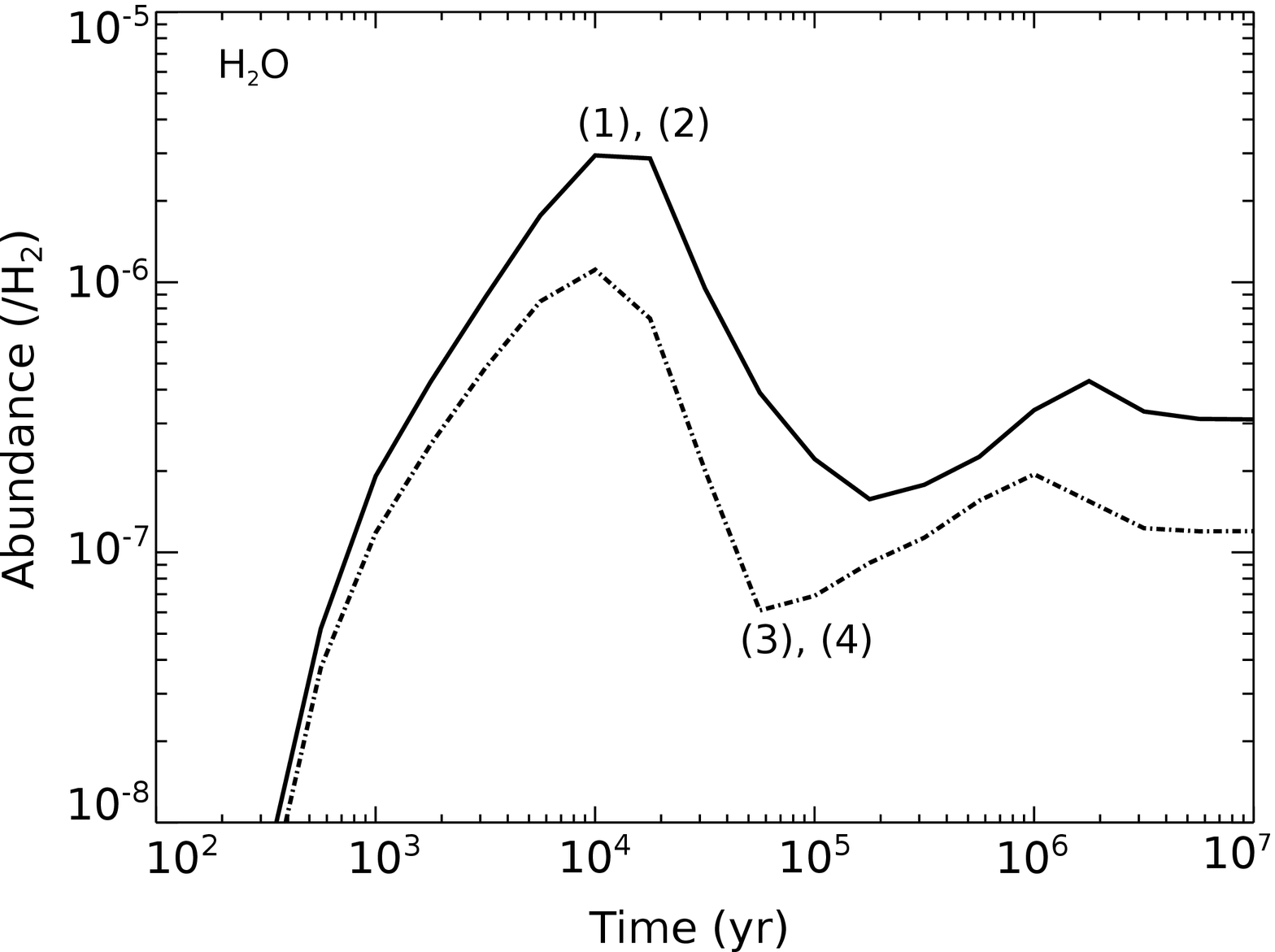}
\includegraphics[width=0.5\linewidth]{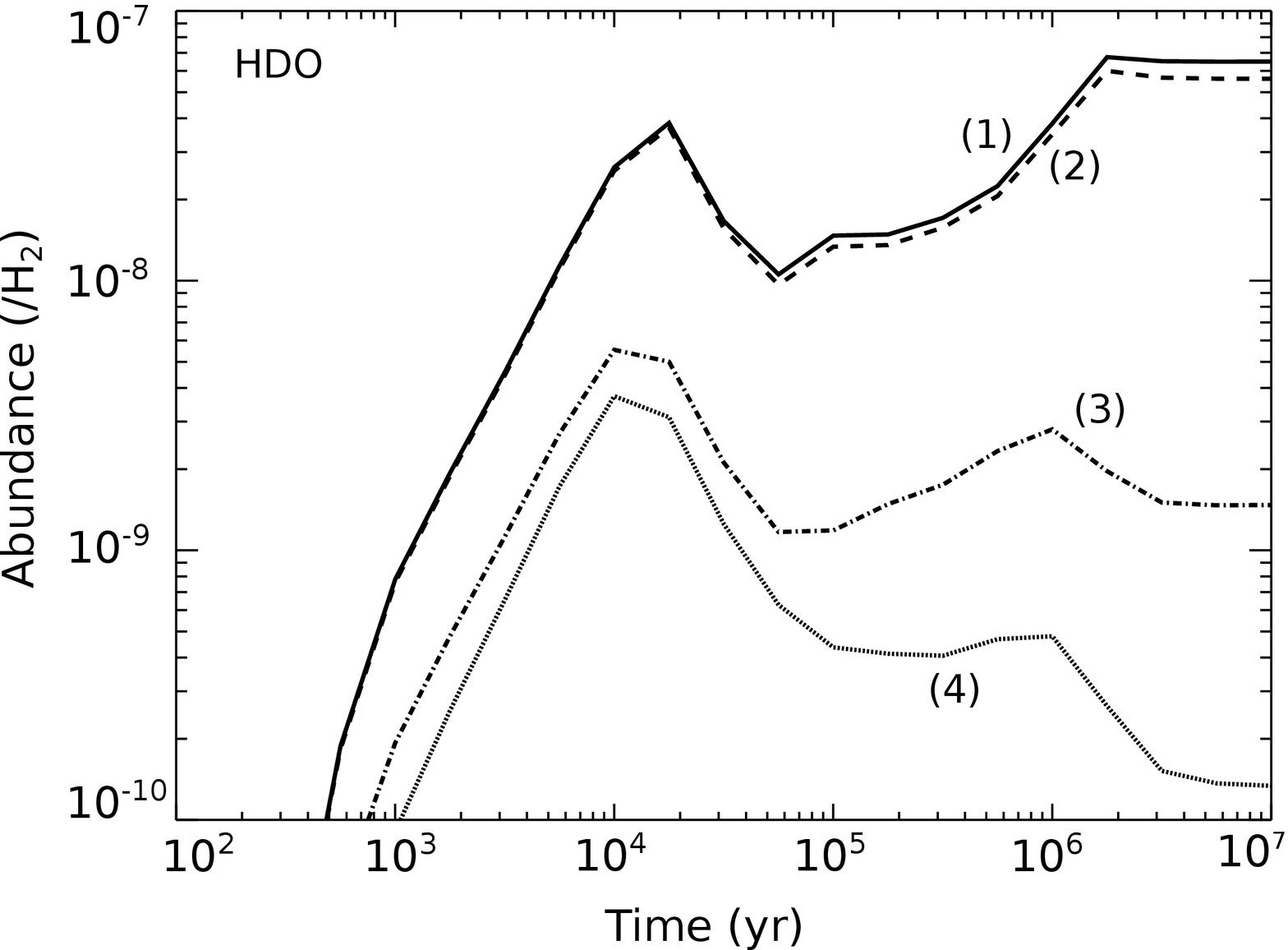}
\includegraphics[width=0.5\linewidth]{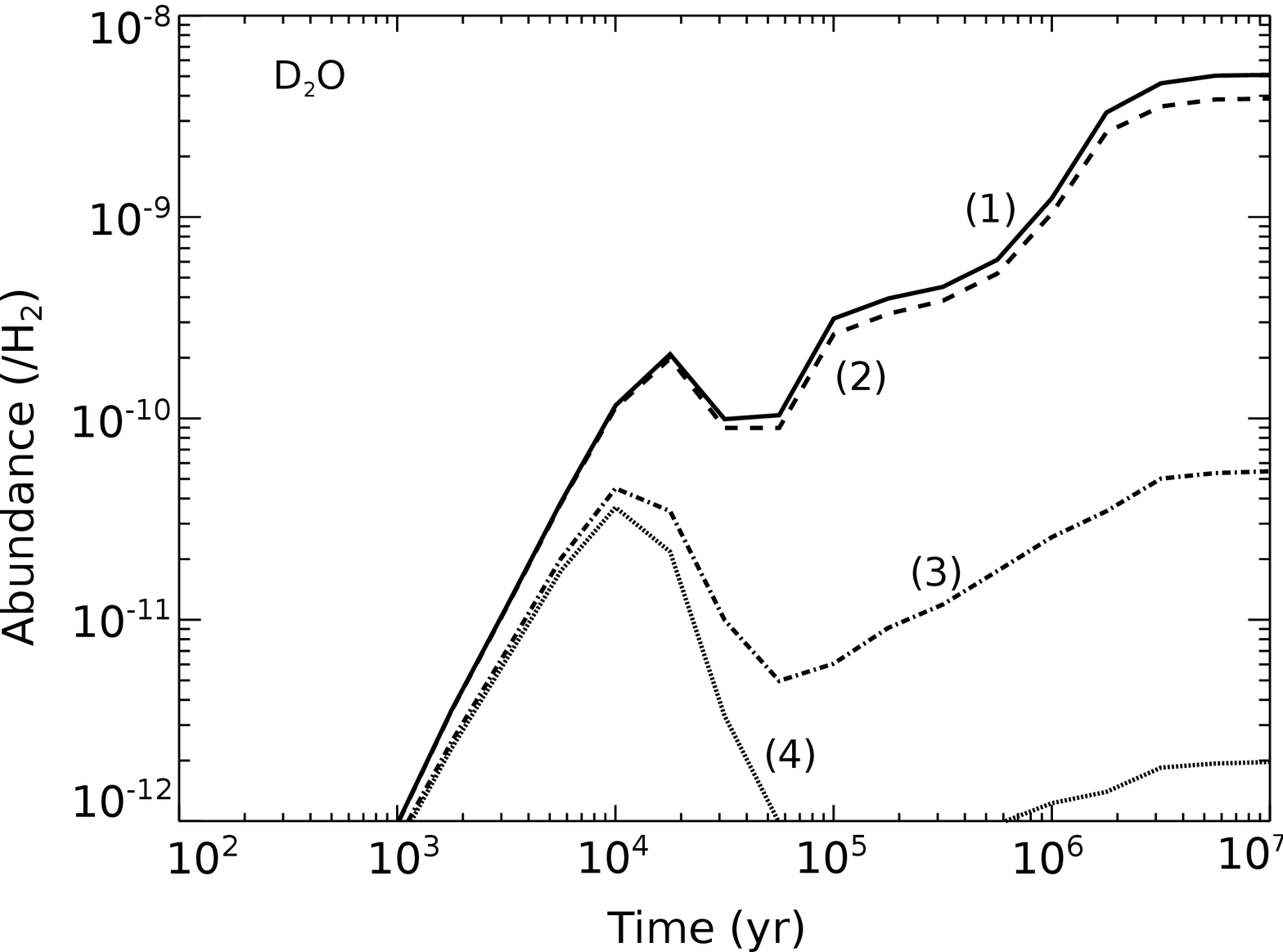}
\caption{Abundances of H$_2$O, HDO and D$_2$O as a function of time, computed in the foreground cloud for different values of the rate coefficient of the reaction H$_2$D$^+$ + H$_2$ $\rightarrow$ H$_3^+$ + HD mimicking the o/p H$_2$ ratio of $10^{-3}$ (curves 2), 0.1 (curves 3) and 3 (curves 4). The reference model, without considering the ortho form of H$_2$ is shown by curves 1. For these models, the gas and dust temperatures are 15~K, the density is $2\times 10^5$~cm$^{-3}$, the visual extinction is 4, the C/O elemental ratio is 0.5, and the cosmic-ray ionization rate is $10^{-16}$~s$^{-1}$. \label{opfigs}}
\end{figure*}

\begin{figure*}
\includegraphics[width=0.5\linewidth]{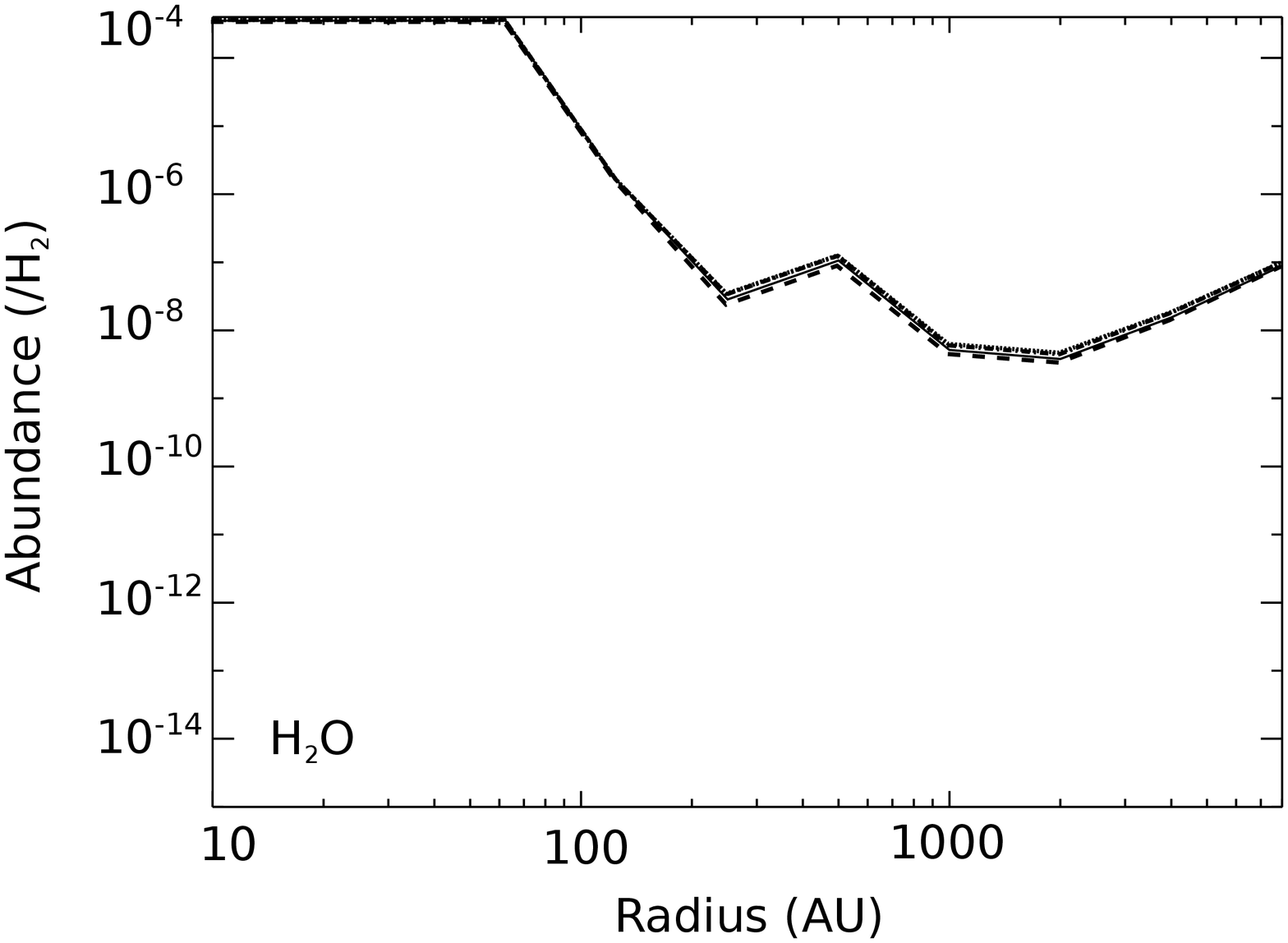}
\includegraphics[width=0.5\linewidth]{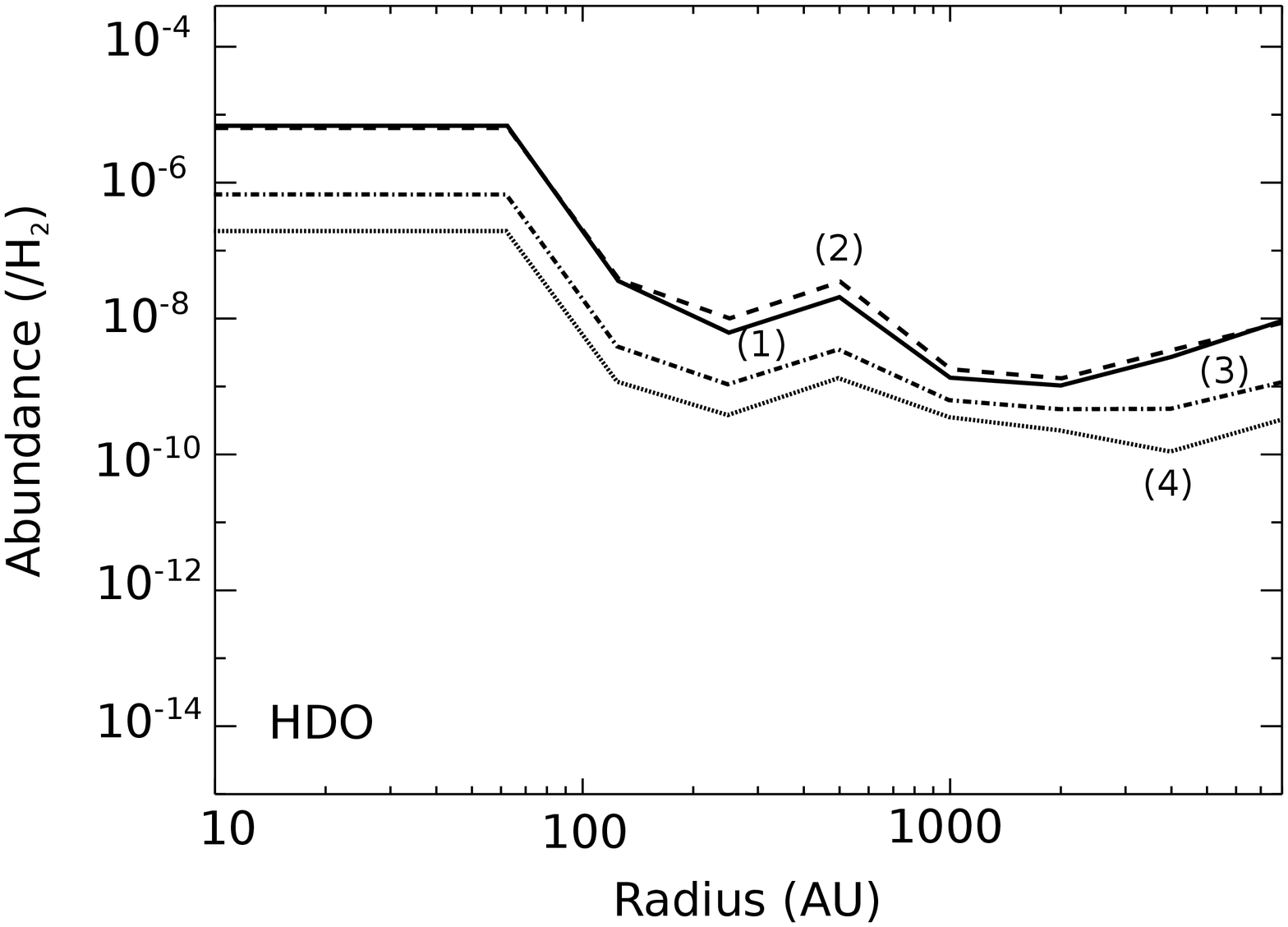}
\includegraphics[width=0.5\linewidth]{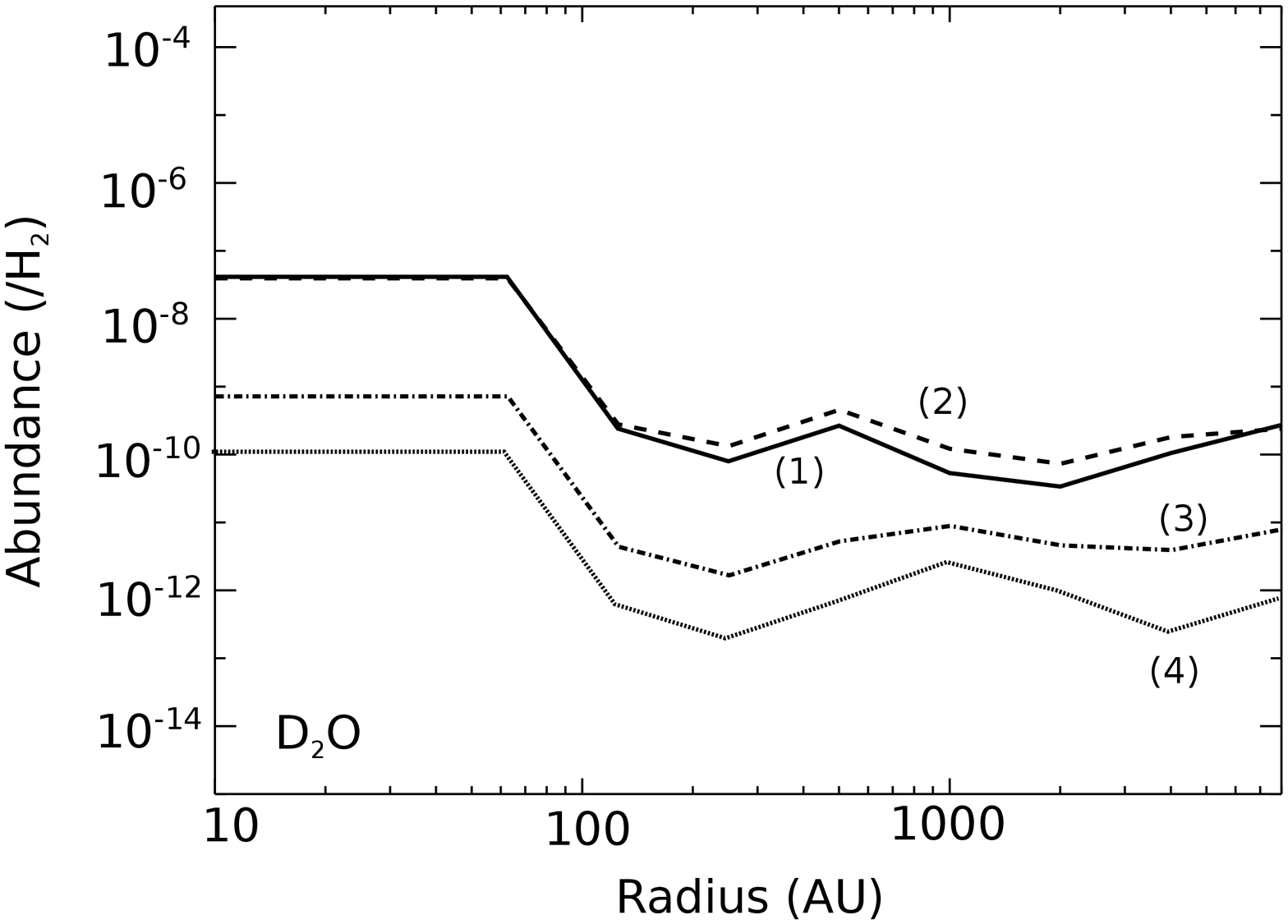}
\caption{Abundances of H$_2$O, HDO and D$_2$O as a function of radius in the protostellar envelope, computed for different values of the rate coefficient of the reaction H$_2$D$^+$ + H$_2$ $\rightarrow$ H$_3^+$ + HD mimicking the o/p H$_2$ ratio of $10^{-3}$ (curves 2), 0.1 (curves 3) and 3 (curves 4). The reference model, without considering the ortho form of H$_2$ is shown by curves 1. For these models, the C/O elemental ratio is 0.5 and the cosmic-ray ionization rate is $10^{-16}$~s$^{-1}$. \label{opfigs_proto}}
\end{figure*}

\end{document}